\documentclass[12pt]{article}

\usepackage{scicite}
\usepackage[top=1.40cm,left=1.40cm,right=1.40cm,bottom=1.40cm,headsep=0pt]{geometry}
\usepackage{times}
\usepackage{color}
\usepackage{amsmath}
\usepackage{amssymb}
\usepackage{bm}
\usepackage{graphicx}
\usepackage{chemarr}
\usepackage{booktabs}
\usepackage{array} 
\usepackage{multirow}
\usepackage{setspace}
\usepackage{enumitem}
\usepackage[labelfont=bf,labelsep=period,justification=justified,font=small,skip=0pt,singlelinecheck=off]{caption}
\usepackage[labelfont=bf,font=normalsize,skip=0pt,singlelinecheck=on]{subcaption}
\usepackage{titlesec}
\usepackage{ulem}
\usepackage{titling}
\usepackage{xurl}
\usepackage{etoolbox}

\newcommand{\ie}{i.e.,\ }
\newcommand{\eg}{e.g.,\ }
\newcommand{\cf}{cf.\ }
\newcommand{\myfigure}[2][0.8]{\includegraphics[width=#1\textwidth,height=#1\textheight,keepaspectratio]{#2}}

\newcommand{\splitatcommas}[1]{%
  \begingroup
  \begingroup\lccode`~=`, \lowercase{\endgroup
    \edef~{\mathchar\the\mathcode`, \penalty0 \noexpand\hspace{0pt plus .1em}}%
  }\mathcode`,="8000 #1%
  \endgroup
}

\newcommand{\bfemph}[1]{\textbf{\textit{#1}}}
\renewcommand{\emph}[1]{\bfemph{#1}}

\definecolor{DarkBlue}{RGB}{0,0,0}
\definecolor{DarkRed}{RGB}{189,22,44}

\newcommand{\hltext}[1]{\textcolor{DarkBlue}{#1}}
\newcommand{\rtwotext}[1]{{\textcolor{black}{#1}}}

\titlespacing*{\section}{0pt}{1.5ex plus 0.5ex minus .5ex}{0.3em} 
\titlespacing*{\subsection}{0pt}{1.5ex plus 0.5ex minus .5ex}{0.3em}  
\titlespacing*{\paragraph}{0pt}{1.5ex plus 0.5ex minus .5ex}{0.3em}  
\titlespacing*{\subsubsection}{0pt}{1.5ex plus 0.5ex minus .5ex}{0.3em}  

\BeforeBeginEnvironment{figure}{\vskip-1ex}
\AfterEndEnvironment{figure}{\vskip-2ex}

\newlength{\bibitemsep}
\newlength{\bibparskip}
\let\oldthebibliography\thebibliography
\renewcommand\thebibliography[1]{%
  \oldthebibliography{#1}%
  \setlength{\parskip}{\bibitemsep}%
  \setlength{\itemsep}{\bibparskip}%
}

\graphicspath{{Figures_Paper/}}

\newenvironment{myquote}%
  {\list{}{\leftmargin=0.25in\rightmargin=0.25in}\item[]}%
  {\endlist}

\newenvironment{sciabstract}{%
\begin{myquote} \bf}
{\end{myquote}}

\title{\Large\bf Analysis of sloppiness in model simulations: unveiling parameter uncertainty when mathematical models are fitted to data \vspace{-10pt}}

\author
{Gloria M. Monsalve-Bravo,$^{1,2,3,\ast}$ Brodie A. J. Lawson,$^{4,5,6,7}$   Christopher   Drovandi,$^{4,5,6}$  \\  Kevin Burrage,$^{5,6,7,8}$ Kevin S. Brown,$^{9,10}$ Christopher M. Baker,$^{11,12,13}$ \\ Sarah A. Vollert,$^{4,5,6}$ Kerrie  Mengersen,$^{4,5,6}$     Eve McDonald-Madden,$^{1,2,\dagger}$ \\ Matthew P. Adams$^{3,4,5,6,\dagger}$
\\
\normalsize{$^{1}$School of Earth and Environmental Sciences,}\\
\normalsize{The University of Queensland, St Lucia, QLD 4072, Australia}\\
\normalsize{$^{2}$Centre for Biodiversity and Conservation Science,}\\
\normalsize{The University of Queensland, St Lucia, QLD 4072, Australia}\\
\normalsize{$^{3}$School of Chemical Engineering, The University of Queensland, St Lucia, QLD 4072, Australia}\\
\normalsize{$^{4}$ Centre for Data Science, Queensland University of Technology, Brisbane, QLD 4001, Australia}\\
\normalsize{$^{5}$ARC Centre of Excellence for Mathematical and Statistical Frontiers,}\\
\normalsize{Queensland University of Technology, Brisbane, QLD 4001, Australia}\\
\normalsize{$^{6}$School of Mathematical Sciences, Queensland University of Technology, Brisbane, QLD 4001, Australia}\\
\normalsize{$^{7}$ARC Centre of Excellence for Plant Success in Nature and Agriculture,}\\
\normalsize{Queensland University of Technology, Brisbane, QLD 4001, Australia}\\
\normalsize{$^{8}$Department of Computer Science, University of Oxford, Oxford OX1 3QD, United Kingdom}\\
\normalsize{$^{9}$Department of Pharmaceutical Sciences, Oregon State University, Corvallis, OR 97331, United States}\\
\normalsize{$^{10}$Department of Chemical, Biological, \& Environmental Engineering,}\\
\normalsize{Oregon State University, Corvallis, OR 97331, United States}\\
\normalsize{$^{11}$School of Mathematics and Statistics,The University of Melbourne, Parkville, VIC 3010, Australia}\\
\normalsize{$^{12}$Melbourne Centre for Data Science, The University of Melbourne, Parkville, VIC 3010, Australia}\\
\normalsize{$^{13}$Centre of Excellence for Biosecurity Risk Analysis,}\\
\normalsize{The University of Melbourne, Parkville, VIC 3010, Australia}\\
\normalsize{$^\ast$Corresponding author. E-mail: g.monsalvebravo@uq.edu.au (G.M.-B)}\\
\normalsize{$^\dagger$These authors contributed equally to this work}
}
\date{}

\begin{document} 


\baselineskip24pt
\renewcommand{\figurename}{Fig.}
\renewcommand\thesubfigure{\Alph{subfigure}}

\maketitle


\begin{sciabstract}\vspace{-42pt}
This work introduces a \hltext{comprehensive} approach to assess the sensitivity of model outputs to changes in parameter values, constrained by the combination of prior beliefs and data. This novel approach identifies stiff parameter combinations strongly affecting the quality of the model-data fit while simultaneously revealing which of these key parameter combinations are informed primarily by the data or are also substantively influenced by the priors. We focus on the very common context in complex systems where the amount and quality of data are low compared to the number of model parameters to be collectively estimated, and showcase the benefits of this technique for applications in biochemistry, ecology, and cardiac electrophysiology. We also show how stiff parameter combinations, once identified, uncover controlling mechanisms underlying the system being modeled and inform which of the model parameters need to be prioritized in future experiments for improved parameter inference from collective model-data fitting.
\end{sciabstract} \vspace{-5pt}

\noindent\textbf{Teaser:} Analysis of model sloppiness unveils key parameter combinations strongly influencing model outputs.

\section*{Introduction}

A single biological cell is itself a complex system, as is an organism made up of such cells, as is an ecosystem of those organisms interacting with one another. Despite the diversity of systems composing our world, many of these share similar structural and functional features that can be unraveled  through computer simulation \cite{Maayan2017,Geary2020,Villaverde2014}. As a result,  modeling and simulation has become increasingly important  to understand and predict underlying behavior of systems across different scales \cite{Villaverde2014,Mouquet2015,Drovandi2011}, including molecules \cite{Schlick2010}, cells \cite{Lawson2018,Johnstone2016}, engineered processes \cite{Velten2009}, through to astrophysical phenomena \cite{Sundberg2010}. Continuous advances in model descriptions of reality together with \hltext{the model fit} to experimental data have improved the fidelity of computer experiments and made them much more predictive \cite{Geary2020,Maayan2017}. However, the cost of this fidelity is an increase in the number of model parameters \cite{Mouquet2015}, and a greater risk that these parameters are difficult or impossible to be uniquely identified \cite{Gutenkunst2007,Brown2003,Brown2004,Villaverde2014}.  \hltext{For statistical models of familiar form, one may be able to formally determine how and to what extent parameters can possibly be identified. Lewbel~\cite{Lewbel2019} provides many such examples. When it comes to complex models defined, for example, in terms of the solution of a set of differential equations, however, a more practical approach will often be required for parameter estimation \cite{Villaverde2014,Gutenkunst2007}.} Unsurprisingly, a significant amount of uncertainty in parameter values  often remains after even a very successful \hltext{fit of the model to data} \cite{Mannakee2016,Transtrum2011,Adams2020}.
 
Sensitivity analysis and uncertainty quantification comprise a whole field dedicated to learning about how model behavior is controlled by their parameters \cite{Marino2008,Saltelli1993,Sobol2001}. These techniques can be used to assess the sensitivity of the model-data fit to changes in parameter values either in a local sense, around a single point (\ie the set of best-fit parameter values), or in a global sense, across all plausible parameter values consistent with the available data \cite{Mannakee2016,Gutenkunst2007,Saltelli2008}. An alternative approach is Bayesian inference \cite{Girolami2008,Gelman2013}; an increasingly used modelling technique that accounts for collective parameter uncertainty constrained by the combination of both data and prior beliefs \cite{Drovandi2011,Adams2020,Lawson2018,Luengo2020,Drovandi2016,Johnstone2016}. However, regardless of the approach taken to characterize the effects of changes in parameter values on model outputs, critical model parameters are often considered  as individuals in terms of their impact on the model behavior \cite{Saltelli1993}. Sensitivity analysis typically considers the derivative of model outputs with respect to the parameters \cite{Mannakee2016,Marino2008,Saltelli2008} while a Bayesian posterior is analyzed predominantly in terms of its marginal distributions \cite{Adams2020,Lawson2018}. When combinations of model parameters are considered, it is largely in terms of crude numerical scores~\cite{Sobol2001,Saltelli1993,Saltelli2008}. Unfortunately, model parameters that are not very constrained by the data are often assumed \emph{not} to have a strong influence on model predictions, even though it is the case of many systems that certain \emph{combinations} of seemingly unconstrained model parameters are more narrowly constrained by the data than any of the individual model parameters \cite{Brown2004,Brown2003,Transtrum2011,Gutenkunst2007}.

In fact, model parameters can act together or against each other, and often must be understood in terms of their combinations  \cite{Mannakee2016,Brown2004}. Parameter combinations that significantly influence model predictions, called \emph{stiff} eigenparameters, essentially act as emergent ``control knobs'' for the model: predictions are possible without precise knowledge of individual parameter values as it is these  stiff eigenparameters that are tightly constrained by the  data \cite{Transtrum2011,Transtrum2015}.  Conversely, the model-data fit  may be also relatively insensitive to some other parameter combinations,   called  \emph{sloppy} eigenparameters \cite{Brown2003,Brown2004}, which hence are poorly constrained by the data. \cite{White2016,Transtrum2011}. Recently, efforts have been made to unravel these connections among parameters through the  expanding literature on model sloppiness  \cite{Brown2003,Transtrum2017,Hagen2013,Apgar2010,Dufresne2018}. Methods to analyze model sloppiness seek to expose the sensitivities of the model-data  fit to changes in sets of parameter values by characterizing the topography of the surface describing how the model-data fit depends on the model parameters in the vicinity of the best-fit parameter values \cite{Mannakee2016,Transtrum2011}. However, thus far such methods  have primarily focused on the field of systems biology where there is little prior knowledge of parameter values \cite{Gutenkunst2007,White2016,Transtrum2015}, and so the sensitivities of the model-data  fit  to changes in parameter values remain to be considered in the context where prior information is also available (\eg from experts or previous studies) to inform parameter values \cite{Gutenkunst2007,Wu2018,Choy2009,Baker2019}. 
  
In this work, we propose a new comprehensive approach to characterize local and global sensitivities of the model-data fit to changes in parameter values. This is achieved by bringing a Bayesian inference perspective \cite{Girolami2008,Gelman2013} to the analysis of sloppiness  that consequently leads to the robust identification of the stiff eigenparameters. In this way, analysis of sloppiness gains the ability to incorporate prior information and to look beyond the curvature at a single point (\ie the set of best-fit parameter values) in an uncertainty-informed way. Meanwhile, Bayesian inference gains a tool to identify well-constrained combinations of parameters that can be otherwise hidden when considering the uncertainty in individual model parameters, critical when the  number of parameters to be estimated is large.  

As part of our comprehensive approach, we \hltext{extend the usage of two well-established Bayesian approaches to dimensionality reduction \cite{Cui2014,Hotelling1933,Spantini2015,Jolliffe2016} to define} the sensitivity matrix that underlies the analysis of model sloppiness, suitably calculated using the posterior samples generated by Bayesian inference. The first definition uses the covariance of the posterior samples to inform parameter space curvature in a non-localized manner~\cite{Brown2003, Rothenberg1971}, with ties to \hltext{classical principal component analysis (PCA)}~\cite{Hotelling1933}. This approach has appeared in works analyzing model sloppiness but only in the context of uninformative priors~\cite{Brown2003, Mannakee2016, Casey2007}. Considering it here in the Bayesian context with informative priors, we identify the need for the second approach, that uses the dimension reduction idea from Cui~et~al.~\cite{Cui2014} to conveniently separate the effect of any prior information from that of the data. Using this novel \hltext{adaptation of Bayesian techniques for dimensionality reduction} to analyze model sloppiness, we illustrate how to identify the combinations of parameters driving model behavior in applications beyond systems biology, and in a manner that acknowledges separately the available information (\eg via expert knowledge \cite{Choy2009}).

 \hltext{We focus our attention on the fit of three deterministic models to data possessing closed-form likelihood functions, although models that do not satisfy this criterion may also be analyzed using some of the methods presented here (further details in the Discussion)}. Thus, we first highlight the advantages of our approach using the well-known Michaelis--Menten model of enzyme kinetics~\cite{Michaelis1913}. We then apply it to a well-studied ecosystem network from Australia (a relatively data-poor system)~\cite{Pech1998}, and a model for the action potential of heart cells (characterized by complex dynamical behavior)~\cite{Beeler1977}. In these latter two applications, different aspects of the interaction between model and data are revealed  by the analysis of sloppiness that are otherwise hidden by the individual techniques we bring together here. Finally, we illustrate how stiff eigenparameters, once identified, can be used to design future experiments to improve parameter inference from collective model-data fittings and identify controlling mechanisms underlying the systems being modeled.

\section*{Results}

\hltext{Our comprehensive analysis of sloppiness identifies the sensitivities of the model-data fit  to changes in parameter values either in the region local to a point of interest in parameter space (Standard approach, see Methods) or in the global sense, across all plausible  parameter values consistent with available information (Bayesian approach, see Methods). Here, our results illustrate the benefits of using both standard and Bayesian approaches together to identify critical parameter combinations (stiff eigenparameters) that readily acknowledge the source of information (\ie prior and/or data). To do so, we first analyze sloppiness in a biochemical model with three parameters (Motivating example), known to suffer from poor parameter identifiability even when an excellent amount and quality of data are used to estimate model parameters \cite{Wieland2021,White2016,Choi2017}. Then, we analyze sloppiness in an ecological four-species dynamic model with twenty parameters (Case study 1), representing a typical dilemma in ecology of having too many parameters to be practically estimated well using noisy time-series data \cite{Wieland2021,Adams2020,Geary2020}. Finally, we analyze sloppiness in a cardiac electrophysiology model with nine parameters (Case study 2), representing complex systems with strong nonlinear dynamics  \cite{Johnstone2016,Lawson2018}.}

\subsection*{Motivating example: the Michaelis--Menten kinetics}

\subsubsection*{Critical parameter combinations are readily identified by the analysis of sloppiness}

The ubiquitous Michaelis--Menten model of biochemistry~\cite{Michaelis1913} is a perfect example to demonstrate both the benefits of understanding parameter dependence through the lens of model sloppiness and of bringing a Bayesian approach to the topic \hltext{(Step i, see Methods)}. This model describes the dependence of an enzyme-catalyzed reaction rate $\upsilon$ on substrate concentration $[S]$ as  \cite{Briggs1925}
\begin{equation}						
	\upsilon=\frac{k_{cat}[E_T][S]}{K_M+[S]}=\frac{k_{cat}[E_T]}{1+K_M/[S]},\label{eq:Michaelis--Menten}
\end{equation}
where parameters $k_{cat}$ and $[E_T]$ together dictate the maximum rate of reaction ($\upsilon_{\max}$), while $K_M$ controls the substrate concentrations at which saturation effects become significant \cite{Choi2017}. 

\begin{figure} [!htpb]
	\centering
	\centering\myfigure[0.40]{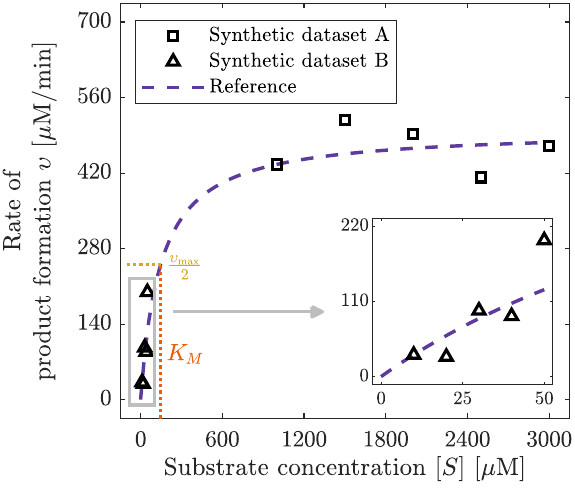}
	\caption{\textbf{Synthetic data generated using  a measurement error of  $\bm{\varepsilon=25\%}$ vs.  noiseless model prediction based on Eq.~\ref{eq:Michaelis--Menten} using reference  parameter values $\bm{k_{cat}=100\;\mathrm{min^{-1}}}$,  $\bm{[E_T]=5\;\mathrm{\mu M}}$ and $\bm{K_M=146.7\;\mathrm{\mu M}}$.} Dataset A is obtained at a relatively high $[S]$ and dataset B is obtained at a relatively low $[S]$. Neither dataset can inform the full rate of reaction $\upsilon$ across the range of $[S]$. (See also Figures' Supplementary Legends.)}
	\label{fig:M-M model data}
\end{figure}

From the right-hand side of Eq.~\ref{eq:Michaelis--Menten}, it is already clear that there are two rate-limiting regimes, one in which the reaction rate simplifies to zero order kinetics with respect to substrate at high $[S]$, and the other one in which the reaction rate simplifies to first order kinetics at low $[S]$~\cite{Michaelis1913,Briggs1925}. To illustrate our methods, we thus consider two noisy synthetic datasets~\hltext{(Step ii, see Methods)} representing these two well-known rate-limiting regimes -- the first dataset (A) consists of five measurements obtained beyond the saturation point, while the second dataset (B) consists of five measurements obtained before saturation has any apparent impact on the model behavior~(Fig.~\ref{fig:M-M model data}). Both datasets fail to describe the full behavior represented by Eq.~\ref{eq:Michaelis--Menten}, and thus suitably highlight the well-known parameter identifiability issues in this model~\cite{White2016,Choi2017}.

In dataset A, measurements only inform the reaction rate at saturation, $\upsilon\approx k_{cat}[E_T]$, and so nothing can be learned about parameter $K_M$. While this  tendency could also be identified by traditional sensitivity analysis~\cite{Marino2008} or by inspecting the posterior variance for this parameter obtained from Bayesian inference~\cite{Tomczak2019,Choi2017}, approaches for model sloppiness go a step further. By identifying key \emph{directions} in the space of the log-parameters, as encoded by the eigenvectors and eigenvalues of a sensitivity matrix, model sloppiness identifies that dataset A only informs the \emph{product} of the remaining two parameters in Eq.~\ref{eq:Michaelis--Menten}, $k_{cat}[E_T]$. Regardless of whether a traditional definition (matrices $\mathbf{H}$ or $\mathbf{L}$, see Methods) or any of the  Bayesian definitions (matrices $\mathbf{P}$ and $\mathbf{G}$, \hltext{see Methods}) of the sensitivity matrix is taken, a single eigenvalue dominates, with parameter combination denoted $\hat{\theta}_1 = k_{cat}[E_T]$ being the corresponding eigenparameter~(Table~\ref{tb:eigenparameters M-M model}, Scenario 1). This is not however visible in the parameter marginals when Bayesian inference is used to \hltext{fit the model to data}, even in this simple problem~(Fig.~S1). 
{ \newcolumntype{C}{>{\centering\arraybackslash}m{0.11\columnwidth}}
	\newcolumntype{Y}{>{\centering\arraybackslash}m{0.24\columnwidth}}			
	\newcolumntype{Z}{>{\centering\arraybackslash}m{0.13\columnwidth}}
	\renewcommand{\arraystretch}{1.2}
	\begin{table} [!htpb]
		\caption{\textbf{Comparison of the stiffest eigenparameter $\bm{\hat{\theta}_1}$ (associated with the largest eigenvalue $\bm{\lambda_1}$) for three different chosen parameter priors to fit the Michaelis--Menten model~(Eq.~\ref{eq:Michaelis--Menten}) to data~(Fig.~\ref{fig:M-M model data}).} Each  $\hat{\theta}_1$ is  identified via Eq.~\ref{eq:eigparameters} after obtaining eigenvalues~(Fig.~S4) and eigenvectors of sensitivity matrices $\mathbf{H}$ (or $\mathbf{L}$), $\mathbf{P}$ and $\mathbf{G}$. Sensitivity matrices return different stiffest eigenparameters $\hat{\theta}_1$ with change of the prior distributions and dataset used to fit the model.}
		\label{tb:eigenparameters M-M model}
		\centering
		\begin{tabular}{CCYZZZ}
			\hline
			\multirow{2}{\linewidth}{\centering\textbf{Synthetic  data}} & \multirow{2}{\linewidth}{\centering\textbf{Scenario}} & \multirow{2}{\linewidth}{\centering\textbf{Prior distribution}} & \multicolumn{3}{c}{\textbf{Stiffest eigenparameter}, \textbf{$\bm{\hat{\theta}_1}$}}\\
			\cline{4-6}
			& & & $\mathbf{H}\;(\text{or}\;\mathbf{L})$   & $\mathbf{P}$ & $\mathbf{G}$ \\ 
			\cline{1-6}
			\multirow{2}{\linewidth}{\centering Dataset A} & $1$ & Uniform & \multirow{2}{\linewidth}{\centering $k_{cat}[E_T]$}    & $k_{cat}[E_T]$  & \multirow{2}{\linewidth}{\centering $k_{cat}[E_T]$}\\
			&  $2$ & Multivariate log-normal & & $K_M$ &      \\ 
			\cline{1-6}
			Dataset B &  $3$ &  Uniform and log-normal    &${k_{cat}[E_T]}/{K_M}$ & $K_M$ & $k_{cat}$ \\
			\hline
		\end{tabular}
\end{table}}

Analogously,  model sloppiness successfully identifies the parameter combination governing the rate of reaction in the non-saturating regime (Fig.~\ref{fig:M-M model data}, Dataset B). Given that this dataset is taken at low substrate concentration ($[S] \ll K_M$), Eq.~\eqref{eq:Michaelis--Menten} reduces to a linear dependence $\upsilon \approx (k_{cat}[E_T]/K_M)\,[S]$  and coefficient $k_{cat}[E_T]/K_M$ is the dominant eigenparameter~(Table~\ref{tb:eigenparameters M-M model}, Scenario~3), which uncovers the nature of the poor parameter identifiability in this model. However, in this scenario as well as the second scenario for dataset A~(Table~\ref{tb:eigenparameters M-M model}), we choose informative priors that cause the Bayesian approaches to model sloppiness (matrices $\mathbf{P}$ and $\mathbf{G}$) to lead to different dominant eigenparameters. We explore the information provided by these approaches that take into account both prior and data to inform model parameters in the following section.

\subsubsection*{A Bayesian perspective reveals whether stiff parameter combinations are informed by the data or are influenced by the prior}

\begin{figure}[!ht]
    \centering\myfigure[1]{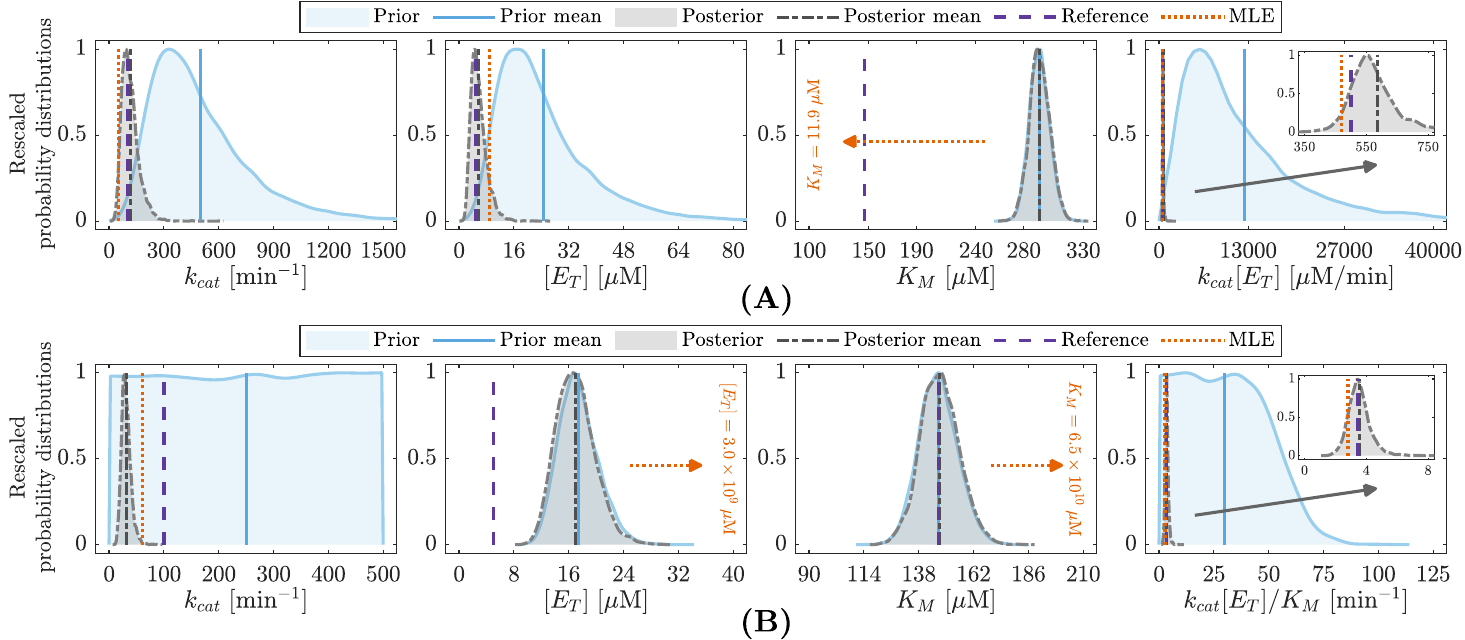}
	{\vspace{-15pt}
    \phantomsubcaption\label{fig:eig_Scenario2_marginals}
    \phantomsubcaption\label{fig:eig_Scenario3_marginals}
    }   
 	\caption{\textbf{Prior and posterior distributions for the model parameters together with the stiffest eigenparameters from Scenarios~2 and~3~(Table~\ref{tb:eigenparameters M-M model}) compared to their associated sets of best-fit values (MLE) and reference values.} (See also insets.)
 	\textbf{(\subref{fig:eig_Scenario2_marginals})} Scenario~2 parameters $k_{cat}$ and $[E_T]$ together with stiffest eigenparameters  ${\hat{\theta}_1=K_M}$ (matrix $\mathbf{P}$) and $\hat{\theta}_1=k_{cat}[E_T]$ (matrices $\mathbf{H}$ or $\mathbf{L}$ and $\mathbf{G}$).
 	\textbf{(\subref{fig:eig_Scenario3_marginals})} Scenario~3 parameter $[E_T]$ together with stiffest eigenparameters
 	${\hat{\theta}_1=k_{cat}}$ (matrix $\mathbf{G}$), ${\hat{\theta}_1=K_M}$ (matrix $\mathbf{P}$) and ${\hat{\theta}_1=k_{cat}[E_T]/K_M}$ (matrices $\mathbf{H}$ or $\mathbf{L}$).
 	Parameter combinations $k_{cat}[E_T]$ and ${k_{cat}[E_T]/K_M}$ are well-constrained by the data in Scenarios~2 and~3, respectively. Parameter $K_M$  is well-constrained by the prior (posterior and prior overlapping) in both scenarios, and parameter $k_{cat}$ is well-constrained by the data relative to the prior in Scenario 3. 
 	The best-fit values for parameter $K_M$ lie far away from its reference values in both scenarios. As parameter combination ${k_{cat}[E_T]/K_M}$ is well-constrained by the data in  Fig.~\ref{fig:eig_Scenario3_marginals}, the posterior distribution for parameter $k_{cat}$ is left-shifted from the reference to compensate for parameter $[E_T]$ that is right-shifted from the reference in Fig.~S3A. 
 	(See also Figures' Supplementary Legends.)}
 	\label{fig:M-M_marginals}
 \end{figure}

 Often, values for model parameters are meaningfully constrained by known feasible ranges or by expert information~\cite{Wu2018,Baker2019,Choy2009}, which can potentially change both the most plausible set of values for the parameters, and the nature of the \emph{new} information provided by the data. To demonstrate how the Bayesian approach to analyzing model sloppiness addresses this, we consider different scenarios where the reaction rate data~(Fig.~\ref{fig:M-M model data}) is now coupled with prior information, and thus highlight how the stiff eigenparameters obtained using our two definitions of the sensitivity matrix (matrices $\mathbf{P}$ and $\mathbf{G}$) together reveal whether parameter values  are informed by the data or are influenced by the prior. We first fit Eq.~\ref{eq:Michaelis--Menten} to dataset A~\hltext{(Steps iii and iv, see Methods)}, considering a multivariate log-normal distribution for the model parameters that sets the value of one parameter ($K_M$) far away from its reference value~(Fig.~S2). As a result, the posterior correctly concentrates around the reference  parameter values used to generate the data~(Fig.~\ref{fig:eig_Scenario2_marginals}, first and second panel), except for the poorly-specified parameter ($K_M$) for which the prior renders it unable to~(Fig.~\ref{fig:eig_Scenario2_marginals}, third panel). Here, prior and posterior distributions for parameter $K_M$ are approximately equivalent (overlapping), thus reflecting that the data collected at saturation is uninformative to this parameter value. However, by examining the curvature of the posterior via its inverse covariance matrix $\mathbf{P}$~\hltext{(Steps v and vi, see Methods)}, this parameter emerges as the stiffest eigenparameter~(Table~\ref{tb:eigenparameters M-M model}, Scenario~2). Thus, as prior and posterior distributions for parameter $K_M$ are overlapped~(Fig.~\ref{fig:eig_Scenario2_marginals}, third panel), this method reveals that the information already contained in the prior is dominating that provided by the data. 

To learn the data informativity on model parameters while simultaneously acknowledging any prior information, we use the likelihood-informed subspace (LIS) method. This approach works by transforming the effects of the prior on the curvature of parameter space~\cite{Cui2014,Cui2016}, leaving only the effects of the data via the likelihood (further details in the Methods). By doing so, the LIS method  produces a sensitivity matrix ($\mathbf{G}$) that identifies the region in parameter space where the informativity of the data  prevails over that of the prior information \cite{Cui2014,Cui2016}. For example, by imposing an informative prior for parameter $K_M$ in this scenario, the method (matrix $\mathbf{G}$) recognizes that no additional information is gained about this parameter from dataset A through the model-data fitting process, and so it returns the same dominant eigenparameter $\hat{\theta}_1=k_{cat}[E_T]$~(Fig.~\ref{fig:eig_Scenario2_marginals}, fourth panel) as the methods (matrix $\mathbf{H}$ or $\mathbf{L}$) that ignore the prior altogether~(Table~\ref{tb:eigenparameters M-M model}, Scenario~2). A natural question is then what does the LIS method provide that is not already given by a standard analysis of sloppiness? The key benefit is that if prior information does change the most plausible (prior-informed) region of parameter space, and the model behaves differently in this region, the LIS method will identify the directions in parameter space where the data are most ``informative'' relative to the prior, as we discuss next.

In Scenario~3~(Table~\ref{tb:eigenparameters M-M model}), we fit Eq.~\ref{eq:Michaelis--Menten} to dataset B~\hltext{(Steps iii and iv, see Methods)} considering a combination of uniform and log-normal prior distributions that  strongly specify values of parameters $[E_T]$ and $K_M$ well and badly~(Fig.~S3), respectively. Given that dataset B only constrains the value of combination of parameters $k_{cat}[E_T]/K_M$~(Fig.~\ref{fig:eig_Scenario3_marginals}, fourth panel), the extreme values of the parameters selected by unconstrained \hltext{maximum likelihood estimation}~(Fig.~\ref{fig:eig_Scenario3_marginals}, MLE in the second and third panel) highlight the importance of specifying plausible ranges for parameters via a Bayesian prior. As for Bayesian inference, the posterior distribution simply fixes the value of the parameter $k_{cat}$~(Fig.~\ref{fig:eig_Scenario3_marginals}, first panel) to a value that constrains well eigenparameter $k_{cat}[E_T]/K_M$~(Fig.~\ref{fig:eig_Scenario3_marginals}, fourth panel). Similar to Scenario 2, model sloppiness, as implied by the posterior covariance method (matrix $\mathbf{P}$), selects one of the parameters strongly specified by the prior, $K_M$~(Fig.~\ref{fig:eig_Scenario3_marginals}, third panel), as the stiffest eigenparameter~(Table~\ref{tb:eigenparameters M-M model}). In this scenario, the LIS method (matrix $\mathbf{G}$) instead identifies that dataset B  acts only to fix the value of parameter $k_{cat}$ and selects it as the dominant eigenparameter. That is, in contrast to the standard analysis of sloppiness only considering the likelihood surface, the LIS method uncovers \emph{new} information provided by the data when there is prior parameter knowledge. Thus, the Bayesian methods together clarify whether the model parameters (or eigenparameters) are informed by the data or are significantly influenced by the prior beliefs.

\subsection*{Case study 1: Ecosystem network}

\subsubsection*{A global perspective to analyzing sloppiness reveals true informativity of the data}

\begin{figure} [!htpb]
	\centering\myfigure[1]{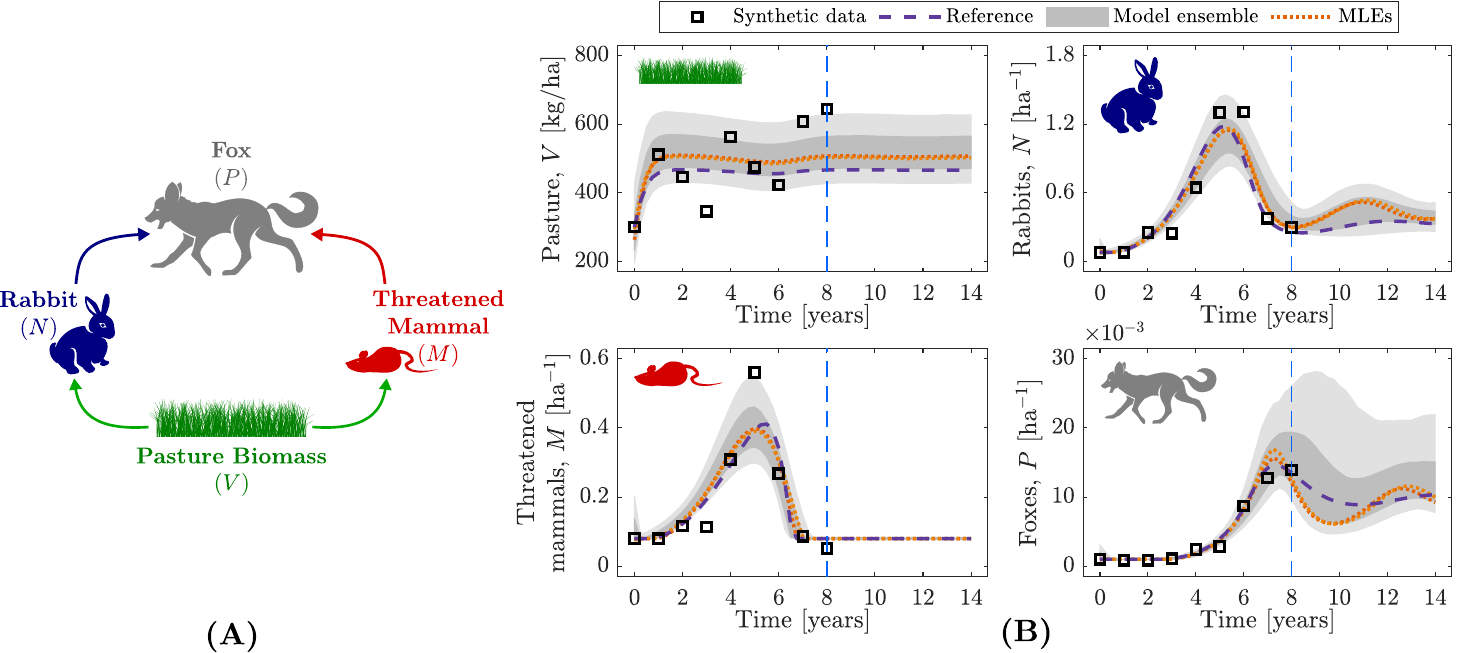}
	{\vspace{-15pt}
    \phantomsubcaption\label{fig:Ecological Network}
    \phantomsubcaption\label{fig:network model data}
    }  
	\caption{\textbf{Ecosystem interaction network,  consisting of pasture ($\bm{V}$), rabbits ($\bm{N}$), foxes ($\bm{P}$), and threatened species ($\bm{M}$).} 
	\textbf{(\subref{fig:Ecological Network})} Ecosystem network in which arrows indicate the direction of the energy transfer associated with the species interaction. 
	\textbf{(\subref{fig:network model data})}  Synthetic time-series data for ecological abundance with measurement error of  $\varepsilon=25\%$  together with noiseless model prediction using reference parameter values~(Table~S2), model predictions using two sets of best-fit parameter values (MLEs), and model ensemble predictions using all plausible parameter values~(Fig.~S5).  
	The ecosystem network model fits the synthetic time-series data, with the model ensemble propagating parameter uncertainty into species abundance predictions. 
	(See also Figures' Supplementary Legends.)}
	\label{fig:Ecological Network and Data}
\end{figure}

 Unlike the simple motivating example considering two well-known rate-limiting regimes that readily unveiled the controlling eigenparameters~(Fig.~\ref{fig:M-M_marginals}, fourth panels), with much larger models, inferring the parameter combinations that are more or less sensitive to the model-data fit can be difficult from a simple model inspection. To illustrate this,  as a more complex case study from ecology, we use a  well-known four-species ecosystem network model \cite{Pech1998} that includes two threat species (foxes and rabbits), one threatened species  (native mammals), and a basal species (pasture), as depicted in Fig.~\ref{fig:Ecological Network}. This ecosystem model consists of four discrete-time equations (based on ordinary differential equations) and eight constitutive equations (Table~S1) whose twenty parameter point estimates~(Table~S2) were inferred from several studies at two semi-arid locations in Australia \cite{Pech1998}. Here, we thus seek to illustrate key benefits of the Bayesian analysis of sloppiness for data-poor systems, characterized by low quality and amount of observed data due to practical limitations \cite{Adams2020,Maayan2017,Brown2003,Geary2020}. \hltext{To do so, we first fit the ecosystem network model~(Table~S1) to noisy synthetic time-series data using both maximum likelihood estimation and Bayesian inference \hltext{(Steps i-iv, see Methods)}},  considering  a multivariate log-normal prior distribution for the model parameters~(Fig.~S5). 
 After \hltext{fitting the model to data}, model predictions~(Fig.~\ref{fig:network model data}) based on a model ensemble~(shaded regions), considering all plausible parameter values~(Fig.~S5), enclose both the simulated noisy data~($\bm{\square}$ symbols) and true ecosystem dynamic behavior (dashed profiles). They also enclose predictions based on two sets of best-fit parameter values~(dotted profiles) obtained from starting the \hltext{maximum likelihood estimation algorithm} at two different initial parameter values \hltext{(Step iii, see Methods)}. Further, parameter marginals~(Fig.~S5) enclose these two separate point estimates and also show that most of the model parameters are poorly constrained by the data.

In addition to quantifying parameter uncertainty, a global perspective to the problem of fitting models to data can benefit the inference of critical parameter combinations that control the quality of the model-data fit. For example, while local changes in the topography of the surface described by the likelihood function in the vicinity of the two sets of best-fit parameter values~(Fig.~S5) mislead inference of stiff eigenparameters through the standard analysis of sloppiness~(\cf $\hat{\theta}_i,\,i=1,2,3$ in Table~\ref{tb:eigenparameters network model} from matrices $\mathbf{H}$ or $\mathbf{L}$, evaluated at the different sets of best-fit values $\bm{\theta}^\star_1$ and $\bm{\theta}^\star_2$), the Bayesian methods (matrices $\mathbf{P}$ and $\mathbf{G}$) fully characterize the structure of this surface by considering all plausible parameter values~\hltext{(Steps v and vi, see Methods)}, informed by the combination of both data and prior beliefs. In this way, differences between dominant eigenparameters from Bayesian sensitivity matrices $\mathbf{P}$ and $\mathbf{G}$~(Table~\ref{tb:eigenparameters network model}) also demonstrate that the prior is influencing  the most plausible region of parameter space, which thus implies that the surface described by the posterior distribution (Eq.~\ref{eq:Bayes}) and the likelihood function   (Eq.~\ref{eq:application likelihood}) are different locally and globally.
{ \newcolumntype{C}{>{\centering\arraybackslash}m{0.18\columnwidth}}
	\newcolumntype{Z}{>{\centering\arraybackslash}m{0.2\columnwidth}}
	\newcolumntype{X}{>{\centering\arraybackslash}m{0.10\columnwidth}}
	\renewcommand{\arraystretch}{1.2}
	\begin{table} [!htpb]
		\caption{\textbf{Comparison of the stiffest eigenparameters $\bm{\hat{\theta}_1}$, $\bm{\hat{\theta}_2}$ and  $\bm{\hat{\theta}_3}$  (associated with the largest eigenvalues $\bm{\lambda_1}$,  $\bm{\lambda_2}$, and $\bm{\lambda_3}$) considering  a multivariate log-normal prior distribution for the parameters to \hltext{fit} the ecosystem model (Table~S1) to data (Fig.~\ref{fig:network model data}).}  Each  $\hat{\theta}_1$, $\hat{\theta}_2$ and $\hat{\theta}_3$ is  identified via Eq.~\ref{eq:eigparameters} after obtaining eigenvalues~(Fig.~S6) and eigenvectors from matrices $\mathbf{H}$ (or $\mathbf{L}$), $\mathbf{P}$ and $\mathbf{G}$. Stiff eigenparameters from matrix $\mathbf{H}$ (or $\mathbf{L}$) are obtained at two sets of best-fit parameter values $\theta^\star_1$ and $\theta^\star_2$~(Fig.~S5). Matrix $\mathbf{H}$ (or $\mathbf{L}$) returns different stiff eigenparameters when evaluated at two distinct sets of best-fit parameter values while matrices $\mathbf{P}$ and $\mathbf{G}$ return different stiff eigenparameters because the prior influences the model-data fit.}
		\label{tb:eigenparameters network model}
		\centering\renewcommand{\arraystretch}{1.2}
		\begin{tabular}{C*{2}{Z}XZ}
			\hline
			\multirow{3}{\linewidth}{\centering\textbf{Eigenparameter} $\bm{\hat{\theta}_i}$} &  \multicolumn{4}{c}{\textbf{Sensitivity matrices}}\\
			\cline{2-5}
			  & \multicolumn{2}{c}{\textbf{$\mathbf{H}\;\text{or}\;\mathbf{L}$ evaluated at}}  & $\mathbf{P}$   & $\mathbf{G}$\\
			\cline{2-3}
			& $\bm{\theta}^\star_1$ & $\bm{\theta}^\star_2$  &  & \\
			\hline
			$1$ & $(c_N/a_N)({a_M}/{c_M})^{0.4}$  & $({c_M}/{a_M})(a_N/c_N)^{0.9}$ & ${a_P^{0.9}}/{c_P}$   & $({c_M}/{a_M})(a_N/c_N)^{0.9}$ \\
			$2$  & $({c_M}/{a_M})(c_N/a_N)^{0.4}$ & $({c_N}/{a_N})(c_M/a_M)^{0.9}$  & ${c_N}/{a_N}$  &   $({c_N}/{a_N})(c_M/a_M)^{0.9}$  \\	
			$3$  & ${c_P}/({a_P^{0.9}d_N^{0.4}})$  & ${c_P}/{a_P^{0.9}}$ & ${c_M}/{a_M}$ &  ${c_P}/{a_P^{0.9}}$  \\			
			\hline
		\end{tabular}
	\end{table}
}

\subsubsection*{Analysis of sloppiness brings new insights to Bayesian parameter inference}

\begin{figure}[!ht]
	\centering\myfigure[1]{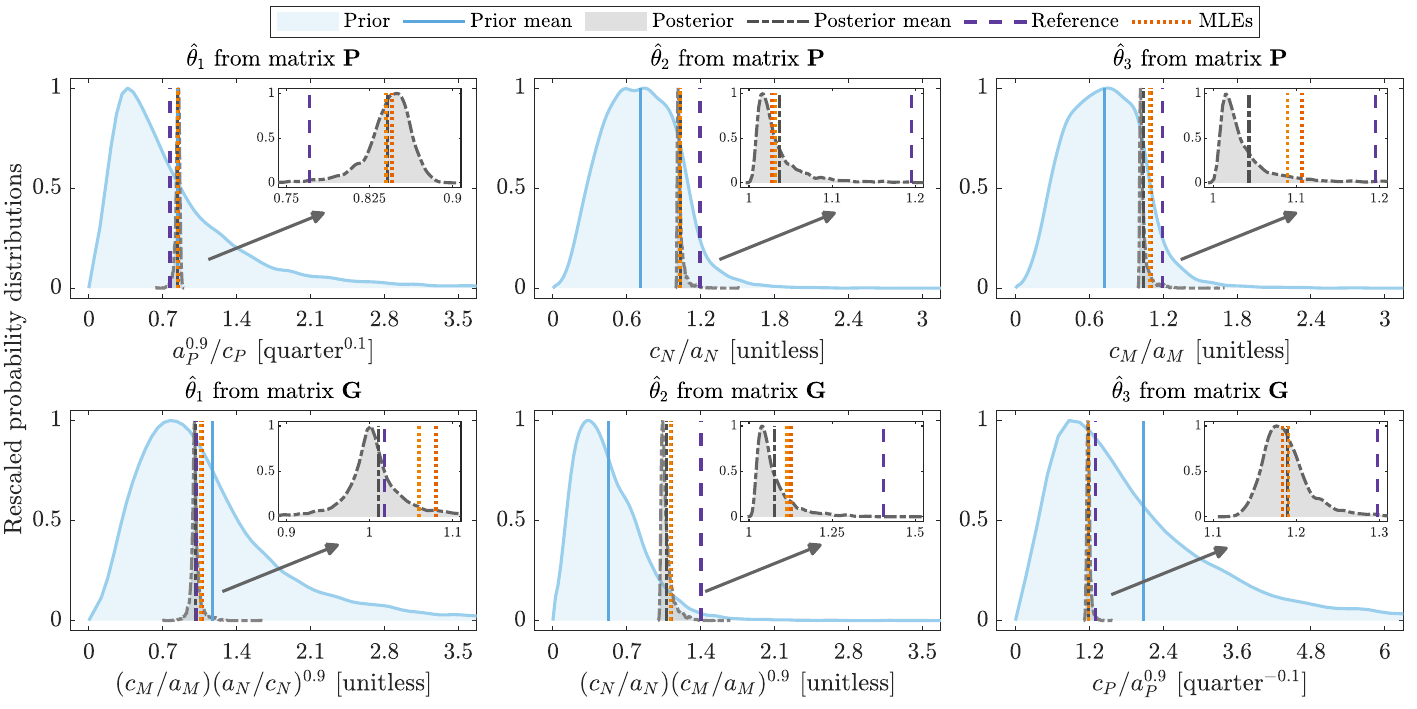}
	\caption{\textbf{Prior and posterior distributions for eigenparameters $\bm{\hat{\theta}_1}$, $\bm{\hat{\theta}_2}$ and $\bm{\hat{\theta}_3}$~(Table~\ref{tb:eigenparameters network model}) compared to their reference values  and associated sets of best-fit values (MLEs).} (See also insets.)  Posterior distributions for all eigenparameters  and their associated MLEs closely match the true values. 
	(See also Figures' Supplementary Legends.)}
	\label{fig:eig_network_marginals}
\end{figure}

Combining model sloppiness together with Bayesian inference reveals critical parameter combinations that can be otherwise lost when only considering the uncertainty in individual model parameters through Bayesian inference. After the fit of the ecological model to data, for example, parameter marginals~(Fig.~S5) illustrate that only a few of the model parameters ($R$, $V_0$ , $D_{III}$ and $\varepsilon$) are well-constrained by the data, which suggests that these parameters have a strong influence on the quality of model-data fit. Instead, the Bayesian analysis of sloppiness (matrices $\mathbf{P}$ and $\mathbf{G}$)  identifies that it is in fact combinations of parameters $c_N$, $a_N$, $c_M$, $a_M$, $c_P$ and $a_P$ that are the most constrained by the available data~(Fig.~\ref{fig:eig_network_marginals}). The prior distribution appears to be weakly informing the three stiffest eigenparameters $\hat{\theta}_1$, $\hat{\theta}_2$ and $\hat{\theta}_3$~(Table~\ref{tb:eigenparameters network model}) since the first eigenparameter $\hat{\theta}_1$ from the posterior covariance method (matrix $\mathbf{P}$) also corresponds to the third eigenparameter $\hat{\theta}_3$ from the LIS method~(matrix $\mathbf{G}$), while the quotient ($\hat{\theta}_3/\hat{\theta}_2$) and product ($\hat{\theta}_2\hat{\theta}_3$) of the second and third eigenparameters $\hat{\theta}_2$ and $\hat{\theta}_3$ from the posterior covariance method~(matrix $\mathbf{P}$) approximate the first and second eigenparameters $\hat{\theta}_1$ and $\hat{\theta}_2$ from the LIS method~(matrix $\mathbf{G}$), respectively.

For this system, the identified stiff eigenparameters~(Fig.~\ref{fig:eig_network_marginals}) do not appear together in single terms within the model~(Table~S1). However, parameter ratios $c_{X}/a_{X}$ (or $a_{X}/c_{X}$) with $X=N,M,P$ arise as part of the dominant eigenparameters~(Table~\ref{tb:eigenparameters network model}) as they appear in separate terms with opposite sign in this model~(Table~S1). As a result, there is a compensation effect between values of parameters $a_{X}$ and $c_{X}$ that has two key implications for the model predictions. Firstly, the model-data fit is highly informative for characterizing growth dynamics ($r_N$, $r_M$, $r_P$) of rabbits  ($a_{N}/c_{N}$), threatened mammals ($a_{M}/c_{M}$) and foxes ($a_{P}/c_{P}$), which is likely to significantly affect animal species abundances ($N$, $M$ and $P$). Secondly,  analysis of sloppiness reveals that by measuring either the maximum rate of decrease ($a_{M}$) or increase ($c_{M}$) of the threatened mammal density (also applies for rabbits and foxes), collective model-data fit will inform values of the other parameter to a similar extent, as we discuss in the next section.

\subsubsection*{Bayesian analysis of sloppiness readily informs future experimental design}

\begin{figure}[!htpb]
	\centering\myfigure[1]{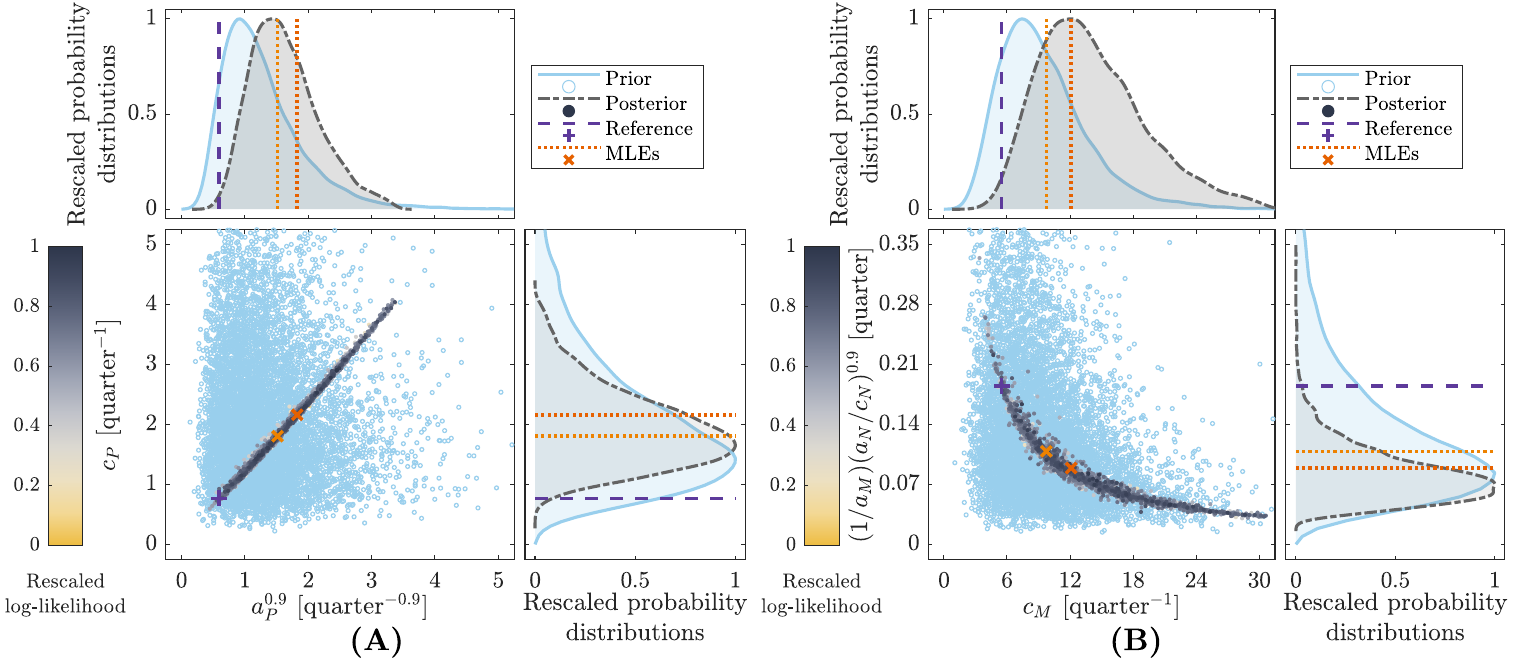}
	{\vspace{-15pt}
    \phantomsubcaption\label{fig:key1-network}
    \phantomsubcaption\label{fig:key2-network}
    }
	\caption{\textbf{Bivariate scatter plots of the prior and posterior distributions (shaded regions) for the first stiffest eigenparameter obtained from the Bayesian methods~(Table~\ref{tb:eigenparameters network model}) compared to their reference parameter values and sets of best-fit values (MLEs).} 
	\textbf{(\subref{fig:key1-network})} $\bm{\hat{\theta}_1}$ and $\bm{\hat{\theta}_3}$ from matrices $\mathbf{P}$ and $\mathbf{G}$, respectively, and
	\textbf{(\subref{fig:key2-network})} $\bm{\hat{\theta}_1}$ from matrix $\mathbf{G}$. Many samples of the posterior distribution yield similar values of the log-likelihood function.
	(See also Figures' Supplementary Legends.)}
	\label{fig:key12-network}
\end{figure}

Bayesian analysis of sloppiness unveils hidden parameter interdependencies  that can help design future experiments for improved parameter inference. For example, given that the posterior covariance method (matrix $\mathbf{P}$) reveals that the ratio of parameters ${a_P^{0.9}}/{c_P}$ is the stiffest eigenparameter~(Table~\ref{tb:eigenparameters network model}), this ratio also indicates that parameters $a_{P}$ and $c_P$ are approximately linearly related, $a_{P}^{0.9}\propto c_P$~(Fig.~\ref{fig:key1-network}). Here, an analogous tendency is seen for the stiffest eigenparameter from the LIS method (matrix $\mathbf{G}$), $({c_M}/{a_M})(a_N/c_N)^{0.9}$~(Table~\ref{tb:eigenparameters network model}), with parameter $c_{M}$ and combination of parameters $({1}/{a_M})(a_N/c_N)^{0.9}$ being instead  inversely related, $c_{M}\propto\left[({1}/{a_M})(a_N/c_N)^{0.9}\right]^{-1}$~(Fig.~\ref{fig:key2-network}). Additionally, many samples from the posterior distributions are seen to lead to the same value of the log-likelihood function (no apparent color change across posterior distribution samples in Fig.~\ref{fig:key12-network}), with the two sets of best-fit parameter values (${\bm{\times}}$ symbols) and reference (true) values ($\bm{+}$ symbols) falling within the corresponding posterior distribution sample. This tendency indicates that every value for the model parameter (or combination of parameters) on one side of the relation (\eg $a_{P}^{0.9}$ and $c_{M}$) has a corresponding  constrained estimate for the parameter (or combination of parameters) on the other side of the relation (\eg $c_P$ and $({1}/{a_M})(a_N/c_N)^{0.9}$). Similar tendencies are seen for the remaining eigenparameters~(Fig.~S7). 

In addition to identifying compensation effects between subsets of parameters~(Figs.~\ref{fig:key12-network} and S7) that lead to similar model outputs~(Fig.~\ref{fig:network model data}), analysis of sloppiness also reveals that prioritizing improvement of the estimates of any of the parameters (or parameter combinations) on one side of the proportionality relationship will immediately improve estimation of parameters (or parameter combinations) on the other side. As an example of this, we considered a multivariate log-normal prior distribution~(Fig.~S8), that is very informative for parameters $a_{N}$, $a_{M}$, $a_{P}$, to fit the ecosystem network model  to data. These prior conditions act as improved estimates for parameters $a_{N}$, $a_{M}$ and $a_{P}$, obtained either from expert elicitation or parameter-specific experiments (\eg spotlight counts\cite{Pech1998}). After the model-data fit~(Fig.~S9), parameters on the other side of the relations ($c_{N}$, $c_{M}$ and $c_{P}$) are also found to be more narrowly constrained. The percentage coefficients of variation ($CV$) for the posterior distributions of parameters $c_{N}$, $c_{M}$ and $c_{P}$ range between $7-8\%$ when a more informative prior distribution is specified for parameters $a_{N}$, $a_{M}$ and $a_{P}$, which are much lower than those obtained (ranging between $30-50\%$) when a vague multivariate log-normal prior distribution is instead specified~(Fig.~S5). Thus, combining Bayesian inference together with the analysis of sloppiness reveals parameter interdependencies that can be strategically exploited to efficiently improve individual parameter inference using less additional data than might be otherwise expected.

\subsection*{Case study 2: Cardiac Electrophysiology}

\subsubsection*{Key controlling mechanisms for complex systems are uncovered by analysis of sloppiness}

The previous section considered an ecological model as an example of a system characterized by a moderately large number of parameters, and poor access to data. A separate class of systems are those for which data are more readily available, but the dynamics that produce the data manifest in complex sensitivities to their controlling parameters. For these systems, the challenge is often how to summarize these nonlinear dynamics in a meaningful, actionable way, and so stiff eigenparameters identified by analyzing model sloppiness have a recognizable potential. However, so far model sloppiness has primarily been considered for models characterized by large numbers of fundamental interactions, such as the Michaelis--Menten kinetics that describe the cell signalling network analyzed in the foundational work of Brown~et~al.~\cite{Brown2003,Brown2004}. Here, we seek to demonstrate the usefulness and purpose of stiff eigenparameters in systems where the constituent dynamics themselves, and not only their interactions, are complex and unwieldy.

As an example of such a system, we consider the Beeler--Reuter (BR) model~\cite{Beeler1977}, which describes the action potential (AP) of a cardiac ventricular myocyte, the pattern of highly regulated ion flow that creates the depolarization and subsequent repolarization governing the heartbeat. This cardiac cell model consists of eight nonlinear ordinary differential equations, six constitutive equations (Table~S3), and nine parameters~(Table~S4). Although an older model, the BR model captures many of the most important electrophysiological features of the ventricular AP~\cite{Krogh-Madsen2020}, and interest remains regarding its sensitivity to changes in its parameters~\cite{Drovandi2016,Johnstone2016}. Cardiac AP models are critical for mechanistically understanding arrhythmia~\cite{Zhou2018},  and the issue of parameter variability is fundamental to understanding the differential effects of antiarrhythmic treatments within a population~\cite{Britton2013} or the cardiotoxicity of other pharmacological agents~\cite{Passini2017}. 

\begin{figure}[!htpb]
	\centering\myfigure[1]{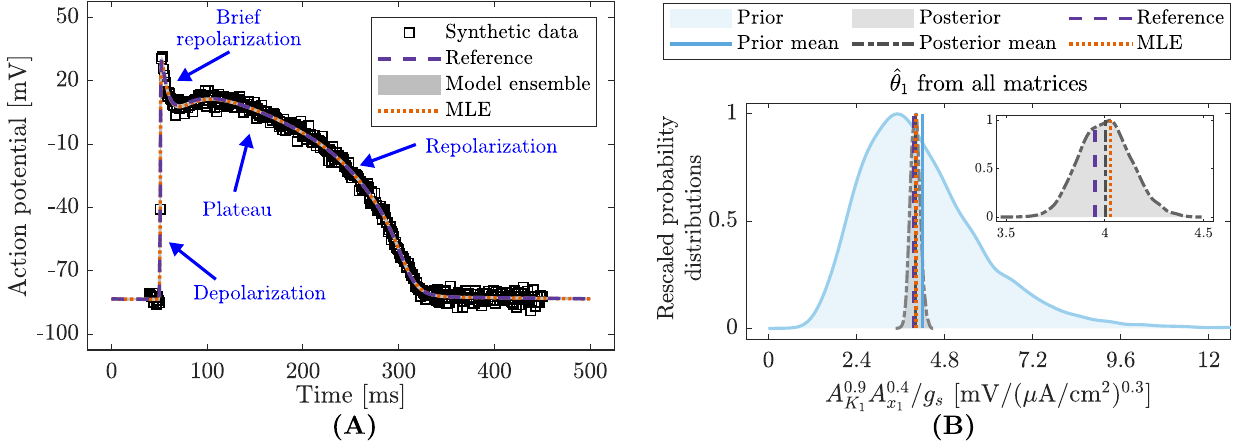}
	{\vspace{-15pt}
    \phantomsubcaption\label{fig:Beeler--Reuter fit}
    \phantomsubcaption\label{fig:Beeler--Reuter Eig}
    }
	\caption{\textbf{BR model fit to synthetic time-series data together with the identified stiffest eigenparameter}.
	\textbf{(\subref{fig:Beeler--Reuter fit})} Synthetic action potential (AP) data with measurement error of  $\sigma=2\mathrm{mV}$ and time resolution of $1\;\mathrm{ms}$ ($1\;\mathrm{kHz}$) together with noiseless model prediction using reference parameter values~(Table~S4), model predictions using the set of best-fit parameter values (MLE), and model ensemble predictions using all plausible parameter values.
	\textbf{(\subref{fig:Beeler--Reuter Eig})} Prior and posterior distributions  for the stiffest eigenparameter $\hat{\theta}_1=A_{K_1}^{0.9}A_{x_1}^{0.4}/g_{s}$ obtained from all matrices compared to their associated set of best-fit values and reference values. (See also inset.)
	Both MLE-based and Bayesian inference-based model predictions overlap the true AP dynamics and synthetic data. All methods lead to the same stiffest eigenparameter. The posterior distribution for the stiffest eigenparameter and their associated best fit value (MLE) closely match the true  eigenparameter value. 
	(See also Figures' Supplementary Legends.)}
	\label{fig:Beeler--Reuter}
\end{figure}

The AP is summarized by the time course of a cell's transmembrane potential in response to stimulation, and can be recorded by an electronic measurement device at good temporal resolution and without much noise (\eg synthetic data in Fig.~\ref{fig:Beeler--Reuter fit}). The complexity in these models rests with the way multiple ion channels, each with its own set of time-adaptive, nonlinear voltage-gated dynamics, combine additively to determine the total ion flow that produces the AP~(Table~S3). The most commonly varied parameters are the relative levels of expression for these different ion channels~\cite{Muszkiewicz2016}, and so rather than describing fundamental quantities such as rates of production or destruction, model parameters in this context describe the extent to which a variety of complex and strongly nonlinear dynamics contribute to the system behavior.

For this system, Bayesian model-data fitting~\hltext{(Steps iii and iv, see Methods)} produces an ensemble of plausible values for model parameters that recapture the data extremely well~(Fig.~\ref{fig:Beeler--Reuter fit}). Most of the individual parameters are well-constrained by the AP data~(as seen from their marginal distributions, Figs.~S10), although none emerge as substantially more important than all others.  Analyzing model sloppiness to consider parameters in terms of their combinations~\hltext{(Steps v and vi, see Methods)}, however, reveals that the combination of parameters $A_{K_1}^{0.9}A_{x_1}^{0.4}/g_{s}$ is the primary driver of the AP dynamics. This eigenparameter's corresponding eigenvalue eclipses the value of the others~(Fig.~S11), and accordingly, its value is extremely well specified by the population of plausible parameter values~(Fig.~\ref{fig:Beeler--Reuter Eig}). This eigenparameter and its relative importance are identified by both the standard and Bayesian approaches for model sloppiness, owing to the use of a relatively uninformative prior and the fact that the data are highly informative about the model parameters. Unlike the Michaelis-Menten kinetics or the ecosystem network model examples, here all approaches for model sloppiness are similarly suitable due to the strong informativeness of the data relative to the prior. 

\begin{figure}[!htpb]
			\centering\myfigure[1]{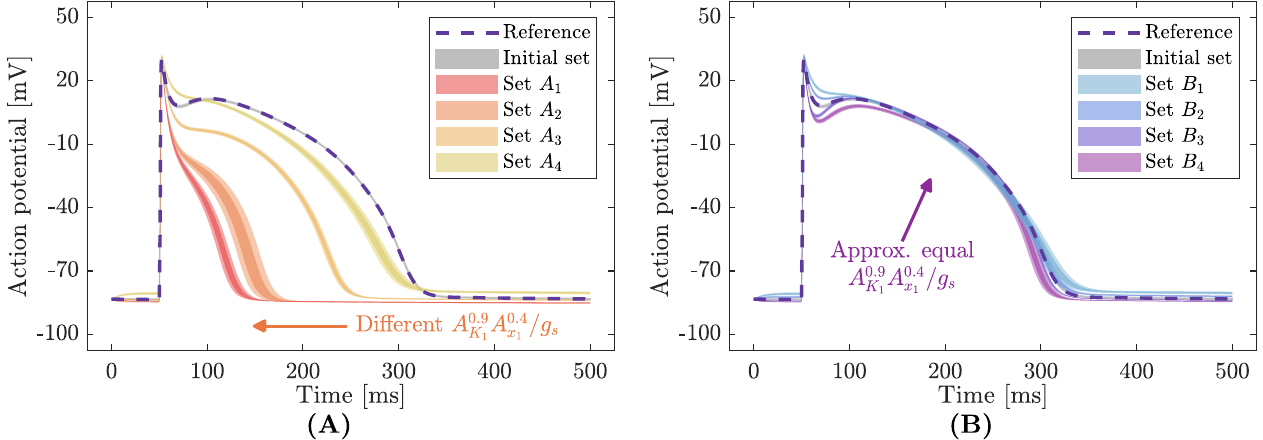}
			{\vspace{-15pt}	
			\phantomsubcaption\label{fig:BRsim1}
            \phantomsubcaption\label{fig:BRsim2}
            }
			\caption{\textbf{Effect of varying parameter values that are part of the stiffest eigenparameter $\bm{\hat{\theta}_1={A_{K_1}^{0.9}A_{x_1}^{0.4}}/{g_s}}$ on the BR model prediction.} 
			\textbf{(\subref{fig:BRsim1})}~Predictions obtained with four sets of parameter values that change the value of $\hat{\theta}_1$~(Fig.~S14, sets $A_1$ to $A_4$). 
			\textbf{(\subref{fig:BRsim2})}~Predictions obtained with four sets of parameter values that keep approximately constant the value of $\hat{\theta}_1$~(Fig.~S14, sets $B_1$ to $B_4$). 
			Similar predictions are obtained with change in the parameter values when the value of $\hat{\theta}_1$ is kept approximately constant while, in the opposite case, predictions strongly differ from those obtained with initial parameter set~(Fig.~S10)what. 
	       (See also Figures' Supplementary Legends.)}
			\label{fig:sim3}
		\end{figure}

The key eigenparameter has a natural  electrophysiological interpretation. Parameters $A_{K_1}$ and $g_s$ describe the relative strengths of the primary outward and inward (\ie counteracting) currents active during the plateau and repolarization phases that compose the bulk of the AP~(Fig.~\ref{fig:Beeler--Reuter fit}), and so they appear in the eigenparameter as a ratio. Here, the third parameter $A_{x_1}$ contributes to the eigenparameter to a lesser extent and appears as a product with parameter $A_{K_1}$, owing to their shared role in describing strengths of the outward potassium currents that drive repolarization. The three currents $I_{K_1}$, $I_{x_1}$ and $I_s$~(Table~S3), associated with these three model parameters ($A_{K_1}$, $A_{x_1}$,  and $g_s$, respectively), exhibit nonlinear dynamics~(Fig.~S12). Thus, it is surprising how well the primary actions of these three currents ($I_{K_1}$, $I_{x_1}$ and $I_s$) can be summarized by a simple product of parameters with exponents ($A_{K_1}^{0.9}A_{x_1}^{0.4}g_s^{-1}$), whose value strongly dictates whether or not the model output reproduces the data~(Fig.~\ref{fig:sim3}). 

 Analysis of model sloppiness naturally uncovers this result, by revealing the precise way in which the three currents $I_{K_1}$, $I_{x_1}$ and $I_s$~(Table~S3) act together and thus highlighting the importance of their balance by assigning a much higher eigenvalue to their eigenparameter than any other. Without considering the curvature of the log-parameters, however, this relationship is not easily observed. The precise relationship between $A_{K_1}$, $A_{x_1}$ and $g_s$ remains hidden from view in standard Bayesian bivariate analysis~\cite{Drovandi2016,Johnstone2016}, and even when directly plotting the values of posterior samples for these three parameters against one another~(Fig.~S13). Such  a relationship is also not obvious from the model definition, where none of the three parameters $A_{K_1}$, $A_{x_1}$ and $g_s$ appear as products or quotients with one another, nor do the coefficients of their addition correspond to the exponents found in the governing eigenparameter.

\subsubsection*{Analysis of sloppiness uncovers knowledge limitations in mathematical models fitted to data}

\begin{figure} [!htpb]
			\centering\myfigure[0.4950]{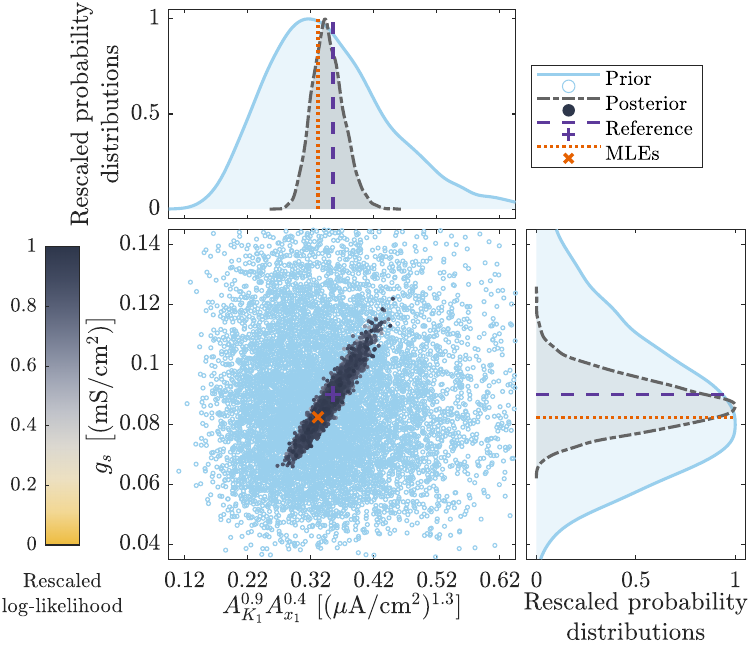}
			\caption{\textbf{Bivariate scatter plot for the prior and posterior distributions (shaded regions) for combination of parameters $A_{K_1}^{0.9}A_{x_1}^{0.4}$ and $g_{s}$ compared to sets of reference parameter values and best-fit values (MLE)}. Many samples of the posterior distribution yield similar values of the log-likelihood function with $A_{K_1}^{0.9}A_{x_1}^{0.4}\propto g_{s}$.
	       (See also Figures' Supplementary Legends.)
			} 
			\label{fig:Beeler--Reuter relations}
		\end{figure}

As  also observed in the ecological application~(Fig.~\ref{fig:key12-network}), the existence of a strong eigenparameter(s) introduces a pronounced structure to the  space of plausible parameter sets~(Fig.~\ref{fig:Beeler--Reuter relations}). The nature of the eigenparameter implies a strong linear interdependency between combination of parameters $A_{K_1}^{0.9} A_{x1}^{0.4}$ and parameter $g_s$, as seen in the posterior samples found by the Bayesian inference~(Fig.~\ref{fig:Beeler--Reuter relations}). Identifying these critical structures introduced by the model-data fitting process is key to understanding the information provided by the data on the model parameters.  Cardiac electrophysiology is a particularly important example as parameter identifiability is a well-established issue for AP models \cite{Johnstone2016,Drovandi2011}. Thus, owing to sloppiness in parameter estimation such as that discovered and quantified here, even perfect AP data~(Fig.~\ref{fig:Beeler--Reuter fit}) can imply multiple different parameter estimations~(Figs.~\ref{fig:BRsim2} and S14), with consequences that then emerge under pathological conditions or in response to drug treatments~\cite{Zaniboni2010}.

 As in many other disciplines, in cardiac electrophysiology, it can be difficult to design further experiments and/or to target experiments to learn specific parameter values. To this end, our comprehensive approach to model sloppiness does uncover the deficiencies in the available information  through the identification of the critical eigenparameters. For example, once these critical eigenparameters are identified, the model can be used to simulate scenarios considering extreme system conditions~(Fig.~S14) that are theoretically still plausible given current data~(Fig.~\ref{fig:BRsim2}). In fact, this concept might be even more applicable where a model's computational runtime limits the feasibility of Bayesian inference. Even when a posterior of plausible parameter value sets cannot be realistically generated, standard analysis of sloppiness can still quickly identify directions in parameter space of poor information. In this way, simulations can be carried out along directions of poor knowledge (\eg perpendicular to the linear relationship depicted in~Fig.~\ref{fig:Beeler--Reuter relations}) to further justify the conclusions of simulation studies against the uncertainty that remains in the  parameters after the \hltext{model-data fit}.

 Indeed, ever-present knowledge limitations about parameter values in cardiac electrophysiology has motivated studies in which virtual populations consisting of many models with varying parameter values are used to explore how populations as a whole, characterized by variable data, respond to different treatments or conditions~\cite{Britton2013,Passini2017}. This has included a Bayesian methodology for calibrating such populations~\cite{Lawson2018}. The Bayesian framework of model sloppiness, that provides a more global sense of parameter space curvature in the plausible region defined by the given data~(Fig.~\ref{fig:Beeler--Reuter relations}), could in fact be applied to the ``posteriors'' of such population-calibration processes, and thus provide a unique way to identify the combinations of parameters that are constrained (or not constrained) by the process of \hltext{fitting} these models to data exhibiting variability.
	
\section*{Discussion}

\subsection*{Recognizing the influence of prior information on the quality model-data fit}

The use of informative priors has been shown to help constrain model parameters when mathematical models are fitted to data in many Bayesian inference applications \cite{Wu2018,Choy2009,Adams2020a,Lawson2018,Drovandi2016}. Despite this advantage, the usage of uninformative priors has predominated in the context of analyzing model sloppiness \cite{Brown2003,Brown2004,Mannakee2016}.  In such a context, vague uniform priors, spanning several orders of magnitude, have been used to prevent potential optimization algorithm failures \cite{Gutenkunst2007,Brown2003,Transtrum2011}, rather than reflecting their true purpose: accounting for pre-existing knowledge about the parameter values \cite{Choy2009,Baker2019}. Here, we introduced how to account for informative priors when analyzing model sloppiness, with our example results illustrating how this approach identifies the relative effect of informative priors on the quality of the model-data fit. Specifically, the LIS method (matrix $\mathbf{G}$) was shown to reveal directions in parameter space where the posterior differs most strongly from the prior (Fig.~\ref{fig:eig_Scenario3_marginals}) while the posterior covariance method  (matrix $\mathbf{P}$) was shown to  reveal directions in parameter space that are strongly informed by  the combination of both  data and priors (Figs.~\ref{fig:eig_Scenario2_marginals} and~\ref{fig:eig_network_marginals}). Additionally, the Bayesian analysis of sloppiness (matrices $\mathbf{P}$ and $\mathbf{G}$) was shown to provide equivalent results to those based on earlier approaches (matrices $\mathbf{H}$ and $\mathbf{L}$) when uninformative (vague) priors are used in the implementation of Bayesian inference (Table~\ref{tb:eigenparameters M-M model}) and  when the data are very informative for the model parameters (Fig.~\ref{fig:Beeler--Reuter} and S11).  Consequently, we have demonstrated that the Bayesian approach to analyzing sloppiness complements earlier approaches \cite{Brown2003,Brown2004} in that the effects of prior beliefs on the quality of the model-data fit can be segregated when all methods are used together. This then clarifies which of the model parameters (or parameter combinations) are predominantly informed by the data or the prior.  

In the motivating Michaelis-Menten kinetics example and the cardiac electrophysiological application, we specifically showed that if stiff eigenparameters obtained from all methods  (matrices $\mathbf{H}$ or $\mathbf{L}$,  $\mathbf{P}$, and $\mathbf{G}$) are similar,  priors are weakly informative and so stiff eigenparameters are largely constrained by the data (Table~\ref{tb:eigenparameters M-M model}, Scenario~1 and Fig.~S11). We also showed in the motivating example that  if stiff eigenparameters obtained from the LIS method (matrix $\mathbf{G}$)  are similar to those obtained from the standard method (matrices $\mathbf{H}$ or $\mathbf{L}$) but different from those obtained from the posterior covariance method (matrix $\mathbf{P}$), then  critical parameter combinations associated with the posterior covariance method (matrix $\mathbf{P}$) are significantly influenced by the priors~(Table~\ref{tb:eigenparameters M-M model}, Scenario~2). Finally, we showed in the same motivating example that if  stiff eigenparameters obtained from the standard method (matrices $\mathbf{H}$ or $\mathbf{L}$) differ from those obtained from the Bayesian methods (matrices $\mathbf{P}$ and $\mathbf{G}$), then priors may also be influencing the \emph{quality} of the model-data fit. Under such conditions, stiff eigenparameters obtained from the LIS method (matrix $\mathbf{G}$) are informed by the data relative to the prior and those from the posterior covariance (matrix $\mathbf{P}$) are mostly constrained by the prior (Table~\ref{tb:eigenparameters M-M model}, Scenario~3). (A topographical interpretation of these findings is also provided in the next section.) In this way, we also demonstrated that our methods are well-suited for applications where there is little prior knowledge about the parameter values \cite{Brown2003,Brown2004,White2016,Mannakee2016} but also for those where prior beliefs can be confidently included as part of the model-data fitting process \cite{Wu2018,Choy2009,Adams2020a,Lawson2018,Drovandi2011}.

\hltext{In our implementation of Bayesian inference, we specifically considered combinations of   vague and informative uniform and/or log-normal prior distributions~(Figs.~S1a-S3a, S5 S8, and S10), as these types of priors are traditionally used in ensemble modeling applications in biochemistry \cite{Brown2003,Gutenkunst2007,Tomczak2019,Casey2007}, ecology \cite{Adams2020a,Adams2020,Baker2019}, and biology \cite{Johnstone2016,Drovandi2016,Lawson2018}. While implementation of Bayesian inference with application-specific prior distributions is beyond the scope of this work, the Bayesian methods (matrices $\mathbf{P}$ and $\mathbf{G}$) to analyzing model sloppiness may not be limited to the types of priors discussed here. We note, however, that for applications using heavy-tailed and/or sparsity-promoting priors in the implementation of Bayesian inference, we anticipate that more and/or better quality data would be required to obtain critical parameter combinations. Under this condition, the posterior covariance method (matrix $\mathbf{P}$) is expected to reveal data-informed stiff parameter combinations at least when the data are considerably more informative than the prior, such as in the Michaelis-Menten kinetics example~(Table~\ref{tb:eigenparameters M-M model}, Scenario 1). However, for implementation of the LIS method  (matrix $\mathbf{G}$), careful estimation of the prior covariance matrix for the logarithms of parameters $\bm{\Omega}$~(Eq.~\ref{eq:LIS-PriorPrecon}) would be required for the successful inference of stiff parameter combinations. Here, an interesting direction for future work would be to apply the prior normalization technique recently introduced by Cui~et~al.~\cite{Cui2022} in the context of Bayesian inverse problems to transform heavy-tailed priors into standard Gaussian distributions, to then implement the LIS method with these prior transformations in the context of analyzing model sloppiness.}

\hltext{Finally, we note that \rtwotext{the sensitivity matrices $\mathbf{P}$ and $\mathbf{G}$~(Eqs.~\ref{eq:cov} and \ref{eq:LIS-PriorPrecon}, respectively) are logarithmically based in this work, since standard methods for analyzing model sloppiness have been usually applied with the doubly logarithmic Hessian~(Eq.~\ref{eq:log-hessian}) owing to their history in analyzing complex systems describing physical processes~\cite{Mannakee2016,Gutenkunst2007,Hagen2013}.} \rtwotext{This is a transformation not typically used in classical implementation of PCA~\cite{Hotelling1933,Jolliffe2016}, however, it is not uncommon to implementation of the LIS method for Bayesian dimensionality reduction applications~\cite{Cui2014,Cui2016}.} While such a conveniently chosen transformation implies the use of single-sign prior distributions (\ie either positive or negative for each parameter) in the implementation of Bayesian inference, it also conveys three key advantages for the Bayesian analysis of model sloppiness: \textit{(i)} it reflects the positivity constraints on model parameters (true of the majority of parameters characterizing physical systems), \textit{(ii)} it prevents inconsistencies in scaling between parameters (due to different orders of magnitude) from impacting the analysis of sloppiness \cite{Gutenkunst2007}, and \textit{(iii)}  it provides a basis to identify stiff eigenparameters as products and/or quotients (combinations) of bare model parameters with different power indices whose magnitude inform the relative parameter importance in the combination~(Eq.~\ref{eq:eigparameters}) \cite{Brown2004}. If despite these advantages,  prior distributions spanning negative to positive values are required for a given application, nonlogarithmic versions of sensitivity matrices $\mathbf{P}$ and $\mathbf{G}$ may be used as part of the analysis of model sloppiness. Under this condition, eigendecomposition on such matrices will instead reveal stiff eigenparameters as linear combinations (summations and subtractions) of bare model parameters pre-multiplied by different coefficients whose magnitude represent the relative parameter importance in the combination. However, large differences in the magnitude of model parameters may mask the true stiff eigenparameters from these nonlogarithmic-based sensitivity matrices, so great care must be taken to use them for analysis of sloppiness.}

\subsection*{Characterizing the topography of the surface described by the model-data fit}

Our  work significantly adds to the literature on sensitivity analysis \cite{Marino2008,Saltelli2008}, which in the context of models fitted to data largely focuses on ``locally''  investigating  the curvature of the surface described by the likelihood function \cite{White2016,Brown2003,Gutenkunst2007,Transtrum2011,Mannakee2016}, around the best-fit parameter values  (MLE).  Thus, a key contribution of the Bayesian approach to analyzing sloppiness is that it accounts for changes in the curvature of this surface ``globally'' upon considering potentially plausible model parameter sets at a finite distance away from the best-fit parameter values \cite{Girolami2008,Lawson2018,Drovandi2016}. Additionally, our implementation of Bayesian inference advances upon earlier such implementations for analyzing model sloppiness \cite{Brown2003,Brown2004}. In these earlier works, a type of Markov Chain Monte Carlo (MCMC)  algorithm, with uninformative priors and started at the set of best-fit parameter values (MLE), was used to characterize the likelihood surface in the vicinity of the pre-identified MLE \cite{Gutenkunst2007,Mannakee2016}. However, we have shown here that different MLEs can lead to completely different locations on the surface describing the likelihood function~(Figs.~\ref{fig:key12-network} and~S7), which can mislead inference of stiff eigenparameters~(Table~\ref{tb:eigenparameters network model}). \hltext{More so, for systems in which the topography of the model-data fit function is very rugged, local optima can misguide the optimization algorithm \cite{Mannakee2016,Brown2004,Transtrum2011}, for example as illustrated by Fern{\'{a}}ndez~Slezak~et~al.~\cite{FernandezSlezak2010} in fitting a model of avascular tumor growth to noisy data using different optimization algorithms. As such, convergence issues become the bottleneck for the identification of stiff eigenparameters via standard approaches to analyzing sloppiness.} Instead, Bayesian inference  as implemented here does \hltext{not rely upon a single set of best-fit parameter values to characterize the surface describing the quality of the model-data fit (see Methods). Rather, all posterior samples contribute to the description of the surface topology. This stochastic exploration of the posterior distribution can reduce the risk of convergence to a local optimum \cite{Luengo2020,Doucet2000}}, with the added value that the Bayesian sensitivity matrices also acknowledge any effect of prior beliefs on the most plausible region in parameter space~(Figs.~\ref{fig:M-M_marginals}, \ref{fig:eig_network_marginals} and~\ref{fig:Beeler--Reuter Eig}).

In our example results, we specifically illustrated that comparison of stiff eigenparameters obtained from both the standard (matrices $\mathbf{H}$ or $\mathbf{L}$)   and Bayesian   (matrices $\mathbf{P}$ or $\mathbf{G}$) methods can reveal whether the topography of the surface described by the likelihood function is globally and locally similar as well as if such a surface is  similar to that described by the posterior distribution. For example, if stiff eigenparameters obtained from all methods (matrices  $\mathbf{H}$ or  $\mathbf{L}$, $\mathbf{P}$ and $\mathbf{G}$)  are similar, the shape of the surface described by the likelihood function  and  posterior distribution are locally and globally similar~(Table~\ref{tb:eigenparameters M-M model}, Scenarios 1 and Fig.~\ref{fig:Beeler--Reuter Eig}). Instead, if stiff eigenparameters obtained from the standard methods (matrices $\mathbf{H}$ or $\mathbf{L}$) are similar to those obtained from the LIS method (matrix $\mathbf{G}$) but differ from those obtained from the posterior covariance method (matrix $\mathbf{P}$), the shape of the surface described by the likelihood function  is locally and globally similar but different from that described by the posterior distribution~(Table~\ref{tb:eigenparameters M-M model}, Scenario 2 and Table~\ref{tb:eigenparameters network model}, with matrix $\mathbf{H}$   or $\mathbf{L}$ evaluated at $\bm{\theta}^\star_2$). Alternatively, if stiff eigenparameters obtained from all methods (matrices  $\mathbf{H}$ or  $\mathbf{L}$, $\mathbf{P}$ and $\mathbf{G}$) are different, the shape of the surface described by the likelihood function   is locally and globally different, but also different from the surface described by the posterior distribution~(Table~\ref{tb:eigenparameters M-M model}, Scenario 3 and Table~\ref{tb:eigenparameters network model}, with matrix $\mathbf{H}$   or $\mathbf{L}$ evaluated at $\bm{\theta}^\star_1$). We note, however, that while differences between the shape of surfaces describing the posterior distribution and likelihood function  are associated with effects of priors on quality of the model data-fit, identifying whether the likelihood is locally and globally similar is crucial when multiple (but also very different) parameter sets lead to similar values of the likelihood function~(Fig.~\ref{fig:key12-network}). This is a situation that is likely to occur when there is limited data to inform model parameters \cite{Mannakee2016,Transtrum2012}, for which the Bayesian sensitivity matrices are thoroughly informed by the data and the prior.

\hltext{We also note  that analysis of model sloppiness, including our new Bayesian perspective on the topic, describes the topography of the likelihood surface using the eigenvectors of the sensitivity matrix. Analogous to the use of PCA for dimension reduction~\cite{Hotelling1933}, this can be viewed in a sense as a linearized description of the topography. However, extending beyond this linearized view would require techniques that produce eigenvectors expressed in terms of the original parameters (as opposed to say, in a reproducing kernel Hilbert space~\cite{Scholkopf1998}). Rather, owing to the connections between the different sensitivity matrices and inverse covariance matrices, methods for improved covariance matrix estimation appear to be a promising direction for extending the way model sloppiness describes topography. For example, the graphical LASSO algorithm estimates sparse inverse covariance matrices that enforce conditional independence between some parameters. This might assist in the identification of stiff eigenparameters similar to how it can assist problems such as classification~\cite{Pavlenko2012}.}

\subsection*{Improving parameter identifiability by designing experiments based on identified parameter interdependencies}

Careful experimental design can improve ambiguous parameter inferences or even biased model predictions~\cite{White2016,Mannakee2016,Casey2007}. In the context of analyzing model sloppiness, much effort  has been devoted to studying effects on parameter identifiability by increasing the quality and quantity of the data used to fit the model \cite{Hagen2013,Tonsing2014,Apgar2010,Transtrum2012}. For example,  Apgar~et~al.~\cite{Apgar2010} carefully designed complementary experiments that constrained parameter values in the model of Brown~et~al.~\cite{Brown2004,Brown2003}. To achieve this, they modified some of the model inputs  to create  different synthetic datasets  that were more informative for some of the model parameters than others, but  when used together, all model parameters could be constrained within $10\%$ of their true values. However, these computational experiments still required considerably more data than those typically obtained in practice \cite{White2016,Mannakee2016}. Instead, we have shown here that the identification of  critical parameter interdependancies may more efficiently improve parameter inference when prior knowledge about related (interdependent) model parameters is strategically improved through expert elicitation or new experiments~(Fig.~S8). 

We also showed in the cardiac electrophysiological application that if experiments are designed to modify parameter values as well as the values of the stiff eigenparameters, these new experiments are likely to provide new information about the system~(Fig.~\ref{fig:BRsim1}). However, if experiments are  designed to change parameter values and instead keep the values of the stiff eigenparameters approximately constant, these  new experiments are unlikely to provide new information about the system~(Fig.~\ref{fig:BRsim2}). We note that if the design of parameter-specific experiments is not practically possible \cite{Hagen2013,Apgar2010,Transtrum2012}, the posterior covariance matrix (inverse of matrix $\mathbf{P}$) can still be used to measure the increase in parameter identifiability obtained by increasing the quantity and quality of data. Furthermore, this technique has been recently used in optimal Bayesian experimental design \cite{Kleinegesse2020}. 

\subsection*{\hltext{Identifying critical parameter combinations in stochastic settings}}

\hltext{In our example results, we identified critical parameter combinations through the analysis of sloppiness for three different deterministic models~(Eq.~\ref{eq:Michaelis--Menten}, and Tables S1 and~S3) fitted to data, in which we also treated the error structure as having been correctly specified by the modeler~(Eq.~\ref{eq:application likelihood}). However, implementing such an approach for stochastic models could be a potential area for future research. In stochastic models, randomness often manifests beyond just noise of known structure being added to a deterministic output~\cite{Beisbart2019}. More so, incorporation of a stochastic component can be used to include effects of random fluctuations into otherwise deterministic models; for example, to represent temporal variations in gene expression in cardiac electrophysiology models~\cite{Dangerfield2012}, similar to the one considered in this work~(Case study 2). However, such models present a challenge for understanding model sloppiness. Although methods suitable for stochastic models (such as minimum distance estimation~\cite{Grazzini2015}) can be used to statistically estimate the values of their parameters, with no closed-form version of the likelihood function available, nor any guarantee of its smoothness, one cannot reasonably evaluate and analyze the Hessian at this point as per the standard approach. Here, the Bayesian perspective on model sloppiness may provide a remedy for these issues. By adopting a likelihood-free method \cite{Sisson2018,Price2018}, posterior samples may still be generated, at which point the posterior covariance method (matrix $\mathbf{P}$) can be used to identify stiff eigenparameters in the context of both data and prior, as we have demonstrated~(Tables~\ref{tb:eigenparameters M-M model} and~\ref{tb:eigenparameters network model}).}

\hltext{In contrast to the posterior covariance method (matrix $\mathbf{P}$), the LIS-based approach presented here (matrix $\mathbf{G}$), separating the analysis of model sloppiness from the effects of the prior, \emph{does} rely upon large numbers of point-wise evaluations of the Hessian matrix~(Eq.~\ref{eq:LIS-PriorPrecon}). To rectify this for stochastic models, one may formulate approximations to the matrix $\mathbf{G}$ that avoid calculation of the Hessian by instead attempting to directly remove the effects of the prior from matrix $\mathbf{P}$. For example, the sensitivity matrices formed by subtracting the posterior inverse covariance $\bm{\Sigma}^{-1}$ from the prior inverse covariance $\bm{\Omega}^{-1}$~\cite{Beskos2018} or by pre-multiplying the posterior covariance $\bm{\Sigma}$ by the inverse prior covariance $\bm{\Omega}^{-1}$~\cite{Muller2012} have been used in the context of Bayesian inference to understand the posterior in connection with the prior. \rtwotext{Although these alternative approaches solve related eigenproblems, we put forward here} the sensitivity matrix  $\mathbf{G}$, formed by pre-multiplication and post-multiplication of the Hessian matrix by the Cholesky  factors~(Eq.~\ref{eq:LIS-PriorPrecon}) of the prior covariance matrix $\bm{\Omega}$. \rtwotext{This matrix factorization leads directly to the eigendirections (parameter combinations) where the data are most informative relative to the prior (further discussion in Methods)}.}

\subsection*{Recognizing knowledge limitations in mathematical models fitted to data}

Regardless of how good a mathematical model is, its predictions are only as useful as its known limitations~\cite{Adams2020a,Geary2020,Milner-Gulland2017}. Here, by recognizing  knowledge limitations in mathematical models fitted to data, our  work also adds to the literature of model development and simulation \cite{Villaverde2014,Geary2020,Velten2009}. For example, the identified stiff parameter combinations in the cardiac electrophysiological application uncovered a hidden controlling mechanism of the system~(Fig.~\ref{fig:Beeler--Reuter Eig}) that dictated the success or failure of the model output accurately reproducing the experimental data~(Fig.~\ref{fig:sim3}). In practical applications, identifying this type of model behavior would inform which of the model parameters are important for model reduction \cite{Transtrum2015} or need to be prioritized in future experimental designs~\cite{Apgar2010,Hagen2013,Casey2007}. Furthermore, the implementation of Bayesian inference \hltext{to fit the model to data} brings the added benefit  of quantifying the uncertainty in both parameter values~(Figs. S1A-S3A, S5, S8 and S10) and model predictions~(Figs.~\ref{fig:network model data}, \ref{fig:Beeler--Reuter fit}, S1B-S3B and S9). As such, this work constitutes an example of how this advanced model-data fitting technique can be exploited to reveal the hidden geometry of  parameter uncertainty and its effects on model predictions -- a topic of growing interest within many fields of science \cite{Geary2020,Mouquet2015,Villaverde2014,Marino2008} that has thus far been hindered due to concerns about system complexity and limited data accessibility \cite{Adams2020a,Geary2020,Milner-Gulland2017}.

\section*{Materials and Methods}

To assist with the subsequent mathematical description, we first summarize how sloppiness of a model is analyzed in its standard, non-Bayesian, form. Then, we describe how it can be analyzed via a Bayesian framework. Finally, we describe the procedure followed in the Results to identify the stiff eigenparameters according to both standard and Bayesian approaches to analyzing model sloppiness.

\subsection*{Standard (non-Bayesian) approach to analyzing sloppiness}

\noindent The standard approach to analyzing sloppiness involves three key steps \cite{Brown2003,Brown2004}:
\begin{enumerate}[nolistsep]
	\item obtaining the best-fit parameter values  $\bm{\theta}^\star$ by fitting the model to  data;
	\item calculating the sensitivity matrix $\mathbf{S}$  evaluated at the best-fit parameter values $\bm{\theta}^\star$, and
	\item identifying the eigenparameters that are more or less sensitive to the model-data fit  through eigendecomposition of the sensitivity matrix $\mathbf{S}$.	 
\end{enumerate}
We detail these three steps for analyzing sloppiness using the standard approach as follows.

\paragraph{Step 1.\ \hltext{Obtaining} values of the model parameters \hltext{by fitting the model to data}:} Let us assume that a \hltext{deterministic model $\bm{y}_{\mathrm{model}}=\bm{f}(\bm{x},\bm{\theta})$, with known structure $\bm{f}$, a known vector of  input conditions $\bm{x}\in\mathbb{R}^{N_x}$ of dimension $N_{x}$ (\eg representing the spatial and/or temporal location at which the model is evaluated, and/or the external conditions that alter the model output), and} parameterized by a vector $\bm{\theta} \in \mathbb{R}^{N_\theta}$ of dimension $N_\theta$, has been proposed to explain a dataset $\bm{y}_{\mathrm{obs}}$ that consists of $N_{\mathrm{obs}}$ independent observations $\bm{y}_{\mathrm{obs}} = (y_{\mathrm{obs},1},\;y_{\mathrm{obs},2},\cdots, y_{\mathrm{obs},N_{\mathrm{obs}}})$ where $y_{\mathrm{obs},k}$ represents the $k$th observation in this dataset, $k\in \left\{1, 2, \cdots,N_{\mathrm{obs}}\right\}$. \hltext{Here, the problem of uniquely estimating values of parameter set $\bm{\theta}$ given data $\bm{y}_{\mathrm{obs}}$ depends on whether the deterministic  model $\bm{y}_{\mathrm{model}}=\bm{f}(\bm{x},\bm{\theta})$ is structurally and, ultimately, practically identifiable, discussed in detail elsewhere \cite{Wieland2021,Rothenberg1971}. However, regardless of the source of parameter unidentifiabilities, the standard approach to model sloppiness considers identifiability of parameters in the context of their best-fit values \cite{Gutenkunst2007,Brown2004}. Typically, a likelihood-based approach is taken, in which the modeler specifies an error structure that then formalizes this notion of best-fit \cite{Dufresne2018,Transtrum2015}. For example, a common choice is to assume that errors are independent and with Gaussian distribution, each having mean zero and a specified standard deviation that could be observation-specific, $\sigma_k$~\cite{Apgar2010,White2016,Mannakee2016,Brown2003}. Under these conditions, the likelihood takes the form \cite{Casey2007,Transtrum2015}}
\begin{equation}
    	\mathcal{L}(\bm{y}_{\mathrm{obs}}|\bm{\theta}) = \displaystyle\prod_{k=1}^{N_{\mathrm{obs}}} \dfrac{1}{\sqrt{2 \pi} \sigma_k} \exp \left[ -\dfrac{1}{2} \left( \dfrac{y_{\mathrm{obs},k} - y_{\mathrm{model},k}(\hltext{\bm{x}},\bm{\theta})}{\sigma_k} \right)^2 \right],
	\label{eq:likelihood}
\end{equation}
where $y_{\mathrm{model},k}(\hltext{\bm{x}},\bm{\theta})$ is the model's prediction of an equivalent noiseless observation for $y_{k}$ given parameters $\bm{\theta}$ \hltext{and input conditions $\bm{x}$}. \hltext{The advantage of this likelihood-based approach is the ability to specify a given error structure that produces a tractable likelihood function, for example, incorporating heteroscedasticity in the data by varying $\sigma_k$ with each observation \cite{Adams2020,Adams2020a} as in Eq.~\ref{eq:likelihood}, or even choosing appropriate error distributions for more specific model-data fitting problems. While appropriate specification of the error structure could potentially depend on domain knowledge, Eq.~\ref{eq:likelihood} serves as a broadly applicable choice~\cite{Gutenkunst2007,Johnstone2016,Hagen2013,Tonsing2014,Adams2020}.}

The values of the model parameter vector $\bm{\theta}$ that maximize the likelihood function $\mathcal{L}(\bm{y}_{\mathrm{obs}}|\bm{\theta})$ are altogether called the maximum likelihood estimator (MLE), here denoted as $\bm{\theta}^\star \equiv \bm{\theta}^\star_{\mathrm{MLE}} = \mathrm{argmax}_{\bm{\theta}}\;\mathcal{L}(\bm{y}_{\mathrm{obs}}|\bm{\theta})$ \cite{Hagen2013}.  We note, however, that a standard least-squares regression may be cast as maximizing a Gaussian likelihood by \hltext{enforcing homoscedastic errors $\sigma_k = \sigma$ and} introducing \cite{Mannakee2016,Brown2003}:
\begin{equation}
	C(\bm{\theta})=-\log\mathcal{L}(\bm{y}_{\mathrm{obs}}|\bm{\theta})=\dfrac{1}{2}N_{\mathrm{obs}}\log\left(2 \pi\right)+\dfrac{1}{2} \hltext{N_{\mathrm{obs}} \log \sigma^2} +\dfrac{1}{2} \displaystyle\sum_{k=1}^{N_{\mathrm{obs}}}  \left( \dfrac{y_{\mathrm{obs},k} - y_{\mathrm{model},k}(\hltext{\bm{x}},\bm{\theta})}{\hltext{\sigma}} \right)^2, \label{eq:log-likelihood}
\end{equation}
where $C(\bm{\theta})$ is the cost function. The first two terms in Eq.~\ref{eq:log-likelihood} are independent of the parameter values, so $C(\bm{\theta})\propto \sum_{k=1}^{N_{\mathrm{obs}}}  \left(y_{\mathrm{obs},k} - y_{\mathrm{model},k}(\hltext{\bm{x}},\bm{\theta}) \right)^2+\mathrm{constant}$. Thus, minimizing the cost function $C(\bm{\theta})$ in Eq.~\ref{eq:log-likelihood} to find the best-fit parameter values is equivalent to maximizing the Gaussian likelihood function in Eq.~\ref{eq:likelihood} to find the MLE. \hltext{Furthermore, as a maximum likelihood estimator in these conditions, the ordinary least squares solution is an \emph{estimator} in the large sample limit, achieving the minimal variance specified by the Cram{\'{e}}r--Rao lower bound~\cite{Newey1994}.}

\paragraph{Step 2.\ Calculating the sensitivity matrix:} The standard approach to analyzing sloppiness obtains the sensitivity matrix $\mathbf{S}$ by investigating how the cost function $C(\bm{\theta})$ in Eq.~\ref{eq:log-likelihood} varies with respect to the parameter vector $\bm{\theta}$ in the vicinity of the maximum likelihood estimator $\bm{\theta}^\star=\bm{\theta}^\star_{\mathrm{MLE}}$. To do so, this matrix is obtained by a Taylor expansion of $C(\bm{\theta})$  around the best-fit parameter values while differentiating with respect to the logarithm of the parameters $(\log\bm{\theta})$, which yields \cite{Mannakee2016,Brown2004,Apgar2010}:
\begin{equation}
	C(\bm{\theta})\approx C(\bm{{\theta}}^\star)+\nabla C(\bm{\theta}^\star)\cdot\left(\log\bm{\theta}-\log\bm{\theta}^\star\right)+\frac{1}{2}\left(\log\bm{\theta}-\log\bm{\theta}^\star\right)^\top\mathbf{H}\left(\log\bm{\theta}-\log\bm{\theta}^\star\right),\label{eq:Taylor}
\end{equation}
where the gradient $\nabla C(\bm{\theta}^\star)$ of the cost function is zero at the best-fit parameter values by definition so that the sensitivity of the model fit to changes in parameter values is characterized by the Hessian matrix $\mathbf{H}$ defined in Eq.~\ref{eq:Taylor},  whose elements are given by \cite{Brown2003,Brown2004}: 
\begin{equation}
	H_{ij}=-\left.\frac{\partial^2 \log \mathcal{L}(\bm{y}_{\mathrm{obs}}|\bm{\theta})}{\partial \log\theta_i \,\, \partial \log\theta_j}\right|_{{\bm{\theta}}={\bm{\theta}^\star}},\label{eq:log-hessian}
\end{equation}
with $i$ and $j$ both taking integer values ranging from $1$ to $N_{\theta}$. Thus, the Hessian matrix describes the quadratic behavior of the cost function $C(\bm{\theta})$ infinitesimally close to the point $\bm{\theta}^\star$, and thus it is considered here as one of the matrices that could be used as the sensitivity matrix $\mathbf{S}$ for analyzing model sloppiness  \cite{Brown2003,Apgar2010,Mannakee2016}.  However, since evaluating second-order derivatives can be computationally expensive, the sensitivity matrix  can also be approximated by the Levenberg-Marquardt Hessian $\mathbf{L}$ at a much lower computational cost,  following \cite{Brown2003,Gutenkunst2007,Mannakee2016}:
\begin{equation}
L_{ij} = \displaystyle\sum_{k=1}^{N_{obs}}\left. \frac{\partial r_k}{\partial \log\theta_i} \dfrac{\partial r_k}{\partial \log\theta_j}\right|_{\bm{\theta}^\star=\bm{\theta}^\star_{\mathrm{MLE}}}, \label{eq:LM}
\end{equation}
where the residual error $r_k$ for the $k$-th observation is calculated via $r_k ={\left[y_{\mathrm{obs},k} - y_{\mathrm{model},k}(\hltext{\bm{x}},\bm{\theta}) \right]}/{\sigma_k}$, and the first derivatives in Eq.~\ref{eq:LM} can be evaluated by first-order finite differences or by integrating sensitivity equations for ordinary differential equation (ODE) models \cite{Gutenkunst2007,Hagen2013}. \rtwotext{The Levenberg-Marquardt Hessian $\mathbf{L}$ corresponds to the Gauss-Newton approximation of the Hessian $\mathbf{H}$ in Eq.~\ref{eq:log-hessian}, guaranteed to be positive semi-definite \cite{Rothenberg1971}.}  Matrix $\mathbf{L}$ is also equal to the observation information matrix evaluated at the MLE, which itself is a sample-based version of the Fisher information matrix (FIM) \cite{Transtrum2015}, whose connections with information theory have been well considered elsewhere \cite{Transtrum2015,Mannakee2016,Villaverde2014}. The  Levenberg-Marquardt Hessian $\mathbf{L}$ thus constitutes a more computationally convenient sensitivity matrix $\mathbf{S}$ for analyzing sloppiness, although as with the Hessian matrix $\mathbf{H}$, only considering the curvature of the likelihood surface infinitesimally close to the MLE.

\paragraph{Step 3.\ Identifying the eigenparameters that are more or less sensitive to the model-data fit:} To identify the stiff eigenparameters, eigenvalues $\lambda_n$ and eigenvectors $\bm{v}_n$ are obtained via eigendecomposition of the sensitivity matrix  $\mathbf{S}$, or via singular value decomposition  if numerical stability is an issue \cite{Brown2003}. Each of the $n=1,2,\ldots,N_\theta$ eigenvectors $\bm{v}_n$  are mutually orthogonal so that eigenparameters can be  conveniently expressed as linear combinations of natural logarithms of model parameters, following \cite{Brown2004}
\begin{equation}
	\alpha_{n}=\displaystyle\sum_{j=1}^{N_\theta}(v_n)_j\log\theta_j,\label{eq:raw eigparameters}
\end{equation}
where $(v_n)_j$ is the $j$th element of the $n$-$th$ eigenvector $\bm{v}_n$ of the sensitivity matrix. Thus,  each eigenparameter $\hat{\theta}_{n}$  can be simply represented  as the product and/or quotient of bare model parameters raised to an index  given by the elements of eigenvector $\bm{v}_n$,  by rewriting  Eq.~\ref{eq:raw eigparameters} as \cite{Brown2004}
\begin{equation}
	\hat{\theta}_{n}:=\exp\left(\alpha_{n}\right)= \displaystyle\prod_{j=1}^{N_\theta}\theta_j^{(v_n)_j},\label{eq:eigparameters}
\end{equation}
with  stiff eigenparameters  $\hat{\theta}_{n}$ associated with the largest eigenvalues $\lambda_n$ and sloppy (soft) eigenparameters  associated with the smallest eigenvalues. The magnitude of each element $(v_n)_j,\,j=1,\ldots,N_\theta$  of the eigenvector $\bm{v}_n$  in Eq.~\ref{eq:eigparameters} therefore indicates the relative contribution of bare parameter $\theta_j$ to eigenparameter $\hat{\theta}_n$. If eigenvectors are normalized, each $(v_n)_j$   takes values between $ $-$1$ and $1$ inclusive, so that all $\hat{\theta}_n$ are products of bare parameters having exponents with magnitudes that do not exceed unity. Any factors $\theta_j^{(v_n)_j}$ in Eq.~\ref{eq:eigparameters} having relatively low magnitudes for $(v_n)_j$ (\eg $\left|(v_n)_j\right| \leq0.2$)  contribute little to the eigenparameter's value, thus these small factors $\theta_j^{(v_n)_j}$ can be practically  excluded from the product  \cite{Brown2003}. Hence, each of the  $n=1,2,\ldots,N_\theta$ eigenparameters $\hat{\theta}_{n}$  may depend strongly on only a few bare parameters that may be importantly related to each other.  

\subsection*{Bayesian approach to analyzing sloppiness} 

In the context of  \hltext{fitting models to data}, Bayesian inference provides a coherent statistical framework to estimate probability distributions for model parameters, constrained by the combination of  data and prior beliefs  \cite{Girolami2008,Choy2009}. Thus, if the model-data \hltext{fitting problem} is recast as a Bayesian inference problem, the final estimates for the probability distribution of parameters $\bm{\theta}$, based on all of the data $\bm{y}_\mathrm{obs}$, is called the posterior distribution $\pi(\bm{\theta}|\bm{y}_\mathrm{obs})$. To apply Bayesian inference, we require definition of both a likelihood function $\mathcal{L}(\bm{y}_{\mathrm{obs}}|\bm{\theta})$  and a prior distribution $p(\bm{\theta})$. An example of the former of these was defined in Eq.~\ref{eq:likelihood}  (\ie Gaussian likelihood)  while the latter of these represents the initial beliefs about the parameter values, which are often based on earlier studies, or in their absence, they are based on expert knowledge \cite{Choy2009,Banner2020}. Once both likelihood function and prior distribution are defined, Bayes' Theorem is then used to obtain the posterior distribution, following  \cite{Gelman2013}
\begin{equation}
	\hltext{\pi(\bm{\theta}|\bm{y}_\mathrm{obs})=\frac{ \mathcal{L}(\bm{y}_{\mathrm{obs}}|\bm{\theta}) p(\bm{\theta})}{\int_{\bm{\Theta}} \mathcal{L}(\bm{y}_{\mathrm{obs}}|\bm{\theta})p(\bm{\theta})\, d\bm{\theta}}}.\label{eq:Bayes}
\end{equation}
\hltext{Here, the denominator is a multi-dimensional integral over the parameter space, $\bm{\Theta}$, that serves as a normalizing constant but is, however, difficult to calculate directly or often intractable \cite{Girolami2008,Gelman2013,Doucet2000}. Therefore, several methods that avoid calculation of this constant have been developed to sample from the posterior distribution}, including Markov Chain Monte Carlo (MCMC) sampling \cite{Robert1999}, Sequential Monte Carlo (SMC) sampling \cite{Doucet2000}, Approximate Bayesian Computation (ABC) \cite{Sisson2018}, Variational Bayesian Inference \cite{Jordan1999}, Laplace Approximation \cite{Tierney1986}, and many others. For the purposes of this section, we simply assume that the posterior has been successfully sampled, and thus we hereafter discuss practical aspects of computing the sensitivity matrix within a Bayesian framework. Thus, analogous to the standard approach to analyzing sloppiness, the Bayesian approach consists of three steps:
\begin{enumerate}[nolistsep]
	\item obtaining an estimate of the posterior distribution $\pi(\bm{\theta}|\bm{y}_\mathrm{obs})$ by \hltext{fitting} the model to data;
	\item calculating a Bayesian-based sensitivity matrix $\mathbf{S}$ from the posterior distribution  $\pi(\bm{\theta}|\bm{y}_\mathrm{obs})$, and
	\item identifying the eigenparameters that are more or less sensitive to the model-data fit  through eigendecomposition of the  Bayesian-based  sensitivity matrix $\mathbf{S}$.
\end{enumerate}
Exact details of the first step above depend on the posterior-computation method chosen while the third step is the same as the third step of the standard approach. Thus, we focus here on the second step, for which we \hltext{adapt} two  \hltext{Bayesian} methods \hltext{for dimensionality reduction} to obtain sensitivity matrices \hltext{for analyzing model sloppiness}. These are described as follows.

\paragraph{a.\ Posterior covariance method:} The posterior covariance method is based on the application of Principal Component Analysis (PCA) \cite{Hotelling1933}. This technique uses eigendecomposition of a sensitivity  matrix (a covariance matrix) to  reduce the dimensionality of large datasets, which thus identifies the dataset components that account for the largest amount of variance  \cite{Jolliffe2016}. In our context, the dataset of interest is a Bayesian ensemble of plausible parameter values, which we obtain from the posterior distribution for the parameters. Thus, if PCA is applied on this specific dataset, eigenvectors and eigenvalues of the posterior covariance matrix inform the variability of the model-data fit to changes in parameter values. However, given that we seek to identify the eigenparameters that are well-constrained by the available data  (\ie  those that have less variability), we  instead calculate the sensitivity matrix $\mathbf{S}$ as the PCA Hessian matrix $\mathbf{P}$ that is based on the inverse of the posterior covariance matrix $\bm{\Sigma}$ \cite{Brown2003},
\begin{equation}
\mathbf{P}= \bm{\Sigma}^{-1},\label{eq:P matrix}
\end{equation}
where the matrix $\bm{\Sigma}$ is calculated in terms of the natural logarithms of model parameters $\log\bm{\theta}$, with this transformation required in  Eq.~\ref{eq:eigparameters} to characterize stiff/sloppy eigenparameters as products or quotients of the bare model parameters. \hltext{This is a key advantage of the posterior covariance method over more sophisticated dimensionally reduction techniques (\eg kernel PCA~\cite{Scholkopf1998} and/or ISOMAP \cite{B.2000}), in which mappings back to original parameter space are not typically sought, and thus the associated eigenparameters describing the lower dimensional parameter space are not readily interpretable}. As such, eigendecomposition of the PCA Hessian matrix  $\mathbf{P}$ identifies which eigenparameters are more or less constrained by the combination of both  data and prior beliefs. Specifically, eigenvectors of matrix  $\mathbf{P}$ with large eigenvalues  indicate stiff eigenparameters, while eigenvectors with  small eigenvalues indicate sloppy eigenparameters.

We note that if Monte Carlo methods such as MCMC sampling \cite{Robert1999}, SMC sampling \cite{Doucet2000}  or ABC \cite{Sisson2018} are used to approximate the posterior as a set of $M$  equally weighted samples $\{\bm{\theta}_m\}_{m=1}^M$, the required posterior covariance matrix $\bm{\Sigma}$, calculated with respect to the natural logarithms of parameters $\log\bm{\theta}$,  can be estimated using the sample covariance matrix  $\hat{\bm{\Sigma}}$,
\begin{equation}
	\bm{\Sigma}\approx \hat{\bm{\Sigma}}= \frac{1}{M-1}\sum_{m=1}^M (\log\bm{\theta}_m - \log\bar{\bm{\theta}})(\log\bm{\theta}_m - \log\bar{\bm{\theta}})^\top, \label{eq:cov}
\end{equation}
where $\log\bar{\bm{\theta}} =  \frac{1}{M}\sum_{m=1}^M\log\bm{\theta}_m$ is the estimated posterior mean for the natural logarithm of parameters. If Monte Carlo  methods are overly computationally expensive, fast approximate methods such as Variational Bayesian Inference or Laplace Approximation \cite{Tierney1986} can be used as an alternative to provide a rapid estimate of the posterior covariance matrix. However, these fast approximate methods provide a rapid, albeit possibly biased, estimate of the posterior covariance matrix \cite{Tierney1986}.

\paragraph{b.\ Likelihood-informed subspace method:} The likelihood-informed subspace method proposed here has its origins in the Bayesian parameter reduction literature; specifically, from the work of Cui~et~al.~\cite{Cui2014} who developed a method for Bayesian inverse problems that identifies the directions in parameter space where the data are most ``informative'' relative to the prior.  The motivation for Cui~et~al.~\cite{Cui2014} was to develop an approximate but accelerated MCMC algorithm that samples over a lower-dimensional subspace, called the likelihood-informed subspace (LIS), to avoid sampling from directions of prior variability that the likelihood does not inform \cite{Cui2016}. The LIS is constructed on the idea that the Hessian of the log-likelihood can be compared to the prior covariance to then identify directions in parameter space along which the posterior distribution differs strongly from the prior, \ie  directions that are likelihood-informed \cite{Cui2016a}. Thus,  we adapt here the approach used by Cui~et~al.~\cite{Cui2014} to construct the LIS to define a sensitivity matrix  in our context. 

Our goal is to make the sensitivity matrix \hltext{dependent} primarily on the   data  and eliminate effects of the prior distribution. To achieve this, we firstly assume that the covariance matrix  $\bm{\Omega}$  of the prior distribution for the logarithms of parameters is known, and that this matrix can be Cholesky factored to a lower triangular matrix $\mathbf{L}_{\mathbf{p}}$ such that $\mathbf{L}_{\mathbf{p}} \mathbf{L}_{\mathbf{p}}^\top=\bm{\Omega}$.  Then, by following Cui~et~al.~\cite{Cui2014}, we define the prior-preconditioned Hessian matrix $\bm{\Psi}(\bm{\theta})$ as:
\begin{equation}
	\bm{\Psi}(\bm{\theta}) =\mathbf{L}_{\mathbf{p}}^\top \mathbf{H}(\bm{\theta}) \mathbf{L}_{\mathbf{p}},\label{eq:LIS-PriorPrecon}
\end{equation}
\hltext{for parameter vector $\bm{\theta}$}, with elements of $\mathbf{H}(\bm{\theta})$ given by  Eq.~\ref{eq:log-hessian}. We  note that Cui~et~al.~\cite{Cui2014} used a multivariate Gaussian prior to define the prior-preconditioned Hessian matrix $\bm{\Psi}(\bm{\theta})$  in Eq.~\ref{eq:LIS-PriorPrecon}, which is needed in that context to approximate the posterior distribution as the product of a lower-dimensional posterior defined on the LIS and the prior distribution marginalized onto a complementary subspace \cite{Cui2016,Cui2016a}. However, given that our purpose is  to identify the directions that are data-informed, and not to approximate a posterior distribution, the LIS definition is not restricted to multivariate Gaussian priors  in our application. Thus, we obtain an expression for the LIS, used here to define the LIS-based sensitivity matrix $\mathbf{G}$,  by integrating over the prior-preconditioned Hessian matrix with respect to the posterior \cite{Cui2014}, which yields 
\begin{equation}
	\mathbf{G} =\displaystyle\int_{\bm{\theta}}\bm{\Psi}(\bm{\theta}) \, \hltext{\pi(\bm{\theta}|\bm{y}_\mathrm{obs})} \, d \bm{\theta}.\label{eq:LIS-Integral}
\end{equation}
Given that Eq.~\ref{eq:LIS-Integral} involves an integral over $N_{\theta}\text{-dimensional space}$, then if the posterior is approximated by a Monte Carlo method (\eg MCMC, SMC or ABC) as a set of $M$ equally weighted samples $\{\bm{\theta}_m\}_{m=1}^M$, the LIS-based sensitivity matrix $\mathbf{G}$  can instead  be  estimated as
\begin{equation}
\mathbf{G}\approx \dfrac{1}{M}  \displaystyle\sum_{m=1}^M \bm{\Psi}(\bm{\theta}_m),\label{eq:LIS matrix}
\end{equation}
where each $\bm{\Psi}(\bm{\theta}_m)$ is calculated via Eq.~\ref{eq:LIS-PriorPrecon} with the Hessian matrix $\mathbf{H}(\bm{\theta}_m)$ of the negative log-likelihood function evaluated at  each posterior sample $\bm{\theta}_m$ via Eq.~\ref{eq:log-hessian} \rtwotext{or  approximated by the Levenberg-Marquardt Hessian $\mathbf{L}(\bm{\theta}_m)$ via Eq.~\ref{eq:LM} to reduce computational cost in the calculation of matrix $\mathbf{G}$ \cite{Cui2014}.} Further, we note that these $M$ matrices $\mathbf{H}(\bm{\theta}_m)$ are all left-multiplied by $\mathbf{L}_{\mathbf{p}}^\top$ and right-multiplied by $\mathbf{L}_{\mathbf{p}}$, with the $M$ resulting $\bm{\Psi}(\bm{\theta}_m)$ matrices averaged to obtain the sensitivity matrix $\mathbf{G}$. As a result, eigendecomposition of this prior-informed sensitivity matrix $\mathbf{G}$ can reveal which eigenparameters are strongly informed by the  data relative to the prior, \ie directions in parameter \emph{logspace} where the posterior differs most strongly from the prior \cite{Cui2016a}.

\subsection*{Demonstrating how to analyze model sloppiness using examples of models fitted to synthetic data}

In this section, we describe the six-step procedure used to  analyze model sloppiness in the examples discussed in the Results. This six-step procedure incorporates both approaches discussed above, \ie the local sensitivity analysis  around the best-fit parameter values (Standard approach) and the global sensitivity analysis considering all plausible parameter values consistent with the available data (Bayesian approach). Each step of the procedure describes specific details of the examples considered in the Results.

\paragraph{Step i.\ Defining the model form:} We consider \hltext{deterministic} models of the  form $\bm{y}_{\mathrm{model}}(\bm{x},\bm{\theta})$  where $\bm{x}\in \mathbb{R}^{N_x}$ is a vector of input conditions, $\bm{\theta} \in \mathbb{R}^{N_\theta}$ is a vector of model parameters, and $\bm{y}_{\mathrm{model}}\in \mathbb{R}^{N_y}$  is a vector of model outputs \hltext{(see Step 1, Standard approach)}. Here, ${N_x}$ and ${N_y}$ are the number of model inputs and outputs, respectively.
 
\paragraph{Step ii.\ Generating synthetic data to fit the model:} We generate measurement data for the motivating example and ecological application by adding heteroscedastic noise with variance proportional to the observation, which follows a truncated normal distribution  $y_{\mathrm{obs},j}(x_i)\sim\mathcal{N}(\mu_{y_{j}}(x_i),\sigma_{y_{j}}(x_i))$ with mean $\mu_{y_{j}}(x_i)=y_{\mathrm{model},j}(x_i,\bm{\theta}_R)$, standard deviation $\sigma_{y_{j}}(x_i)=\varepsilon y_{\mathrm{model},j}(x_i,\bm{\theta}_R)$ and lower truncation bound of zero  on each of the synthetic observations $y_{\mathrm{obs},j}(x_i)$ \cite{Adams2020}. Here,  $\bm{\theta}_R$ is the vector containing the reference (true) values for the model parameters, $\varepsilon$ is a user-defined measurement error  ranging between $0-100\%$, and noise is added to the $j\text{-th}$ model output associated with the $i\text{-th}$ set of input conditions. Alternatively, we generate  measurement data for the  cardiac electrophysiological application by adding homoscedastic noise, which follows a normal distribution $y_{\mathrm{obs},j}(x_i)\sim\mathcal{N}(\mu_{y_{j}}(x_i),\sigma_{y_{j}}(x_i))$ with mean $\mu_{y_{j}}(x_i)=y_{\mathrm{model},j}(x_i,\bm{\theta}_R)$ and constant standard deviation $\sigma_{y_{j}}(x_i)=\sigma$ \cite{Johnstone2016}. In each \hltext{case}, measurement error and sampling frequency (number of measurements) are chosen according to typical experimental conditions. \hltext{As later discussed in detail in Step iv, the choice of error structure used for synthetic data generation is also used to define the form of the likelihood function for each case. That is, the error structure is treated as having been correctly specified by the modeler.}

\paragraph{Step iii.\ Defining the vector of unknown model parameters and their prior distributions:} We define the vector of unknown model parameters $\bm{\theta}$ consisting of: \textit{(i)}  the model parameters, \textit{(ii)}  model initial conditions (only considered in the ecological model), and \textit{(iii)} measurement error $\varepsilon$ or standard deviation $\sigma$ following  the type of noise added to the synthetic data. Then, we  specify  prior distributions for the parameters $p(\bm{\theta})$ using either positive uniform or multivariate log-normal probability distributions \cite{Johnstone2016,Drovandi2016,Adams2020a,Baker2019,Tomczak2019}, as  follows.

In the Michaelis-Menten kinetics example, three different joint prior distributions $p(\bm{\theta})\equiv p(\splitatcommas{k_{cat},[E_T],K_M,\varepsilon})$ are used for the three  parameters $k_{cat}$, $[E_T]$ and $K_M$ of the model and measurement error $\varepsilon$.  The first joint prior  consist of a uniform prior for each  parameter; the second  joint prior consists of  multivariate log-normal priors for all parameters, with the prior of parameter $K_M$ being badly specified;  and the third joint prior consist of a uniform prior for $k_{cat}$, a badly specified log-normal prior for $[E_T]$, and well-specified log-normal prior for $ K_{M}$ and $\varepsilon$. All joint priors assume independence between the model parameters and measurement  error, so $p(\bm{\theta})\equiv p(k_{cat}) p([E_T]) p(K_M) p(\varepsilon)$.  \hltext{In this work}, a badly specified prior for the  $n$-$th$ parameter $\theta_n$  means that this parameter's true value $\theta_{R,n}$ has \hltext{little support under the prior distribution (\ie it lies in the tails of the prior). This is a condition referred to as prior-data conflict that occurs when informative prior beliefs are inconsistent with the information revealed by the data \cite{Evans2006}},  \hltext{even though the model is correctly specified as is assumed here (see Ref.~\cite{Walker2013} for discussion of appropriateness of Bayesian inference when the  model is misspecified)}. Alternatively, a well-specified prior \hltext{in this work} means that the true parameter value $\theta_{R,n}$ \hltext{is well supported by the prior distribution, \ie it lies within the bulk of the prior distribution so that prior beliefs are consistent with the information given by the data}. 

In the ecological application, two different joint prior distributions are used for the twenty parameters  of the ecosystem network model~(Table~S2) and measurement error $\varepsilon$. The first joint prior distribution is chosen to be a product of vague log-normal distributions for each parameter so that this joint prior possesses zero covariance. Alternatively, the second joint prior distribution is chosen to be a product of well-specified log-normal distributions for parameters $a_M$, $a_N$ and $a_P$ and vague log-normal distributions for each of the remaining parameters, including the measurement error $\varepsilon$. As discussed in the Results, well-specified priors for parameters $a_M$, $a_N$ and $a_P$ are chosen based on the stiff eigenparameters, identified from the analysis of sloppiness for the case considering  vague log-normal distributions for each parameter. 

In the cardiac electrophysiological application, a well-specified multivariate log-normal prior distribution is used for the nine parameters of the Beeler--Reuter (BR) model~(Table~S4) and the standard deviation $\sigma$. This joint prior distribution is centered at the reference parameter values and assumes zero covariance between the nine parameters and the standard deviation $\sigma$. Stimulation conditions $A_s$, $t_{on}$ and $t_{dur}$, membrane capacitance $C_m$, and initial conditions $V_m(0)$, $[Ca]_i(0)$, $x_1(0)$, $m(0)$,  $h(0)$, $j(0)$, $d(0)$ and $f(0)$ are set to their reference values~(Table~S4), and are not estimated via our model-data fitting techniques. 

\paragraph{Step iv.\ \hltext{Fitting the model to data:}} We use two approaches to fit each example model to  data, with the first being maximum likelihood estimation (MLE) and the second being Bayesian inference. To implement these two approaches, we conveniently rewrite the Gaussian likelihood function defined in Eq.~\ref{eq:likelihood}, as
\begin{equation}
	\mathcal{L}(\bm{y}_\mathrm{obs}|\bm{\theta}) = \displaystyle\prod_{j=1}^{N_{\mathrm{y}}} \prod_{i=1}^{N_{\mathrm{x}}}\dfrac{1}{\sqrt{2 \pi} \sigma_{y_{j}}(x_i)} \exp \left[ -\dfrac{1}{2} \left( \dfrac{y_{\mathrm{obs},j}(x_i) - y_{\mathrm{model},j}(x_i,\bm{\theta})}{\sigma_{y_{j}}(x_i)} \right)^2 \right],
	\label{eq:application likelihood}
\end{equation}
where $\sigma_{y_{j}}(x_i)=\varepsilon y_{\mathrm{model},j}(x_i,\bm{\theta})$ when heteroscedastic noise is used to generate the synthetic data and $\sigma_{y_{j}}(x_i)=\sigma$ when homoscedastic noise is instead used. Then, we use this Gaussian likelihood function  and specified prior distributions~(Step iii) to approximate  the joint posterior distribution $\pi(\bm{\theta}|\bm{y}_\mathrm{obs})$ via Bayes' Theorem (Eq.~\ref{eq:Bayes}) by implementing the SMC sampling algorithm adapted from Adams~et~al.~\cite{Adams2020,Adams2020a}. In our implementation of this  posterior sampling algorithm, we use a sample size of $M=10000$, Metropolis-Hastings acceptance  fraction of $C=0.95$ and effective sample size reduction  target of $\Delta=0.001$. These settings were sufficient for reproducible sampler performance: results did not vary in independent runs of the  sampling algorithm  using a smaller sample size of $M=5000$  and larger effective sample size reduction target of $\Delta=0.005$~(\hltext{Figs.~S15-S21}).

Once the joint posterior probability distributions $\pi(\bm{\theta}|\bm{y}_\mathrm{obs})$ are obtained for each example, we estimate the best-fit parameter values $\bm{\theta}^\star$ (maximum likelihood estimator) by minimizing  the cost-function $C(\bm{\theta})=-\log\mathcal{L}(\bm{y}_\mathrm{obs}|\bm{\theta})$ with  $\mathcal{L}(\bm{y}_\mathrm{obs}|\bm{\theta})$ given by Eq.~\ref{eq:application likelihood} while using the posterior mean as the initial guess  to  start the optimization. Here, the sets of best-fit parameter values $\bm{\theta}^\star$, $\bm{\theta}_1^\star$ and $\bm{\theta}_2^\star$  are only used to  calculate the sensitivity matrices ($\mathbf{H}$ or $\mathbf{L}$) via the standard approach  while the already obtained prior and posterior distributions are used to calculate the sensitivity matrices ($\mathbf{P}$ and $\mathbf{G}$) based on the Bayesian approach. 

\paragraph{Step v.\ Calculating the sensitivity matrix:} Eqs.~\ref{eq:log-hessian}, \ref{eq:LM}, \ref{eq:P matrix}  and \ref{eq:LIS matrix} are used to calculate the  Hessian $\mathbf{H}$,  the Levenberg-Marquardt Hessian $\mathbf{L}$, the  PCA Hessian $\mathbf{P}$, and the likelihood-informed subspace $\mathbf{G}$, respectively.  Each of these matrices acts as a sensitivity matrix for the purpose of analyzing sloppiness. In the absence of analytical derivatives, we use central finite differences \cite{Hoffman2018} to approximate first and second order derivatives of the log-likelihood function with respect to the logarithm of parameters, with a step size  $\Delta\theta_i=\delta\times\theta_i,i=1,\ldots,N_\theta$  where $\delta$ is a small scalar between $10^{-4}$ and $10^{-2}$. \hltext{Finite differencing is the most widely used technique for numerical differentiation in physical applications \cite{Hoffman2018}, including approximations of the sensitivity matrix ($\mathbf{H}$ or $\mathbf{L}$) in standard analysis of model sloppiness \cite{Brown2004,Apgar2010,Mannakee2016}. However, for more complex models than those considered here, this technique can become computationally expensive, as it requires multiple model evaluations for approximating derivatives. As an alternative, more sophisticated methods such as automatic differentiation \cite{Bucker2006} may also be used (where appropriate)  in conjunction with the analysis of model sloppiness \cite{Gutenkunst2007,Transtrum2017}.}

Rows and columns of each sensitivity matrix characterizing the sensitivity of the model-data fit with respect to the measurement error  \hltext{(represented by $\varepsilon$ or $\sigma$ in Eq.~\ref{eq:application likelihood})} are not calculated,  \hltext{thus preventing the measurement error from appearing in the parameter combinations that are identified through the analysis of model sloppiness. Here, the measurement error is effectively treated as a nuisance parameter, that is, it is involved with the model-data fitting procedure but does not provide information to identify relevant parameter combinations. Additionally, for likelihood functions of the form given by Eq.~\ref{eq:application likelihood}, small changes to the measurement error are expected to affect only the degree of overall curvature of the model-data fit surface, but not the directions of high or low curvature.} As a result, the dimension of the square symmetric sensitivity matrices  $\mathbf{H}$, $\mathbf{L}$, $\mathbf{P}$, and $\mathbf{G}$  obtained here is equal to the number of model parameters, excluding the measurement error \hltext{($\varepsilon$ or $\sigma$ in Eq.~\ref{eq:application likelihood})}, \ie we obtain $3\times 3$ sensitivity matrices in the Michaelis-Menten kinetics example,  $20\times 20$ in the ecological application, and $9\times 9$ in the cardiac electrophysiological application. 

\paragraph{Step vi.\ Identifying stiff eigenparameters:}  Eigenvalues and eigenvectors of the sensitivity matrices $\mathbf{H}$, $\mathbf{L}$, $\mathbf{P}$ and $\mathbf{G}$ are calculated via singular value decomposition \cite{Brown2003}. Then, eigenparameters $\hat{\bm{\theta}}$ are obtained via Eq.~\ref{eq:eigparameters}, in which we consider the contribution of  parameter $\theta_j$ to eigenparameter $\hat{\theta}_n$ only when element $j$ of the normalized $n\text{-th}$ eigenvector $(v_n)_j$ satisfies $\left|(v_n)_j\right|\geq0.2$ (see Step 3, Standard approach). We also rescale exponents $(v_n)_j$ of the bare parameters $\theta_j$ associated with each eigenparameter $\hat{\theta}_n$, so that the magnitude of the largest/smallest index $(v_n)_j$ for every eigenvector $\bm{v}_n$ is $1$. Here, eigenvalues are ordered from  largest to smallest, so that the corresponding eigenparameters are also ordered from stiffest to sloppiest.
    
\subsection*{Trade-offs of locally and globally analyzing  model sloppiness}

To summarize these methods, we have \hltext{proposed a unified framework} to  obtain locally and globally the key quantity for analyzing model sloppiness -- the sensitivity matrix $\mathbf{S}$ (\eg $\mathbf{H}$, $\mathbf{L}$,  $\mathbf{P}$, and $\mathbf{G}$).  This approach accurately estimates uncertainty in parameter values, constrained by the combination of prior information and data, with the key benefit of robustly identifying the relative effect of this prior information in the inference of critical parameter combinations that control the quality of the model-data fit. This is a key achievement of this work as it extends the application of the analysis of sloppiness beyond systems where there is little prior knowledge about the model parameter values \cite{Brown2003,Gutenkunst2007} to those where prior information is more readily available \cite{Baker2019,Wu2018,Choi2017,Adams2020a}, and thus can be confidently incorporated as part of the Bayesian model-data fitting process to constrain parameter values \cite{Lawson2018,Drovandi2016,Johnstone2016,Adams2020}.

 In the implementation of this framework, the local (standard) methods  to analyzing sloppiness (matrices $\mathbf{H}$ or $\mathbf{L}$) were found to be computationally inexpensive in comparison to the Bayesian methods (matrices $\mathbf{P}$ and $\mathbf{G}$). Thus, standard methods can be very useful in model-data fitting applications where computationally expensive models make implementation of Bayesian inference impractical.  Nevertheless, since local analysis of sloppiness considers a single point estimation in parameter space (\ie the best-fit parameter values), this local approach can only accurately quantify the model sensitivity to parameter changes when the likelihood function maximum (or cost function minimum) is well defined \cite{Transtrum2012,Gutenkunst2007}. Unfortunately, if the likelihood surface is relatively complicated (\eg with ridges), this method can mislead inference of stiff eigenparameters (see, for example, Table~\ref{tb:eigenparameters network model}) \cite{Mannakee2016,FernandezSlezak2010}. \hltext{Careful selection of the optimization algorithm is thus needed to avoid convergence to local optima \cite{Transtrum2011,FernandezSlezak2010,Brown2003}.}  \hltext{In addition to this limitation, both standard methods to analyzing sloppiness require a closed-form likelihood function to calculate  sensitivity matrices $\mathbf{H}$ and $\mathbf{L}$, such as the Gaussian likelihood functions~(\eg Eqs.~\ref{eq:likelihood} and \ref{eq:application likelihood}) considered here.}  
 
 Alternatively, the global (Bayesian) analysis of sloppiness  looks beyond the curvature of the likelihood function surface at a single point  while fully exploring the topography of this surface by  using an ensemble of plausible parameter values to characterize the sensitivities of the model-data fit to changes in parameter values.  As part of this global approach, \hltext{we exploited PCA~\cite{Hotelling1933} to implement the posterior covariance method (matrix $\mathbf{P}$)} that assesses the data informativity about the critical parameter combinations while accounting for (including) any prior information about parameter values. This method  does not require approximating gradients of the log-likelihood~(Eq.~\ref{eq:P matrix}), and so calculating sensitivity matrix $\mathbf{P}$ is computationally inexpensive after the posterior distribution is obtained via Bayesian Inference. However, since the posterior covariance method (matrix $\mathbf{P}$) assumes that the posterior structure is well captured by a covariance matrix  $\bm{\Sigma}$ of the logarithm of parameters $\log\bm{\theta}$, it is also restricted to applications where the posterior distribution for $\log\bm{\theta}$  is approximately multivariate normally distributed. Despite this limitation, the posterior covariance method has the added benefit of being readily applicable to all kinds of statistical models, even to those with intractable likelihood functions where ``likelihood-free" Bayesian methods, such as Approximate Bayesian Computation \cite{Sisson2018} and Bayesian Synthetic Likelihood \cite{Price2018}, are prevalent. 

As part of the global approach, we also \hltext{implemented} the likelihood-informed subspace (LIS) method \hltext{for Bayesian dimensionality reduction \cite{Cui2014,Spantini2015} to analyze model sloppiness. Following its origins, the LIS method (matrix $\mathbf{G}$) adapted here assesses} the data informativity about the critical parameter combinations while also acknowledging  and excluding any prior information. \hltext{As with the standard methods (matrices $\mathbf{H}$ or $\mathbf{L}$), the LIS method requires a closed-form likelihood function~(\eg Gaussian likelihood functions in Eqs.~\ref{eq:likelihood} and \ref{eq:application likelihood}) to obtain the sensitivity matrix $\mathbf{G}$~(Eq.~\ref{eq:LIS matrix}). More so, given that the LIS method also involves calculation  of the Hessian matrix at a posterior sample~(Eq.~\ref{eq:LIS-PriorPrecon}), approximating second-order derivatives for all posterior samples can become computationally expensive via finite differencing for models with time-consuming solutions}. Despite these limitations, as the LIS method does not assume a given shape for the posterior distribution, it has the added benefit of being readily applicable to systems with non-Gaussian posterior distributions. \hltext{On the other hand, where the posterior distribution  \rtwotext{\emph{is close}} to a Gaussian, one may replace the LIS's average over Hessians~(Eq.~\ref{eq:LIS matrix}) with the posterior covariance}, 
\begin{equation}
  \rtwotext{\mathbf{K}} \approx \rtwotext{\mathbf{L}_{\mathbf{p}}^\top} \bm{\Sigma}^{-1}  \rtwotext{\mathbf{L}_{\mathbf{p}}}\;\;\;\;\;\rtwotext{\text{or}\;\;\;\;\; \mathbf{K}^{-1}\approx\mathbf{L}_{\mathbf{p}}^{-1}\bm{\Sigma}\mathbf{L}_{\mathbf{p}}^{-\top}}
  \label{eq:aproxLIS}
\end{equation}
\rtwotext{with $\mathbf{L}_{\mathbf{p}} \mathbf{L}_{\mathbf{p}}^\top=\bm{\Omega}$ (see LIS method),}  hence formulating a likelihood-free approximation to matrix $\mathbf{G}$. This makes the LIS applicable for stochastic models with intractable likelihoods, and greatly reduces the extra computational cost of Hessian calculation at all posterior samples. This idea is similar to other approaches comparing prior and posterior covariance matrices to understand the posterior in the context of the prior~\cite{Muller2012,Beskos2018}, \rtwotext{arising from the generalized eigenproblem $\mathbf{H}\mathbf{v}\approx\bm{\Sigma}^{-1}\mathbf{v}=\bm{\lambda}\bm{\Omega}^{-1}\mathbf{v}$ upon approximating the Hessian with the inverse covariance matrix $\bm{\Sigma}^{-1}$ for Gaussian settings~\cite{Spantini2015}. Matrix $\mathbf{K}$ can also be obtained by transforming this generalized eigenproblem into a standard eigenproblem, for which the eigenvectors are then readily interpretable for analyzing model sloppiness. Indeed, as the eigenvectors of matrix $\mathbf{\Sigma}$ in Eq.~\ref{eq:cov} are equivalent to those of matrix $\mathbf{H}$ in Eq.~\ref{eq:LIS-PriorPrecon}   for a Gaussian posterior distribution, matrix  $\mathbf{G}$ in Eq.~\ref{eq:LIS matrix} and matrix $\mathbf{K}$ (or $\mathbf{K}^{-1}$) in Eq.~\ref{eq:aproxLIS} share the same eigenvectors (see Ref.~\cite{Spantini2015} for discussion of eigenproperties of matrices $\mathbf{H}$ and $\mathbf{\Sigma}$ under Gaussian settings).} 

\hltext{Beyond the Bayesian methods discussed here, for applications in which sampling the posterior distribution is infeasible or simply impractical, forward sensitivity analysis methods such as the active subspace~\cite{Constantine2016} could potentially be used as an alternative to assess sensitivities of the model-data fit function to changes in parameter values. \rtwotext{The active subspace (AS) method has} the advantage of evaluating a similar sensitivity matrix at a prior sample (referred to as matrix $\mathbf{C}$ by Constantine~\cite{Constantine2016}), \rtwotext{which makes its implementation less computationally expensive than that of the LIS method since a posterior sample is not needed}. \rtwotext{Similar to the LIS, the AS identifies a set of important (stiff) directions in the space of all parameters \cite{Cui2021,Constantine2016}}. \rtwotext{However, the method also has been recently shown to be not completely analogous to the LIS method for both Gaussian and non-Gaussian settings in the context of Bayesian dimensionality reduction~\cite{Zahm2022}. Consequently,}  for analyzing model sloppiness, eigenparameters obtained from the AS method are expected to have a different interpretation than that of eigenparameters obtained from matrices $\mathbf{P}$ and $\mathbf{G}$ in relation to acknowledging the source of information (\ie prior and/or data). Thus, exploring how the AS method compares to the Bayesian methods discussed here could be an interesting direction for future work.} 

Hence, given the great flexibility of the techniques discussed here to unveil sensitivities of the model-data fit to changes in parameter values,  our comprehensive approach to analyzing model sloppiness does comprise a suitable set of tools to aid understanding of many of nature's systems, ranging from a single cell in the human body \cite{Lawson2018,Johnstone2016} and the myriad of microorganisms found almost everywhere \cite{Villaverde2014,Mouquet2015,Drovandi2011} to large ecosystem networks  \cite{Adams2020,Geary2020} and beyond \cite{Sundberg2010,Velten2009}, through the simultaneous usage of experimental data, mathematical models, and computer simulation. 

\section*{List of Supplementary Materials:}
Figures' Supplementary Legends
\\
Figs.\ S1 to S21
\\
Tables S1 to S4
\nocite{Bowman1997,Jones1993,Dokos2017}
\bibliography{references}

\begin{thebibliography}{10}

\bibitem{Maayan2017}
A.~Ma'ayan, {\it Journal of The Royal Society Interface\/} {\bf 14}, 20170391
  (2017).

\bibitem{Geary2020}
W.~L. Geary, {\it et~al.\/}, {\it Nature Ecology \& Evolution\/}  (2020).

\bibitem{Villaverde2014}
A.~F. Villaverde, J.~R. Banga, {\it Journal of The Royal Society Interface\/}
  {\bf 11}, 20130505 (2014).

\bibitem{Mouquet2015}
N.~Mouquet, {\it et~al.\/}, {\it Journal of Applied Ecology\/} {\bf 52}, 1293
  (2015).

\bibitem{Drovandi2011}
C.~C. Drovandi, A.~N. Pettitt, {\it Biometrics\/} {\bf 67}, 225 (2011).

\bibitem{Schlick2010}
T.~Schlick, {\it {Molecular modeling and simulation: an interdisciplinary
  guide}\/}, vol.~2 (Springer, 2010).

\bibitem{Lawson2018}
B.~A.~J. Lawson, {\it et~al.\/}, {\it Science Advances\/} {\bf 4}, e1701676
  (2018).

\bibitem{Johnstone2016}
R.~H. Johnstone, {\it et~al.\/}, {\it Journal of Molecular and Cellular
  Cardiology\/} {\bf 96}, 49 (2016).

\bibitem{Velten2009}
K.~Velten, {\it {Mathematical modeling and simulation : introduction for
  scientists and engineers}\/} (Wiley-VCH, Weinheim Germany, 2009).

\bibitem{Sundberg2010}
M.~Sundberg, {\it Science, Technology, \& Human Values\/} {\bf 37}, 64 (2010).

\bibitem{Gutenkunst2007}
R.~N. Gutenkunst, {\it et~al.\/}, {\it PLoS Computational Biology\/} {\bf 3},
  e189 (2007).

\bibitem{Brown2003}
K.~S. Brown, J.~P. Sethna, {\it Physical Review E\/} {\bf 68}, 021904 (2003).

\bibitem{Brown2004}
K.~S. Brown, {\it et~al.\/}, {\it Physical Biology\/} {\bf 1}, 184 (2004).

\bibitem{Lewbel2019}
A.~Lewbel, {\it Journal of Economic Literature\/} {\bf 57}, 835 (2019).

\bibitem{Mannakee2016}
L.~Geris, D.~Gomez-Cabrero, {\it Uncertainty in Biology: A Computational
  Modeling Approach\/} (Springer International Publishing, 2016).

\bibitem{Transtrum2011}
M.~K. Transtrum, B.~B. Machta, J.~P. Sethna, {\it Physical Review E\/} {\bf
  83}, 36701 (2011).

\bibitem{Adams2020}
M.~P. Adams, {\it et~al.\/}, {\it Ecology Letters\/} {\bf 23}, 607 (2020).

\bibitem{Marino2008}
S.~Marino, I.~B. Hogue, C.~J. Ray, D.~E. Kirschner, {\it Journal of Theoretical
  Biology\/} {\bf 254}, 178 (2008).

\bibitem{Saltelli1993}
A.~Saltelli, T.~H. Andres, T.~Homma, {\it Computational Statistics \& Data
  Analysis\/} {\bf 15}, 211 (1993).

\bibitem{Sobol2001}
I.~M. Sobol', {\it Mathematics and Computers in Simulation\/} {\bf 55}, 271
  (2001).

\bibitem{Saltelli2008}
A.~Saltelli, {\it et~al.\/}, {\it {Global sensitivity analysis : the primer}\/}
  (John Wiley \& Sons, Ltd, 2008).

\bibitem{Girolami2008}
M.~Girolami, {\it Theoretical Computer Science\/} {\bf 408}, 4 (2008).

\bibitem{Gelman2013}
A.~Gelman, {\it et~al.\/}, {\it {Bayesian data analysis}\/} (Chapman and
  Hall/CRC, New York, 2013), third edn.

\bibitem{Luengo2020}
D.~Luengo, L.~Martino, M.~Bugallo, V.~Elvira, S.~S{\"{a}}rkk{\"{a}}, {\it
  EURASIP Journal on Advances in Signal Processing\/} {\bf 2020}, 25 (2020).

\bibitem{Drovandi2016}
C.~C. Drovandi, {\it et~al.\/}, {\it Journal of The Royal Society Interface\/}
  {\bf 13}, 20160214 (2016).

\bibitem{Transtrum2015}
M.~K. Transtrum, {\it et~al.\/}, {\it The Journal of Chemical Physics\/} {\bf
  143}, 010901 (2015).

\bibitem{White2016}
A.~White, {\it et~al.\/}, {\it PLOS Computational Biology\/} {\bf 12}, e1005227
  (2016).

\bibitem{Transtrum2017}
M.~K. Transtrum, A.~T. Sari{\'{c}}, A.~M. Stankovi{\'{c}}, {\it IEEE
  Transactions on Power Systems\/} {\bf 32}, 2243 (2017).

\bibitem{Hagen2013}
D.~R. Hagen, J.~K. White, B.~Tidor, {\it Interface Focus\/} {\bf 3}, 20130008
  (2013).

\bibitem{Apgar2010}
J.~F. Apgar, D.~K. Witmer, F.~M. White, B.~Tidor, {\it Molecular BioSystems\/}
  {\bf 6}, 1890 (2010).

\bibitem{Dufresne2018}
E.~Dufresne, H.~A. Harrington, D.~V. Raman, {\it Journal of Algebraic
  Statistics\/} {\bf 9}, 30 (2018).

\bibitem{Wu2018}
P.~P.-Y. Wu, M.~J. Caley, G.~A. Kendrick, K.~McMahon, K.~Mengersen, {\it
  Applied Statistics\/} {\bf 67}, 417 (2018).

\bibitem{Choy2009}
S.~L. Choy, R.~O'Leary, K.~Mengersen, {\it Ecology\/} {\bf 90}, 265 (2009).

\bibitem{Baker2019}
C.~M. Baker, {\it et~al.\/}, {\it Ecological Applications\/} {\bf 29}, e01811
  (2019).

\bibitem{Cui2014}
T.~Cui, J.~Martin, Y.~M. Marzouk, A.~Solonen, A.~Spantini, {\it Inverse
  Problems\/} {\bf 30}, 114015 (2014).

\bibitem{Hotelling1933}
H.~Hotelling, {\it Journal of Educational Psychology\/} {\bf 24}, 417 (1933).

\bibitem{Spantini2015}
A.~Spantini, {\it et~al.\/}, {\it SIAM Journal on Scientific Computing\/} {\bf
  37}, A2451 (2015).

\bibitem{Jolliffe2016}
I.~T. Jolliffe, J.~Cadima, {\it Philosophical Transactions of the Royal Society
  A: Mathematical, Physical and Engineering Sciences\/} {\bf 374}, 20150202
  (2016).

\bibitem{Rothenberg1971}
T.~J. Rothenberg, {\it Econometrica\/} {\bf 39}, 577 (1971).

\bibitem{Casey2007}
F.~P. Casey, {\it et~al.\/}, {\it IET systems biology\/} {\bf 1}, 190 (2007).

\bibitem{Michaelis1913}
L.~Michaelis, M.~Menten, {\it Biochem. Z\/} {\bf 49}, 333 (1913).

\bibitem{Pech1998}
R.~P. Pech, G.~M. Hood, {\it Journal of Applied Ecology\/} {\bf 35}, 434
  (1998).

\bibitem{Beeler1977}
G.~W. Beeler, H.~Reuter, {\it The Journal of Physiology\/} {\bf 268}, 177
  (1977).

\bibitem{Wieland2021}
F.-G. Wieland, A.~L. Hauber, M.~Rosenblatt, C.~T{\"{o}}nsing, J.~Timmer, {\it
  Current Opinion in Systems Biology\/} {\bf 25}, 60 (2021).

\bibitem{Choi2017}
B.~Choi, G.~A. Rempala, J.~K. Kim, {\it Scientific Reports\/} {\bf 7}, 17018
  (2017).

\bibitem{Briggs1925}
G.~E. Briggs, J.~B. Haldane, {\it The Biochemical journal\/} {\bf 19}, 338
  (1925).

\bibitem{Tomczak2019}
J.~M. Tomczak, E.~W{\c{e}}glarz-Tomczak, {\it FEBS Letters\/} {\bf 593}, 2742
  (2019).

\bibitem{Cui2016}
T.~Cui, Y.~Marzouk, K.~Willcox, {\it Journal of Computational Physics\/} {\bf
  315}, 363 (2016).

\bibitem{Krogh-Madsen2020}
T.~Krogh-Madsen, D.~J. Christini, eds., {\it {Modeling and Simulating Cardiac
  Electrical Activity}\/} (IOP Publishing, 2020).

\bibitem{Zhou2018}
X.~Zhou, A.~Bueno-Orovio, B.~Rodriguez, {\it Current Opinion in Physiology\/}
  {\bf 1}, 95 (2018).

\bibitem{Britton2013}
O.~J. Britton, {\it et~al.\/}, {\it Proceedings of the National Academy of
  Science\/} {\bf 110}, E2098 (2013).

\bibitem{Passini2017}
E.~Passini, {\it et~al.\/}, {\it Frontiers in Physiology\/} {\bf 8}, 668
  (2017).

\bibitem{Muszkiewicz2016}
A.~Muszkiewicz, {\it et~al.\/}, {\it Progress in Biophysics and Molecular
  Biology\/} {\bf 120}, 115 (2016).

\bibitem{Zaniboni2010}
M.~Zaniboni, I.~Riva, F.~Cacciani, M.~Groppi, {\it Mathematical Biosciences\/}
  {\bf 228}, 56 (2010).

\bibitem{Adams2020a}
M.~P. Adams, {\it et~al.\/}, {\it Environmental Modelling {\&} Software\/} {\bf
  130}, 104717 (2020).

\bibitem{Cui2022}
T.~Cui, X.~Tong, O.~Zahm, {\it arXiv preprint arXiv:2202.00074\/}  (2022).

\bibitem{FernandezSlezak2010}
D.~{Fern{\'{a}}ndez Slezak}, C.~Su{\'{a}}rez, G.~A. Cecchi, G.~Marshall,
  G.~Stolovitzky, {\it PLOS ONE\/} {\bf 5}, e13283 (2010).

\bibitem{Doucet2000}
A.~Doucet, S.~Godsill, C.~Andrieu, {\it Statistics and Computing\/} {\bf 10},
  197 (2000).

\bibitem{Transtrum2012}
M.~K. Transtrum, P.~Qiu, {\it BMC Bioinformatics\/} {\bf 13}, 181 (2012).

\bibitem{Scholkopf1998}
B.~Sch{\"{o}}lkopf, A.~Smola, K.-R. M{\"{u}}ller, {\it Neural Computation\/}
  {\bf 10}, 1299 (1998).

\bibitem{Pavlenko2012}
T.~Pavlenko, A.~Bj{\"{o}}rkstr{\"{o}}m, A.~Tillander, {\it Journal of Applied
  Statistics\/} {\bf 39}, 1643 (2012).

\bibitem{Tonsing2014}
C.~T{\"{o}}nsing, J.~Timmer, C.~Kreutz, {\it Physical Review E\/} {\bf 90},
  23303 (2014).

\bibitem{Kleinegesse2020}
S.~Kleinegesse, C.~Drovandi, M.~U. Gutmann, {\it Bayesian Analysis\/} pp. 1 --
  30 (2021).

\bibitem{Beisbart2019}
C.~Beisbart, N.~J. Saam, {\it {Computer simulation validation: Fundamental
  concepts, methodological frameworks, and philosophical perspectives}\/}
  (Springer, 2019).

\bibitem{Dangerfield2012}
C.~E. Dangerfield, D.~Kay, K.~Burrage, {\it Physical Review E\/} {\bf 85},
  051907 (2012).

\bibitem{Grazzini2015}
J.~Grazzini, M.~Richiardi, {\it Journal of Economic Dynamics and Control\/}
  {\bf 51}, 148 (2015).

\bibitem{Sisson2018}
S.~A. Sisson, Y.~Fan, M.~A. Beaumont, {\it Handbook of Approximate Bayesian
  Computation\/} (Chapman \& Hall / CRC Press, Boca Raton, Florida, 2018).

\bibitem{Price2018}
L.~F. Price, C.~C. Drovandi, A.~Lee, D.~J. Nott, {\it Journal of Computational
  and Graphical Statistics\/} {\bf 27}, 1 (2018).

\bibitem{Beskos2018}
A.~Beskos, A.~Jasra, K.~Law, Y.~Marzouk, Y.~Zhou, {\it SIAM/ASA Journal on
  Uncertainty Quantification\/} {\bf 6}, 762 (2018).

\bibitem{Muller2012}
U.~K. M{\"{u}}ller, {\it Journal of Monetary Economics\/} {\bf 59}, 581 (2012).

\bibitem{Milner-Gulland2017}
E.~J. Milner-Gulland, K.~Shea, {\it Journal of Applied Ecology\/} {\bf 54},
  2063 (2017).

\bibitem{Newey1994}
W.~K. Newey, D.~McFadden, {\it Handbook of Econometrics\/} (Elsevier, 1994),
  vol.~4, pp. 2111--2245.

\bibitem{Banner2020}
K.~M. Banner, K.~M. Irvine, T.~J. Rodhouse, {\it Methods in Ecology and
  Evolution\/} {\bf 11}, 882 (2020).

\bibitem{Robert1999}
C.~P. Robert, G.~Casella, {\it Monte Carlo Statistical Methods\/}
  (Springer-Verlag, New York, 1999).

\bibitem{Jordan1999}
M.~I. Jordan, Z.~Ghahramani, T.~S. Jaakkola, L.~K. Saul, {\it Machine
  Learning\/} {\bf 37}, 183 (1999).

\bibitem{Tierney1986}
L.~Tierney, J.~B. Kadane, {\it Journal of the American Statistical
  Association\/} {\bf 81}, 82 (1986).

\bibitem{B.2000}
T.~J. B., S.~V. De, L.~J. C., {\it Science\/} {\bf 290}, 2319 (2000).

\bibitem{Cui2016a}
T.~Cui, K.~J.~H. Law, Y.~M. Marzouk, {\it Journal of Computational Physics\/}
  {\bf 304}, 109 (2016).

\bibitem{Evans2006}
M.~Evans, H.~Moshonov, {\it Bayesian Analysis\/} {\bf 1}, 893 (2006).

\bibitem{Walker2013}
S.~G. Walker, {\it Journal of Statistical Planning and Inference\/} {\bf 143},
  1621 (2013).

\bibitem{Hoffman2018}
J.~D. Hoffman, S.~Frankel, {\it {Numerical methods for engineers and
  scientists}\/} (CRC press, 2018).

\bibitem{Bucker2006}
H.~M. B{\"{u}}cker, G.~Corliss, P.~Hovland, U.~Naumann, B.~Norris, {\it
  {Automatic Differentiation: Applications, Theory, and Implementations}\/},
  vol.~50 (Springer-Verlag Berlin Heidelberg, 2006).

\bibitem{Constantine2016}
P.~G. Constantine, C.~Kent, T.~Bui-Thanh, {\it SIAM Journal on Scientific
  Computing\/} {\bf 38}, A2779 (2016).

\bibitem{Cui2021}
T.~Cui, X.~T. Tong, {\it arXiv preprint: 2101.02417\/}  (2021).

\bibitem{Zahm2022}
O.~Zahm, T.~Cui, K.~Law, A.~Spantini, Y.~Marzouk, {\it Mathematics of
  Computation [arXiv preprint: 1807.03712]\/}  (2022).

\bibitem{Bowman1997}
A.~W. Bowman, A.~Azzalini, {\it {Applied smoothing techniques for data
  analysis: the kernel approach with S-Plus illustrations}\/}, vol.~18 (Oxford
  University Press Inc, 1997).

\bibitem{Jones1993}
M.~C. Jones, {\it Statistics and Computing\/} {\bf 3}, 135 (1993).

\bibitem{Dokos2017}
S.~Dokos, {\it {Modelling organs, tissues, cells and devices: using MATLAB and
  COMSOL multiphysics}\/} (Springer, Sydney, Australia, 2017).

\end{thebibliography}
\bibliographystyle{Science}

\section*{Acknowledgments}

\noindent\textbf{Funding:} This work has been supported by a Research Stimulus (RS) Postdoctoral Fellowship from the University of Queensland (GMM-B), the Australian Research Council (ARC)  Centre of Excellence for Mathematical and Statistical Frontiers Grant No CE140100049 (BAJL), the ARC Centre of Excellence for Plant Success in Nature and Agriculture Grant No CE200100015 (KB), a National Science Foundation (NSF) Grant MCB No 1715342 (KSB), an ARC Linkage Grant No LP160100496~(EM-M), and an ARC Discovery Early Career Researcher Award No DE200100683~(MPA).

\noindent\textbf{Author contributions:}  All authors contributed to the conceptualization, original draft and, review \& editing of the manuscript. GMM-B, BAJL, CD, KB and MPA  contributed to the design of numerical experiments. GMM-B and  BAJL performed numerical experiments. All authors contributed to the analysis of results.

\noindent\textbf{Competing interests:} The authors declare that they have no competing interests.

\noindent\textbf{Data and materials availability:} All data needed to evaluate the conclusions in the paper are present in the paper and/or the Supplementary Materials. MATLAB code associated with this paper is available from the Figshare Repository: \url{https://doi.org/10.6084/m9.figshare.19228008}.
\end{document}

\invisiblesection{Supplementary Material}
	
	\renewcommand{\figurename}{Fig.}
	\renewcommand\thesubfigure{\Alph{subfigure}}
	\renewcommand\thepage{\arabic{page}}
	\setcounter{page}{1}
	\renewcommand\thefigure{S\arabic{figure}}
	\renewcommand\thetable{S\arabic{table}}
	\setcounter{figure}{0}
	\setcounter{table}{0} 

\begin{center}	
\Large{Supplementary Materials for}\vspace{10pt}\\
\large{\textbf{Analysis of sloppiness in model simulations: unveiling  parameter uncertainty when mathematical models are fitted to data}}\vspace{10pt}\\
\normalsize{Gloria M. Monsalve-Bravo,$^{\ast}$  Brodie~A.~J.~Lawson,   Christopher~Drovandi, Kevin~Burrage, Kevin~S.~Brown, Christopher~M.~Baker, Sarah~A.~Vollert,  Kerrie~Mengersen, Eve~McDonald-Madden,  Matthew~P.~Adams}\\
{\footnotesize$^\ast$Corresponding author. E-mail: g.monsalvebravo@uq.edu.au}
\end{center}
	
\noindent\normalsize{\textbf{This PDF file includes:}}
\vspace{5pt}
\begin{addmargin}[1em]{0pt}
Figures' Supplementary Legends
\\
Figs.\ S1 to S21
\\
Tables S1 to S4
\end{addmargin}	
	
\newpage

\noindent{\large\textbf{Figures' Supplementary Legends}}\vspace{30pt}

\noindent{\textbf{Additional legend information for Figs.~2, 4, 5, 6B and Figs.~S1A to S3A, S5, S7, S8, S10,  S13, S14, S15A to S17A, S18 and S20}} 

\begin{itemize}
    \item \textbf{Shaded regions (\swatch{ColorPrior}, \swatch{ColorPosterior}, \swatch{ColorPosterior_1!50!white}, \swatch{ColorPosterior_2!50!white}, \swatch{ColorPosterior_3!50!white}, \swatch{ColorPosterior_4!50!white}, \swatch{ColorPosterior_5!50!white}, \swatch{ColorPosterior_6!50!white}, \swatch{ColorPosterior_7!50!white}, \swatch{ColorPosterior_8!50!white}):} Prior and posterior   distributions for the model parameters. Probability distributions are obtained via kernel density estimation \CiteBowman based on the initial sample of the prior and the sample of the posterior obtained from our Sequential Monte Carlo (SMC) sampling algorithm assuming a reflecting boundary correction to account for prior bounds \CiteJones. The $y$-axes correspond to relative (rather than absolute) probabilities, with all density functions rescaled between  $0-1$.
    \item \textbf{Solid vertical lines (\PriorLine):} Prior distribution mean for the model parameters, estimated from from the prior distribution sample.
    \item \textbf{Dash-dotted vertical lines (\PosteriorLine, \PosteriorLineA, \PosteriorLineB, \PosteriorLineC, \PosteriorLineD, \PosteriorLineE, \PosteriorLineF, \PosteriorLineG, \PosteriorLineH):} Posterior distribution mean for the model parameters, estimated from the posterior distribution sample.
    \item \textbf{Open circle symbols (${\bm{\textcolor{PriorDot}{\circ}}}$):} Prior distribution samples.
    \item \textbf{Closed circle symbols ($\bm{{\textcolor{Like1}{\bullet}}}\bm{{\textcolor{Like2}{\bullet}}}\bm{{\textcolor{Like3}{\bullet}}}$):} Posterior distribution samples, in which the color change across posterior samples represents changes in the values of the rescaled log-likelihood function with $\bm{{\textcolor{Like3}{\bullet}}}=0$ and $\bm{{\textcolor{Like1}{\bullet}}}=1$.
     \item \textbf{Dashed vertical lines (\RefLine) and plus sign symbols ($\bm{\textcolor{RefColor}{+}}$):} Set of reference (true) parameter values.
    \item \textbf{Dotted vertical lines (\OptLine) and cross symbols ($\bm{\textcolor{OptColor}{\times}}$):} Set of best-fit parameter values obtained via maximum likelihood estimation (MLE).
\end{itemize}\vspace{20pt}

\noindent{\textbf{Additional legend information for Figs.~1, 3B, 6A, 7 and Figs.~S1B to S3B, S9, S12, S15B to S17B, S19 and S21}}

\begin{itemize}
    \item \textbf{Open triangle ($\bm{\triangle}$) and square ($\bm{\square}$) symbols:} Synthetic data
    \item \textbf{Dark (\swatch{EnsembleColor65}, \swatch{EnsembleColor65_1!60!white}, \swatch{EnsembleColor65_2!60!white}, \swatch{EnsembleColor65_3!60!white}, \swatch{EnsembleColor65_4!60!white}, \swatch{EnsembleColor65_5!60!white}, \swatch{EnsembleColor65_6!60!white}, \swatch{EnsembleColor65_7!60!white}, \swatch{EnsembleColor65_8!60!white}) and light  (\swatch{EnsembleColor95}, \swatch{EnsembleColor95_1!30!white}, \swatch{EnsembleColor95_2!30!white}, \swatch{EnsembleColor95_3!30!white}, \swatch{EnsembleColor95_4!30!white}, \swatch{EnsembleColor95_5!30!white}, \swatch{EnsembleColor95_6!30!white}, \swatch{EnsembleColor95_7!30!white}, \swatch{EnsembleColor95_8!30!white}) shaded areas:}  $68\%$ and $95\%$ central credible intervals  for the model ensemble prediction, obtained from posterior simulation considering all plausible parameter values estimated via Bayesian inference  implemented using our SMC sampling algorithm. 
    \item \textbf{Dashed lines (\RefLine, \RefLineA, \RefLineB, \RefLineC, \RefLineD):} Model predictions based on the set of reference (true) parameter values. 
    \item \textbf{Dotted lines (\OptLine):} Model predictions based on the set of best-fit parameter values obtained maximum likelihood estimation (MLE).
    \item \textbf{Dashed vertical lines (\DataLine):} Start time of model forecasts (only applicable for the ecological network model).
\end{itemize}

\begin{figure}[p]
	\centering
	\begin{subfigure}[b]{0.576\textwidth}
		\centering
		\centering\myfigure[1]{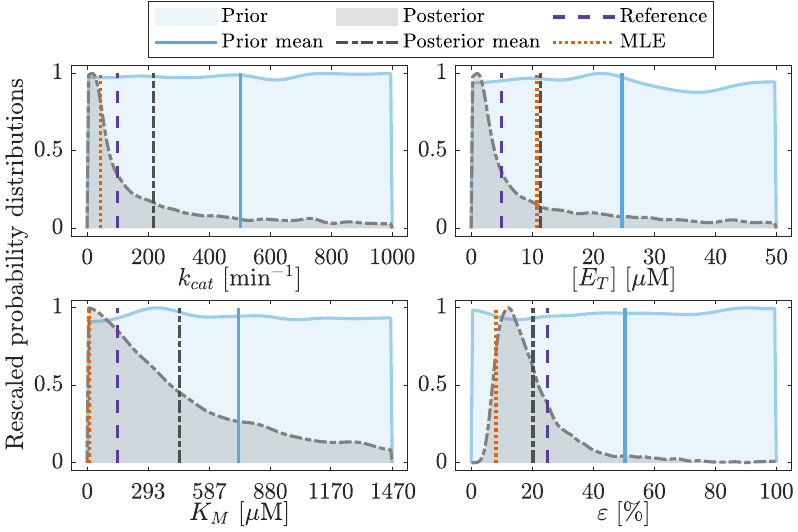}
		\caption{}
		\label{fig:MM-uniform-priors}
	\end{subfigure}
	\hfill
	\begin{subfigure}[b]{0.414\textwidth}
		\centering
		\centering\myfigure[1]{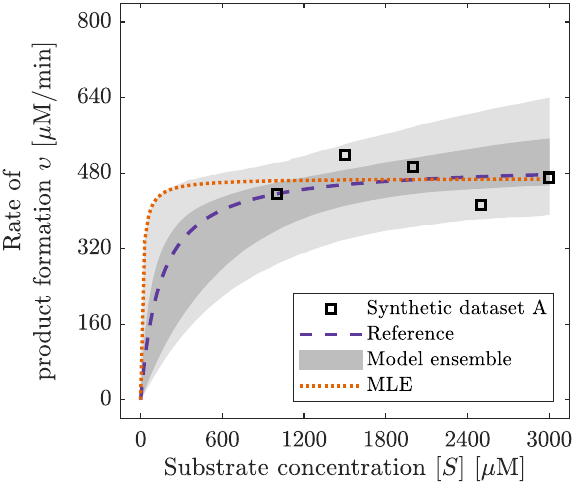}
		\caption{}
		\label{fig:MM-uniform-fit}
	\end{subfigure}
	\caption{\textbf{Estimated parameter values together with fit of the Michaelis--Menten model to the noisy synthetic dataset A (\ie  high substrate concentration with $\bm{[S]\gg K_M}$) considering uniform prior distributions for all parameters.}
		\textbf{(A)} Prior and posterior   distributions for the parameters together with reference parameter values  and best-fit parameter values.
		\textbf{(B)} Dataset A together with noiseless model prediction using reference parameter values, model predictions using two sets of best-fit parameter values, and model ensemble predictions using all plausible parameter values obtained via Bayesian inference. 
		While the  model ensemble fits synthetic dataset A, the data  is uninformative for parameter $K_M$.
		(See also Figures' Supplementary Legends.)
		}
		\label{fig:MM-uniform}
\end{figure}

	\begin{figure}[p]
		\centering
		\begin{subfigure}[b]{0.576\textwidth}
			\centering
			\centering\myfigure[1]{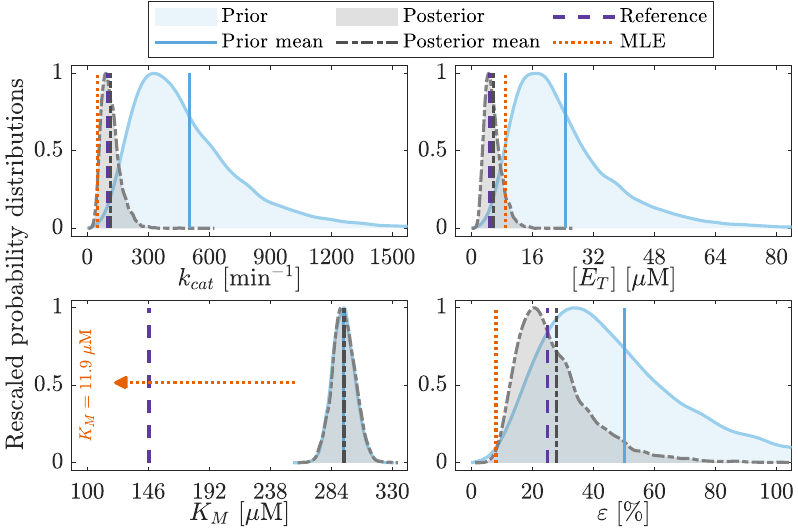}
			\caption{}
			\label{fig:MM-lognormal-priors}
		\end{subfigure}
		\hfill
		\begin{subfigure}[b]{0.414\textwidth}
			\centering
			\centering\myfigure[1]{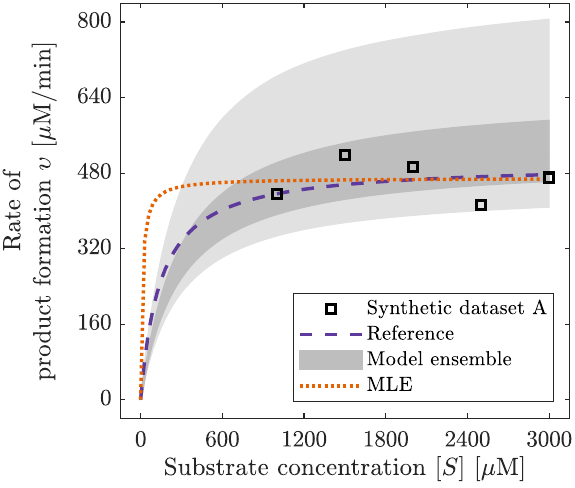}
			\caption{}
			\label{fig:MM-lognormal-fit}
		\end{subfigure}
		\caption{\textbf{Estimated parameter values together with fit of the Michaelis--Menten model to the noisy synthetic dataset A (\ie  high substrate concentration with $\bm{[S]\gg K_M}$)  considering a multivariate log-normal prior distribution for all parameters with that of parameter $\bm{K_M}$ badly specified with $\bm{p(K_M=146.7\;\mathrm{\mu M})\approx0}$.}
		    \textbf{(A)} Prior and posterior   distributions for the parameters together with reference parameter values  and best-fit parameter values.
		    \textbf{(B)} Dataset A together with noiseless model prediction using reference parameter values, model predictions using two sets of best-fit parameter values, and model ensemble predictions using all plausible parameter values obtained via Bayesian inference. 
			 While the  model ensemble fits synthetic dataset A, posterior   distribution and best-fit values for $K_M$ lie far away from the  reference value.
	        (See also Figures' Supplementary Legends.)}
		\label{fig:MM-lognormal}
	\end{figure}

	\begin{figure}[p]
		\centering
		\begin{subfigure}[b]{0.576\textwidth}
			\centering\myfigure[1]{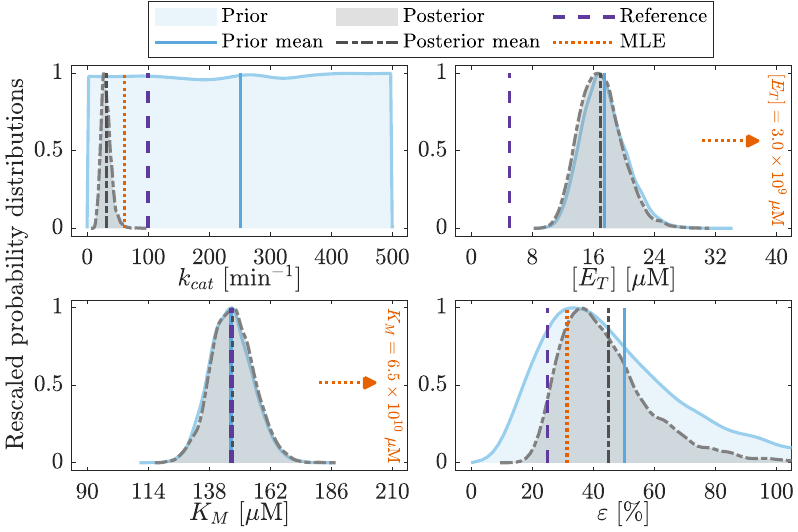}
			\caption{}
			\label{fig:MM-custom-priors}
		\end{subfigure}
		\hfill
		\begin{subfigure}[b]{0.414\textwidth}
			\centering\myfigure[1]{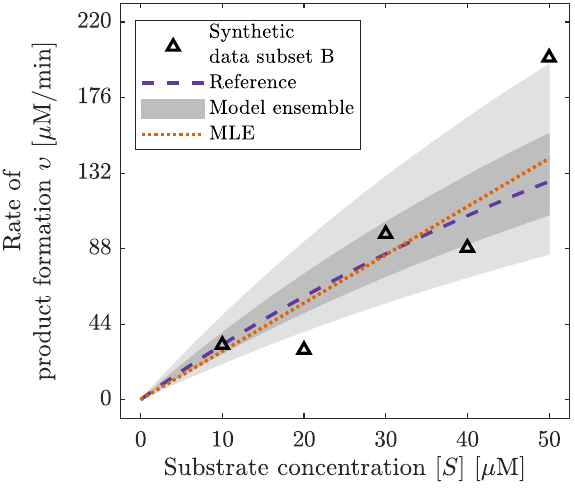}
			\caption{}
			\label{fig:MM-custom-fit}
		\end{subfigure}
		\caption{\textbf{Estimated parameter values together with fit of the Michaelis--Menten model to the noisy synthetic dataset B (\ie at low substrate concentration with $\bm{[S]\ll K_M}$) considering a uniform prior for $\bm{k_{cat}}$, a badly specified log-normal prior for $[E_T]$ with $\bm{p([E_T]=5\;\mathrm{\mu M})\approx0}$, a well-specified log-normal prior for $\bm{K_{M}}$ with $\bm{p(K_M=146.7\;\mathrm{\mu M})\approx1}$ and a log-normal prior for $\bm{\sigma}$.}
		    \textbf{(A)} Prior and posterior   distributions for the parameters together with reference parameter values  and best-fit parameter values.
		    \textbf{(B)} Dataset B together with noiseless model prediction using reference parameter values, model predictions using two sets of best-fit parameter values, and model ensemble predictions using all plausible parameter values obtained via Bayesian inference. 
			While the  model ensemble fits synthetic dataset B, best-fit values for parameters $[E_T]$ and $[K_M]$ lie far away from their  reference values.
			(See also Figures' Supplementary Legends.)}
		\label{fig:MM-custom}
	\end{figure}

	\begin{figure}[p]
		\centering
		\begin{subfigure}{0.3275\textwidth}
			\centering\myfigure[1]{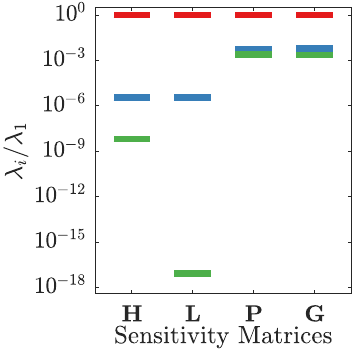}
			\caption{}
			\label{fig:uniform-eig-MM}
		\end{subfigure}
		\hfill
		\begin{subfigure}{0.3275\textwidth}
			\centering\myfigure[1]{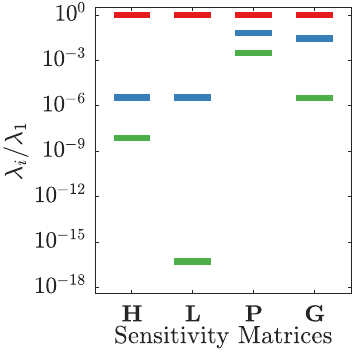}
			\caption{}
			\label{fig:lognormal-eig-MM}
		\end{subfigure}
		\hfill
		\begin{subfigure}{0.3275\textwidth}
			\centering\myfigure[1]{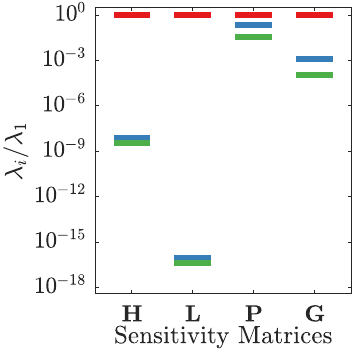}
			\caption{}
			\label{fig:custom-eig-MM}
		\end{subfigure}
		\caption{\textbf{Sensitivity matrix eigenvalue spectra (based on SVD) for the Michaelis--Menten model fit to noisy synthetic data}, considering 
		\textbf{(A)} uniform prior distributions for all parameters (Scenario~1, Fig.~\ref{fig:MM-uniform}), 
		\textbf{(B)}  multivariate log-normal prior distribution for all parameters with that of parameter $K_M$ badly specified (Scenario~2, Fig.~\ref{fig:MM-lognormal}), and 
		\textbf{(C)}  a uniform prior for $k_{cat}$, a badly specified log-normal prior for $[E_T]$, a well-specified log-normal prior for $ K_{M}$ (Scenario~3, Fig.~\ref{fig:MM-custom}). 
		Largest eigenvalue $\lambda_1$ is used to rescale eigenvalues between $0-1$.}
		\label{fig:eig-MM}
	\end{figure}

\begin{figure}[p]
	\centering
	\centering\myfigure[0.88]{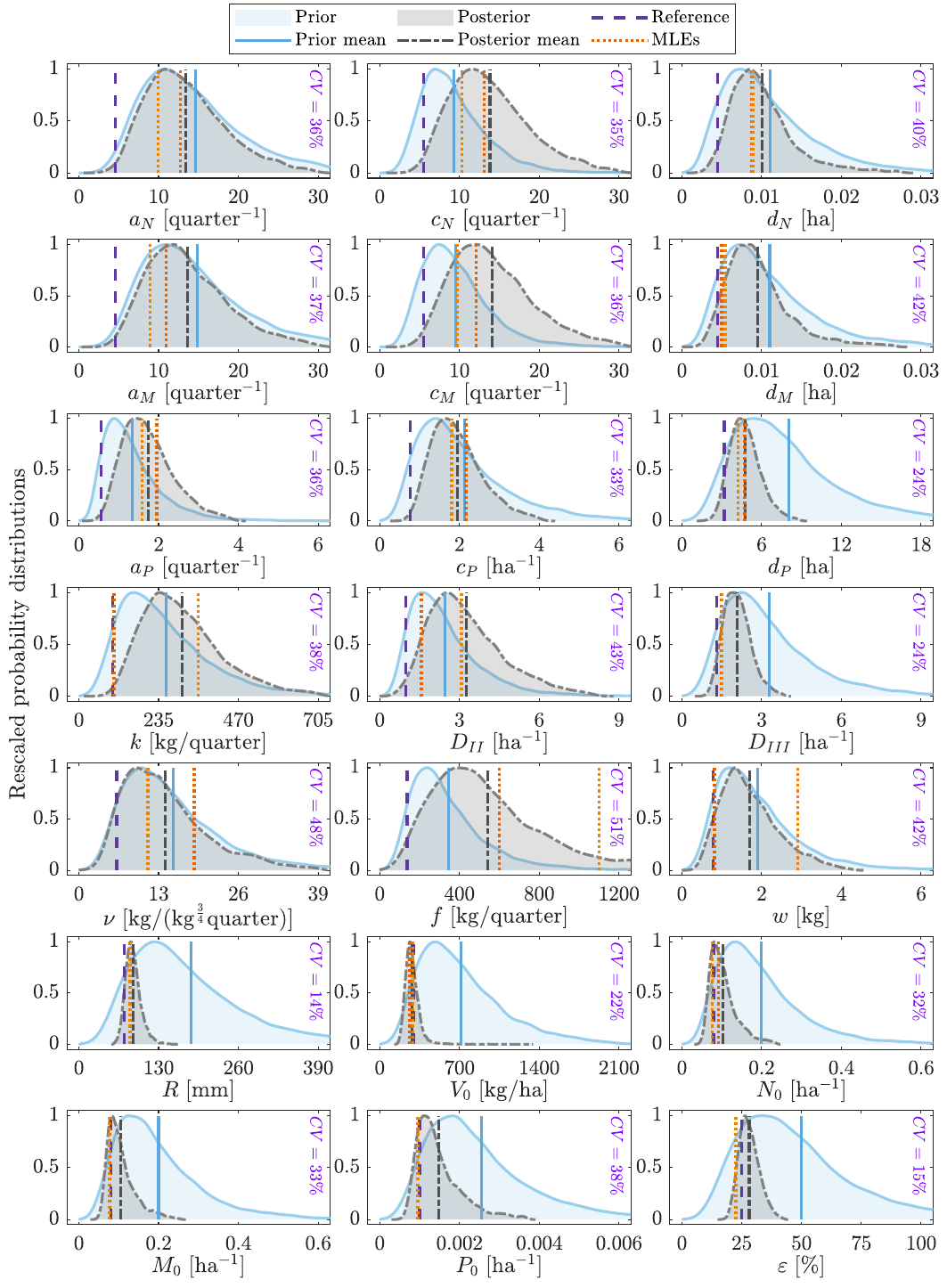}
	\caption{\textbf{Prior and posterior   distributions  for the parameters of the ecosystem network model~(Table~S1) together with the sets of reference parameter values  and best-fit parameter values (MLEs) considering a vague multivariate log-normal prior for all parameters.}  
	Percentage coefficient of variation for the posterior distribution sample of each parameter ($CV$) is reported within each panel (compare to Fig.~\ref{fig:priors_network_new}). 
	Most of the model parameters are poorly constrained by the data.
	(See also Figures' Supplementary Legends.)}
	\label{fig:priors_network}
\end{figure}

		\begin{figure} [p]
			\centering\myfigure[0.3275]{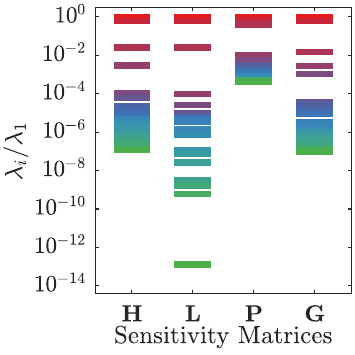}
			\caption{\textbf{Sensitivity matrix eigenvalue spectra (based on SVD) for the ecosystem network model in Table~S1 fit to noisy synthetic data, considering a multivariate log-normal prior distribution for all parameters~(Fig.~\ref{fig:priors_network}).} Largest eigenvalue $\lambda_1$ is used to rescale eigenvalues between $0-1$.} 
			\label{fig:eig_network}
		\end{figure}

			\begin{figure}[p]
			\centering
			\begin{subfigure}{0.4950\textwidth}
				\centering\myfigure[1]{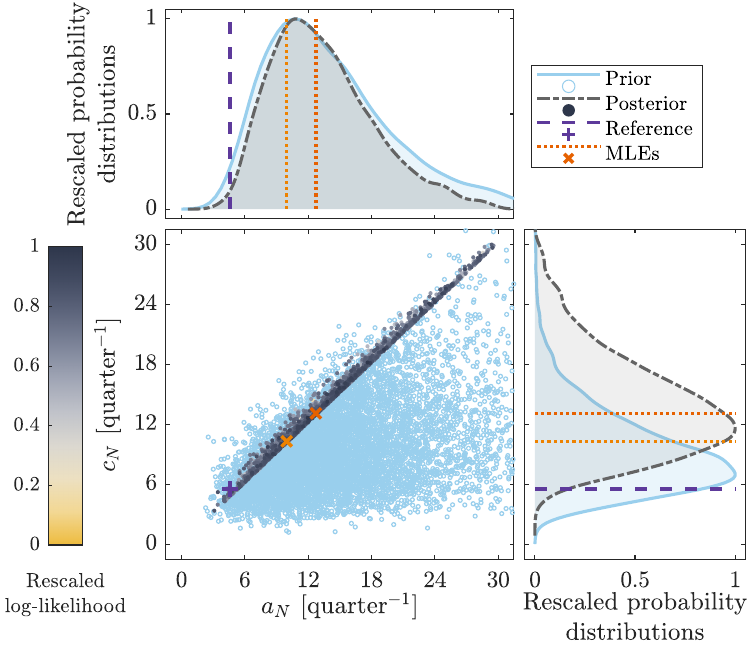}
				\caption{}
				\label{fig:pmaxtrix-key2-network}
			\end{subfigure} 
			\hfill
			\begin{subfigure}{0.4950\textwidth}
			\centering\myfigure[1]{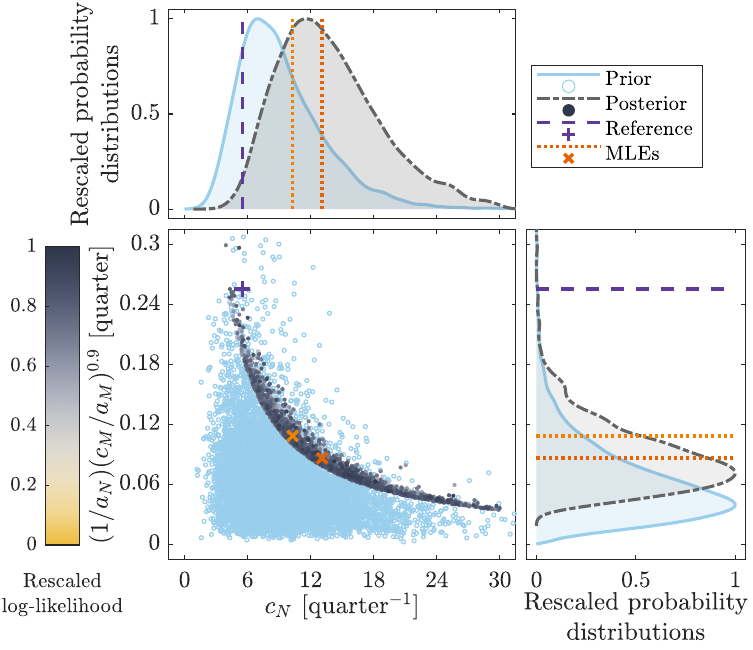}
			\caption{}
			\label{fig:pmaxtrix-key3-network}
		\end{subfigure}
			\caption{\textbf{Posterior   distributions for the remaining stiff eigenparameters obtained from the Bayesian methods~(Table~2)  compared to the sets of reference values and best-fit values (MLEs).}
			\textbf{(A)} $\hat{\theta}_2$ from matrix $\mathbf{P}$ and 
			\textbf{(B)} $\hat{\theta}_2$ from matrix $\mathbf{G}$.
			Similar tendencies are seen for  $\hat{\theta}_2$ and $\hat{\theta}_3$ from matrix $\mathbf{P}$, thus $\hat{\theta}_3$ is not shown.
			Many samples of the posterior distribution yield similar values of the log-likelihood function, with $a_{N}\propto c_{N}$, $a_{M}\propto c_{M}$ and ${c_N}\propto\left[({1}/{a_N})(c_M/a_M)^{0.9}\right]^{-1}$.
	        (See also Figures' Supplementary Legends.)}
			\label{fig:pkey-network}
		\end{figure}

		\begin{figure}[p]
			\centering\myfigure[0.88]{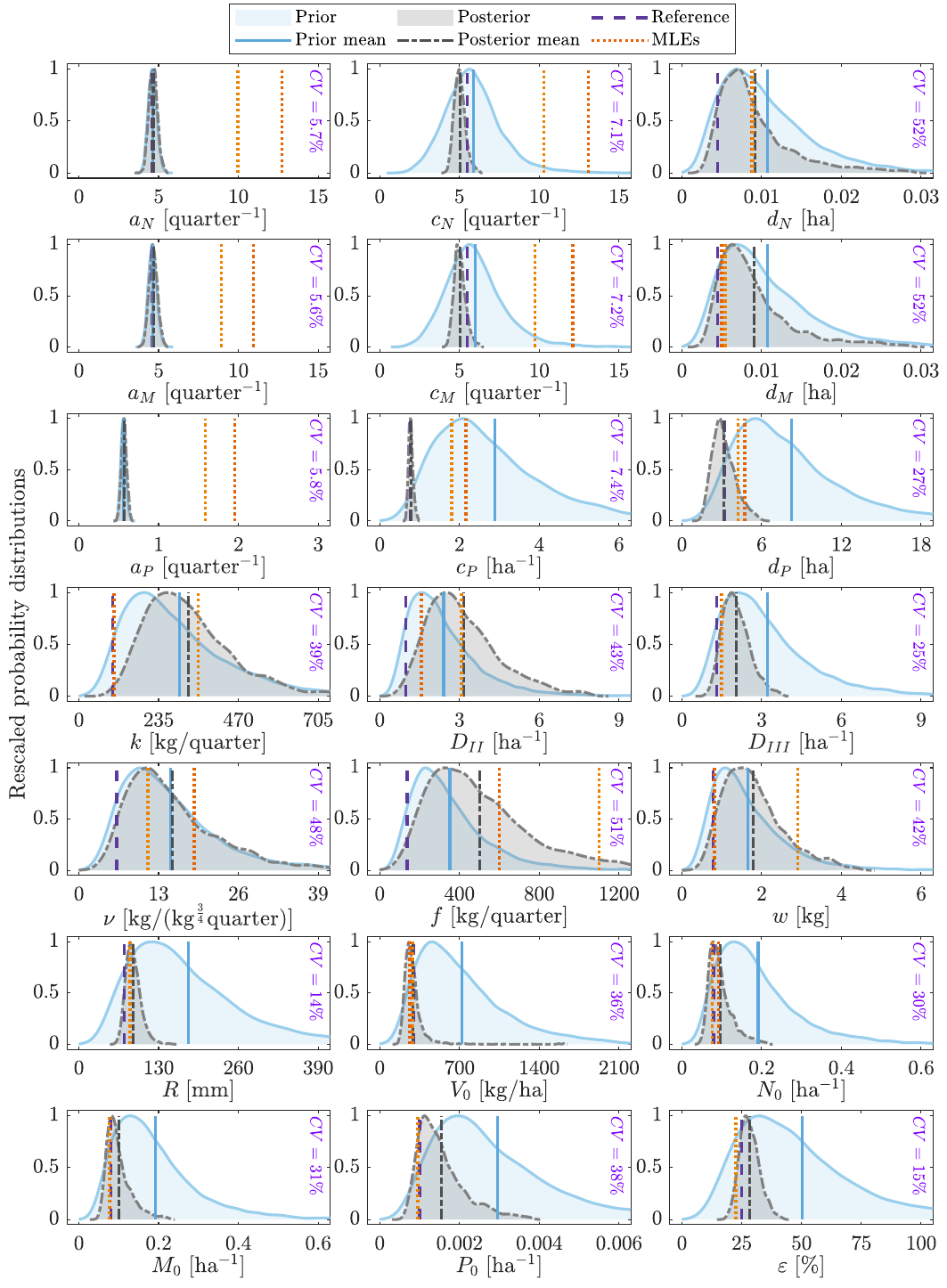}
			\caption{\textbf{Prior and posterior   distributions  for the parameters of the ecosystem network model~(Table~S1) together with the set of reference parameter values  and best-fit parameter values (MLEs) considering a  more informative multivariate log-normal prior for all parameters.} 
			Percentage coefficient of variation for the posterior sample of each  parameter ($CV$) is reported within each panel (compare to Fig.~\ref{fig:priors_network}).
			Improved priors for parameters $a_{N}$, $a_{M}$, $a_{P}$ also constrain parameters $c_{N}$, $c_{M}$ and $c_{P}$ due to the identified parameter relations.
	        (See also Figures' Supplementary Legends.)}
			\label{fig:priors_network_new}
		\end{figure}

	\begin{figure}[p]
		\centering\myfigure[0.7650]{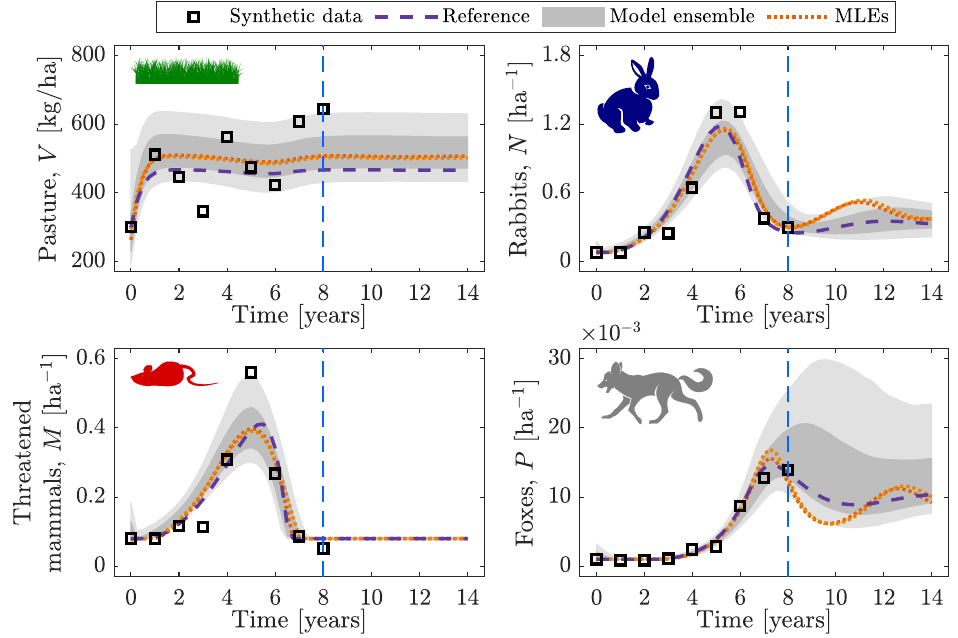}
		\caption{\textbf{Ecosystem network model fit to time-series data considering a  more informative multivariate log-normal prior for all parameters in the implementation of Bayesian inference.}
		Synthetic time-series data for ecological abundance with measurement error of  $\varepsilon=25\%$  together with noiseless model prediction using reference parameter values~(Table~S2), model predictions using two sets of best-fit parameter values (MLEs), and model ensemble predictions using all plausible parameter values~(Fig.~\ref{fig:priors_network_new}).
		The ecosystem network model fits the synthetic time-series data, with the model ensemble propagating parameter uncertainty into species abundance predictions.
	   (See also Figures' Supplementary Legends.)}
		\label{fig:fit_network_new}
	\end{figure}

	\begin{figure}[p]
	\centering\myfigure[0.95]{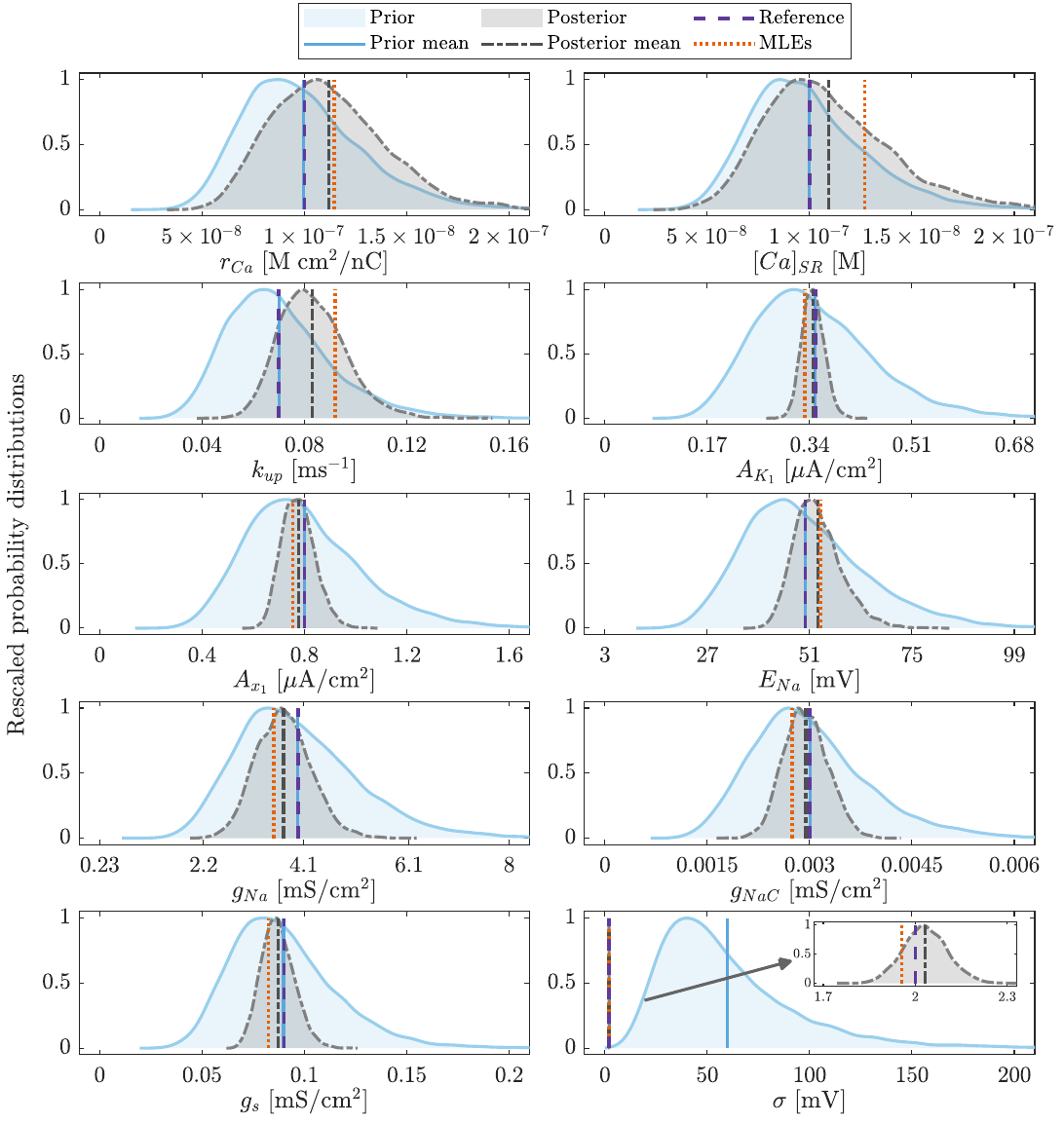}
	\caption{\textbf{Prior and posterior distributions  for the parameters of the Beeler-Reuter   model~(Table~S3) together with reference parameter values and best-fit parameter values considering a multivariate log-normal prior for all parameters.}
	Most of the model parameters are well-constrained by the data.
	(See also Figures' Supplementary Legends.)}
	\label{fig:priors_Beeler}
\end{figure}

		\begin{figure} [p]
			\centering\myfigure[0.3275]{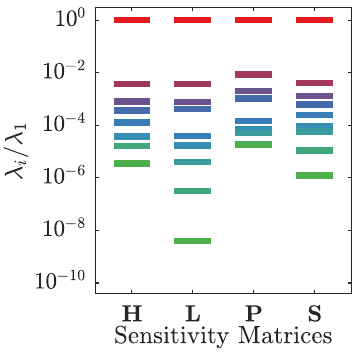}
			\caption{\textbf{Sensitivity matrix eigenvalue spectra (based on SVD) for the Beeler-Reuter model in Table~S3 fit to synthetic data, considering a multivariate log-normal prior distribution for all parameters.} Largest eigenvalue $\lambda_1$ is used to rescale eigenvalues between $0-1$.} 
			\label{fig:eig_Beeler}
		\end{figure}
	
		\begin{figure}[p]
		\centering\myfigure[0.800]{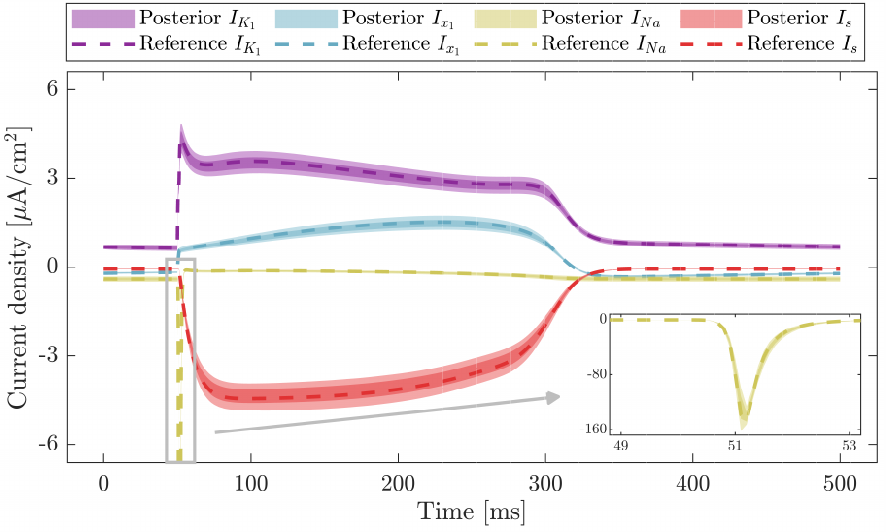}
		\caption{\textbf{Dynamics of the individual currents together composing the Beeler-Reuter model action potential.} Time courses of the four currents that sum to form the action potential, displaying both those observed for the reference model (dashed lines) used to generate the synthetic data  and the $68\%$ and $95\%$ confidence intervals (dark and light shaded regions, respectively) among the posterior population of parameter values obtained via Bayesian inference. The inset displays the sharp spike of the $Na^{+}$ current that initiates the action potential. All currents exhibit clear nonlinear behavior, and interact with each other only indirectly via the cell's overall polarization level.
	(See also Figures' Supplementary Legends.)}
		\end{figure}

	\begin{figure}[p]
	\centering\myfigure[1]{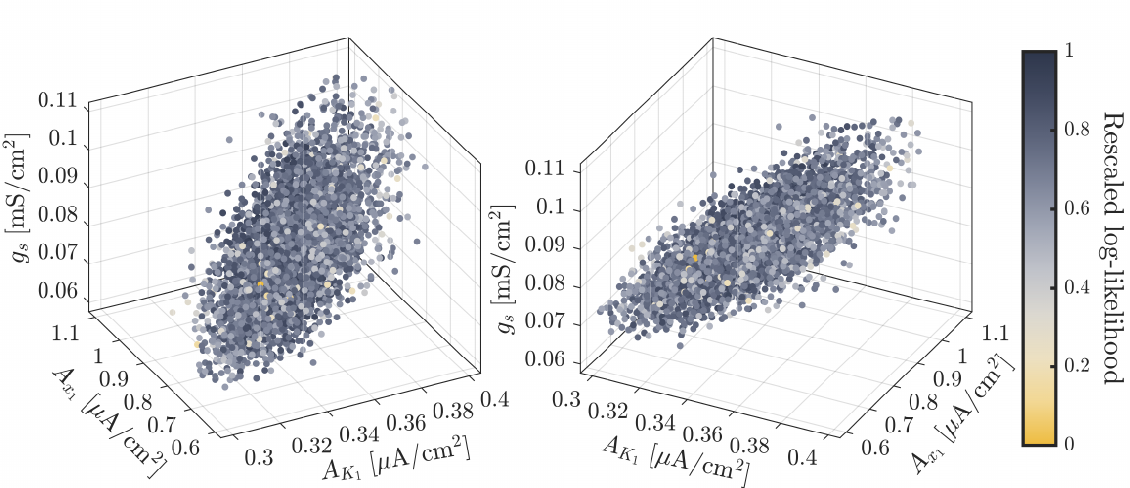}
	\caption{\textbf{Trivariate scatter plot of the posterior distribution for  parameters $\bm{A_{K_1}}$,  $\bm{A_{x_1}}$ and $\bm{g_s}$, with side view}.  Unlike Fig.~6B, relationship between combination of parameters $A_{K_1}$ and $A_{x_1}$ with parameter $g_s$ is not easily visible from this trivariate scatter plot.
	(See also Figures' Supplementary Legends.)}
	\label{fig:trivariate-key-Beeler}
\end{figure}

	\begin{figure}[p]
	\centering\myfigure[1]{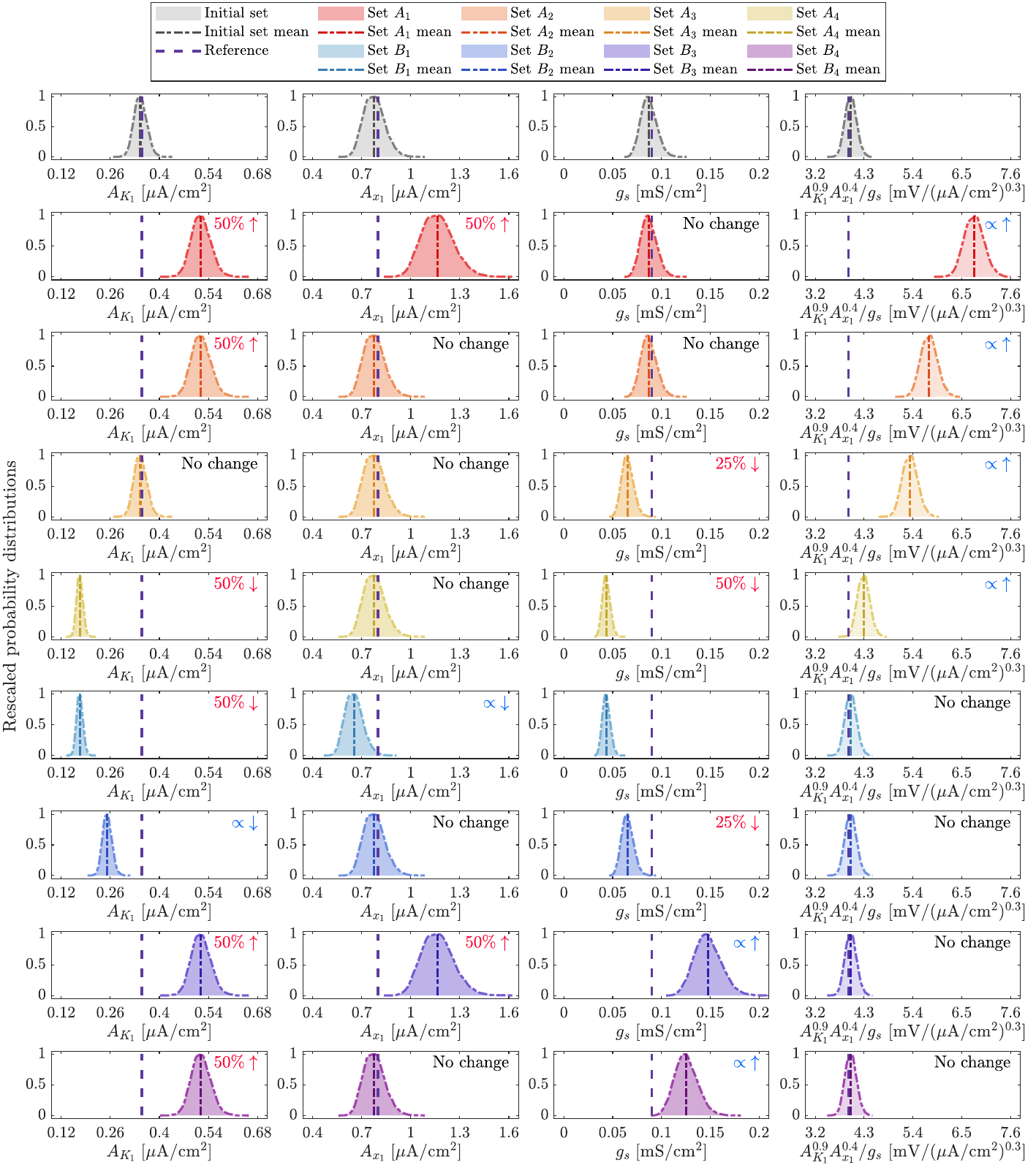}
	\caption{\textbf{Variation of parameters $\bm{A_{K_1}}$,  $\bm{A_{x_1}}$ and $\bm{g_s}$ to change or keep approximately constant the value of stiffest eigenparameter $\bm{\hat{\theta}_1={A_{K_1}^{0.9}A_{x_1}^{0.4}}/{g_s}}$.} The first row of panels depicts the initial set of parameter values, obtained via Bayesian inference while the remaining rows of panels depict eight sets of specified parameter values $A_1$ to $A_4$ and $B_1$ to $B_4$ which change or keep constant the value of the stiffest eigenparameter, respectively. Changes in the parameter values relative to their estimated values (shown in the top four panels) are indicated at the top right corner of each panel, with $\propto$ symbol indicating a proportional change in the value of the parameter or eigenparameter  while $\uparrow$ and $\downarrow$ symbols indicating an increase and decrease of the parameter value, respectively.
	(See also Figures' Supplementary Legends.)}
	\label{fig:BRparam}
\end{figure}

\begin{figure}[p]
	\centering
	\begin{subfigure}[b]{0.576\textwidth}
		\centering
		\centering\myfigure[1]{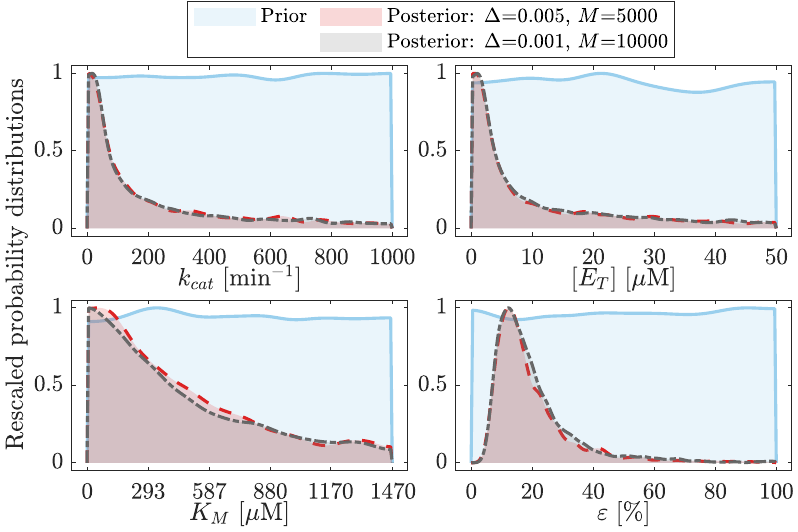}
		\caption{}
		\label{fig:Reproducibility-MM-uniform-priors}
	\end{subfigure}
	\hfill
	\begin{subfigure}[b]{0.414\textwidth}
		\centering
		\centering\myfigure[1]{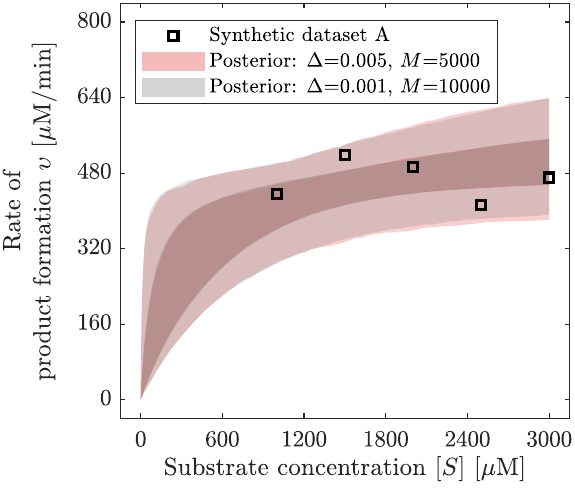}
		\caption{}
		\label{fig:Reproducibility-MM-uniform-fit}
	\end{subfigure}
	\caption{\textbf{Estimated parameter values and fit of the Michaelis--Menten model to the noisy synthetic dataset A (\ie  high substrate concentration with $\bm{[S]\gg K_M}$) by implementing the posterior sampling algorithm with two different combinations of the effective sample size reduction target $\bm{\Delta}$ and sample size $\bm{M}$ considering uniform prior distributions for all parameters.}
		\textbf{(A)}  Prior and posterior distributions for the parameters, with the posterior distributions overlapped.
		\textbf{(B)} Dataset A together with the model ensemble predictions using all plausible parameter values obtained via Bayesian inference (also overlapped). 
		Both posterior distribution and model ensemble prediction considering $\Delta=0.005$ and $M=5000$  reproduce those considering $\Delta=0.001$ and $M=10000$.
		(See also Figures' Supplementary Legends.)
		}
		\label{fig:Reproducibility-MM-uniform}
\end{figure}

	\begin{figure}[p]
		\centering
		\begin{subfigure}[b]{0.576\textwidth}
			\centering
			\centering\myfigure[1]{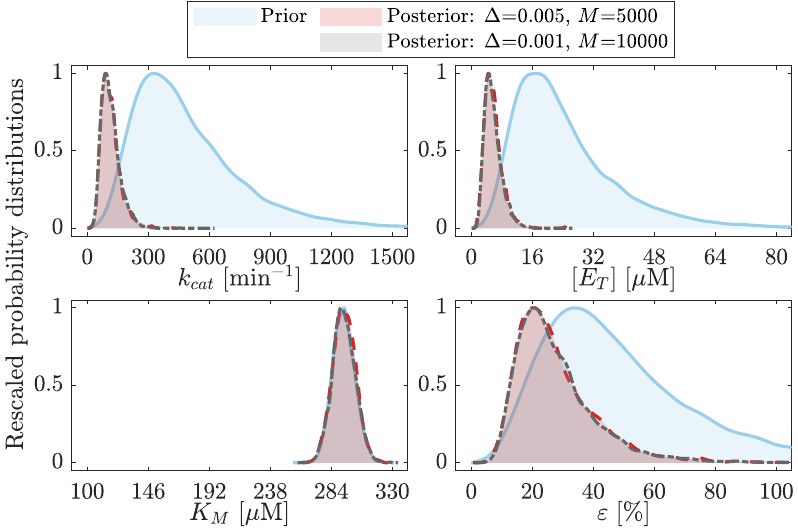}
			\caption{}
			\label{fig:Reproducibility-MM-lognormal-priors}
		\end{subfigure}
		\hfill
		\begin{subfigure}[b]{0.414\textwidth}
			\centering
			\centering\myfigure[1]{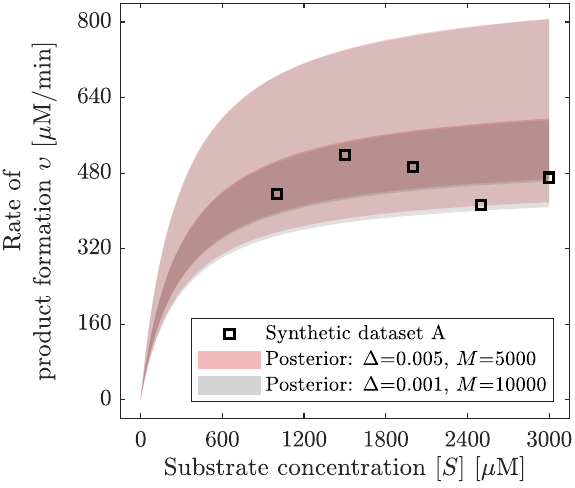}
			\caption{}
			\label{fig:Reproducibility-MM-lognormal-fit}
		\end{subfigure}
		\caption{\textbf{Estimated parameter values and fit of the Michaelis--Menten model to the noisy synthetic dataset A (\ie  high substrate concentration with $\bm{[S]\gg K_M}$) by implementing the posterior sampling algorithm with two different combinations of the effective sample size reduction target $\bm{\Delta}$ and sample size $\bm{M}$ considering a multivariate log-normal prior distribution for all parameters with that of parameter $\bm{K_M}$ badly specified with $\bm{p(K_M=146.7\;\mathrm{\mu M})\approx0}$.}
		    \textbf{(A)} Prior and posterior distributions for the parameters, with the posterior distributions overlapped.
		    \textbf{(B)} Dataset A together with the model ensemble predictions using all plausible parameter values obtained via Bayesian inference (also overlapped). 
            Both posterior distribution and model ensemble prediction considering $\Delta=0.005$ and $M=5000$ reproduce those considering $\Delta=0.001$ and $M=10000$.
	        (See also Figures' Supplementary Legends.)}
		\label{fig:Reproducibility-MM-lognormal}
	\end{figure}

	\begin{figure}[p]
		\centering
		\begin{subfigure}[b]{0.576\textwidth}
			\centering\myfigure[1]{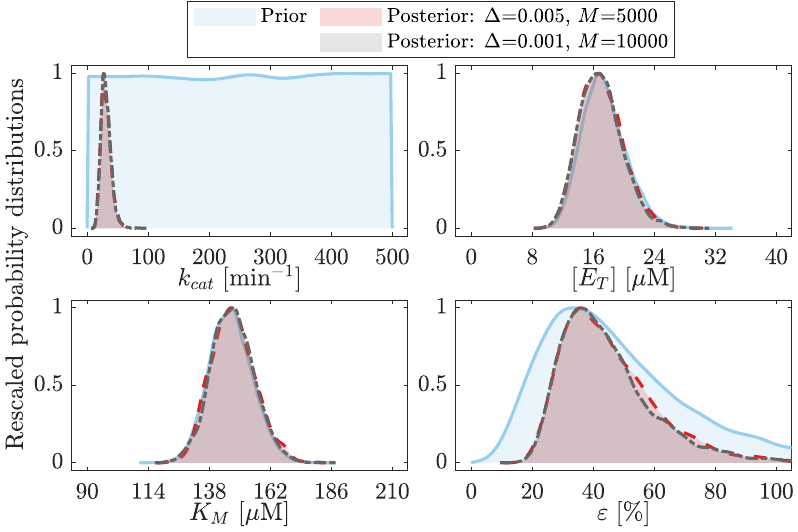}
			\caption{}
			\label{fig:Reproducibility-MM-custom-priors}
		\end{subfigure}
		\hfill
		\begin{subfigure}[b]{0.414\textwidth}
			\centering\myfigure[1]{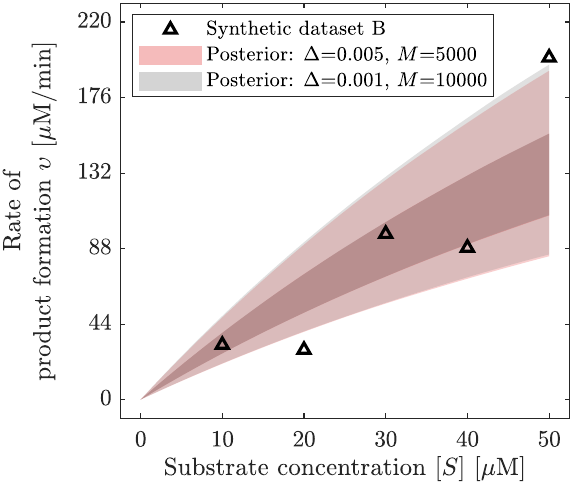}
			\caption{}
			\label{fig:Reproducibility-MM-custom-fit}
		\end{subfigure}
		\caption{\textbf{Estimated parameter values and fit of the Michaelis--Menten model to the noisy synthetic dataset B (\ie at low substrate concentration with $\bm{[S]\ll K_M}$) by implementing the posterior sampling algorithm with two different combinations the effective sample size reduction target $\bm{\Delta}$ and sample size $\bm{M}$ considering a uniform prior for $\bm{k_{cat}}$, a badly specified log-normal prior for $[E_T]$ with $\bm{p([E_T]=5\;\mathrm{\mu M})\approx0}$, a well-specified log-normal prior for $\bm{K_{M}}$ with $\bm{p(K_M=146.7\;\mathrm{\mu M})\approx1}$ and a log-normal prior for $\bm{\sigma}$.}
		    \textbf{(A)} Prior and posterior distributions for the parameters, with the posterior distributions overlapped.
		    \textbf{(B)} Dataset B together with the model ensemble predictions using all plausible parameter values obtained via Bayesian inference (also overlapped). 
            Both posterior distribution and model ensemble prediction considering $\Delta=0.005$ and $M=5000$ reproduce those considering $\Delta=0.001$ and $M=10000$.
			(See also Figures' Supplementary Legends.)}
		\label{fig:Reproducibility-MM-custom}
	\end{figure}

\begin{figure}[p]
	\centering
	\centering\myfigure[0.88]{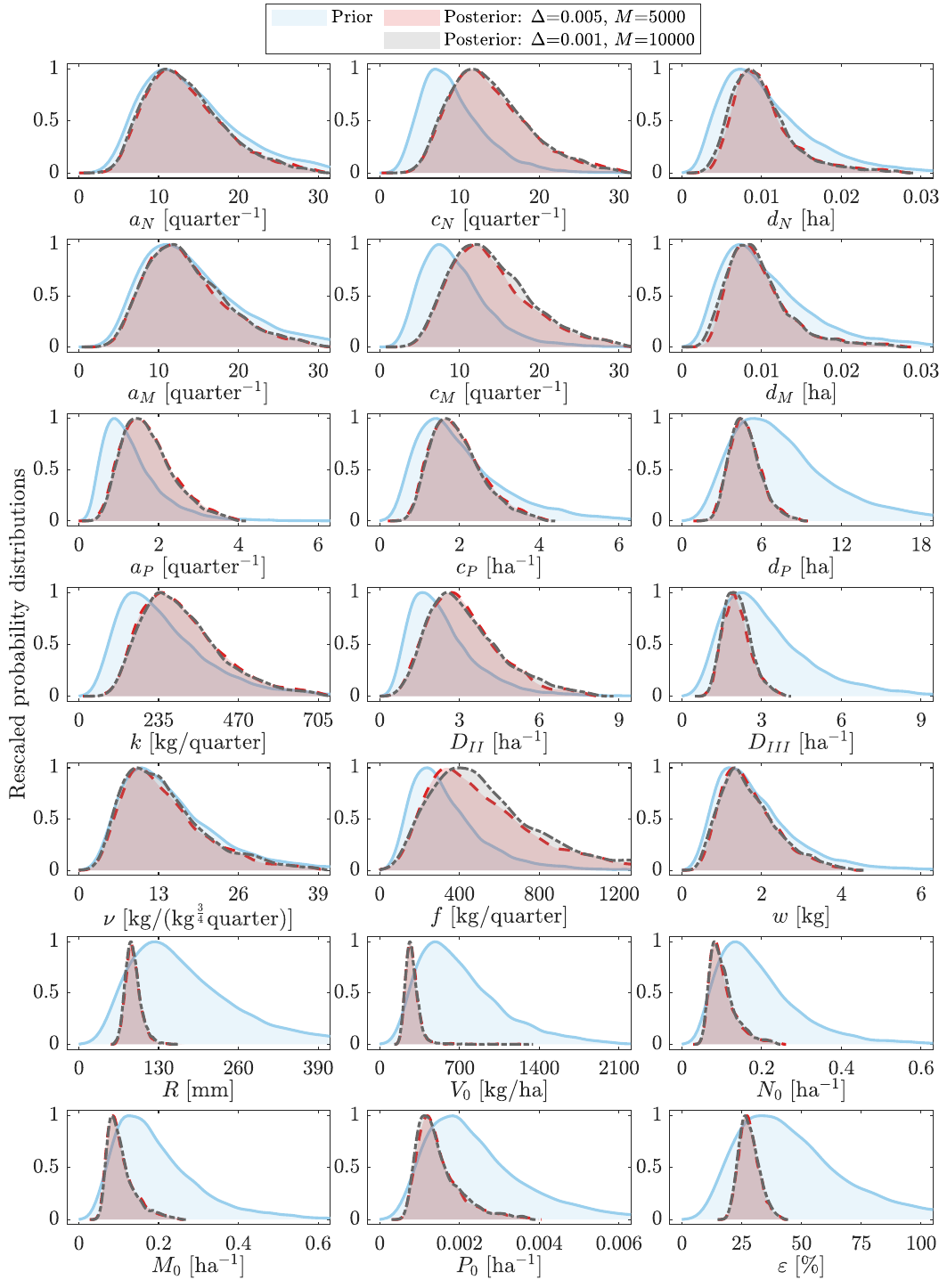}
	\caption{\textbf{Prior and posterior   distributions  for the parameters of the ecosystem network model~(Table~S1) by implementing the posterior sampling algorithm with two different combinations of the effective sample size reduction target $\bm{\Delta}$ and sample size $\bm{M}$ considering a vague multivariate log-normal prior for all parameters.}  
    Posterior distribution considering $\Delta=0.005$ and $M=5000$ reproduces that considering $\Delta=0.001$ and $M=10000$.
	(See also Figures' Supplementary Legends.)}
	\label{fig:Reproducibility-priors_network}
\end{figure}

	\begin{figure}[p]
		\centering\myfigure[0.7650]{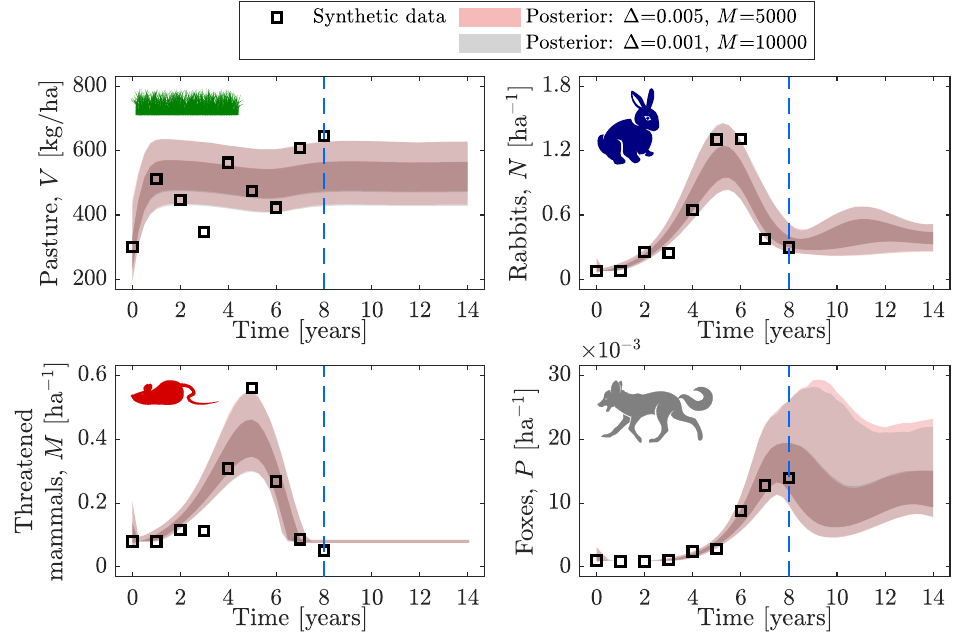}
		\caption{\textbf{Ecosystem network model fit to time-series data by implementing the posterior sampling algorithm with two different combinations of the effective sample size reduction target $\bm{\Delta}$ and sample size $\bm{M}$  considering a vague multivariate log-normal prior for all parameters.}
		Synthetic time-series data for ecological abundance with measurement error of  $\varepsilon=25\%$  together with the model ensemble predictions using all plausible parameter values~(Fig.~\ref{fig:Reproducibility-priors_network}).
        Model ensemble predictions considering $\Delta=0.005$ and $M=5000$ reproduce those considering $\Delta=0.001$ and $M=10000$.
	   (See also Figures' Supplementary Legends.)}
		\label{fig:Reproducibility-fit_network}
	\end{figure}
	
	\begin{figure}[p]
	\centering\myfigure[0.95]{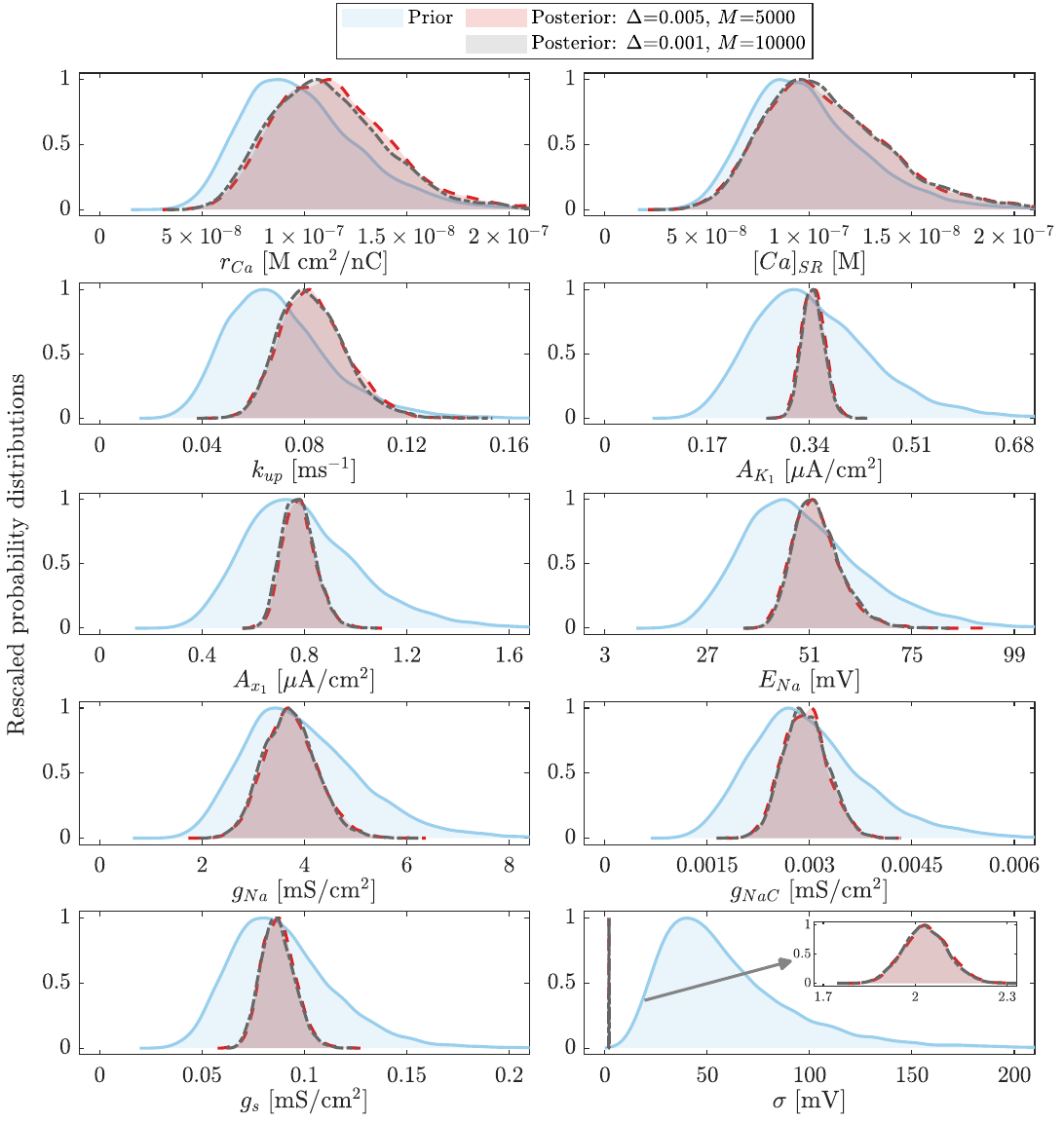}
	\caption{\textbf{Prior and posterior distributions  for the parameters of the Beeler-Reuter   model~(Table~S3) by implementing the posterior sampling algorithm with two different combinations of the effective sample size reduction target $\bm{\Delta}$ and sample size $\bm{M}$ considering a multivariate log-normal prior for all parameters.}
	Posterior distribution considering $\Delta=0.005$ and $M=5000$ reproduces that considering $\Delta=0.001$ and $M=10000$.
	(See also Figures' Supplementary Legends.)}
	\label{fig:Reproducibility-priors_Beeler}
\end{figure}

	\begin{figure}[p]
	\centering\myfigure[0.60]{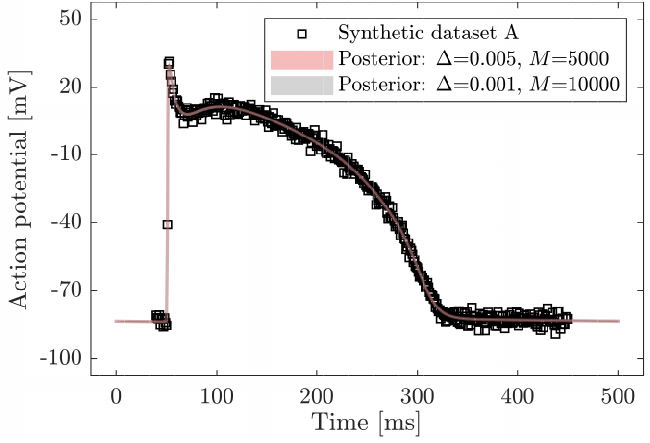}
	\caption{\textbf{Beeler-Reuter model fit to synthetic time-series data by implementing the posterior sampling algorithm with two different combinations of the effective sample size reduction target $\bm{\Delta}$ and sample size $\bm{M}$  considering a multivariate log-normal prior for all parameters.}
	Synthetic action potential (AP) data with measurement error of  $\sigma=2\mathrm{mV}$ and time resolution of $1\;\mathrm{ms}$ ($1\;\mathrm{kHz}$) together with the model ensemble predictions using all plausible parameter values~(Fig.~\ref{fig:Reproducibility-priors_Beeler}).
    Model ensemble predictions considering $\Delta=0.005$ and $M=5000$ reproduce those considering $\Delta=0.001$ and $M=10000$.
	(See also Figures' Supplementary Legends.)}
	\end{figure}

\FloatBarrier	

\invisiblesubsection{Ecosystem network model}

	{\newcolumntype{C}{>{\raggedright\arraybackslash}m{0.45\textwidth}}
		\newcolumntype{Y}{>{\centering\arraybackslash}m{0.09\textwidth}}			
		\newcolumntype{Z}{>{\arraybackslash}m{0.38\textwidth}}
		\begin{table}[p]
			\begin{tabular}{CYZ}
				\toprule
				\textbf{Equation} &  & \textbf{Description}  \\ 
				\toprule
				\multicolumn{3}{l}{\textbf{Pasture biomass ($\bm{V}$):}}\hspace{0.015\textwidth}\\
				$  \ufrac{\Delta V}{\Delta t}=
				r_V(t)-g_{N}(t)N-h_{M}(t)M$ & \tagarray\label{Eq. Pasture} & 	Pasture biomass at time $t$  \\
				 $r_V(t)=-55.12-0.0153V-0.00056V^2+2.5R$ & \tagarray\label{Eq. V change} & Pasture biomass growth rate   \\
				 $g_{N}(t)=h_M(t)=\nu w^{{3}/{4}}\left[1-\exp\left\lbrace{-\ufrac{V+r_V(t)\Delta t}{f}}\right\rbrace \right]$&  \tagarray\label{Eq. V consumption by N} & Pasture consumed by rabbits \& threatened species\\
				\midrule
				\multicolumn{3}{l}{\textbf{Rabbit density   ($\bm{N}$):}}\hspace{0.015\textwidth}\\
				$\ufrac{\Delta N}{\Delta t}=r_{N}(t)N-g_{P}(t)P,\;\;\;N\geq 0.08\;\mathrm{ha^{-1}}$ & \tagarray\label{Eq. Rabbit dynamics} &  Rabbit density at time $t$ \\
				 $r_{N}(t)=-a_{N}+c_{N}\left[1-\exp\left\lbrace -d_NV\right\rbrace \right]$&  \tagarray\label{Eq. N growth} & Numerical response of rabbits\\
				 $g_{P}(t)=\ufrac{\left[{k}/{w}\right]N^2}{N^2+D_{III}^2}$&  \tagarray\label{Eq. N death by P} & Functional response of foxes to rabbits\\
				\midrule
				\multicolumn{3}{l}{\textbf{Threatened species density ($\bm{M}$):}}\hspace{0.015\textwidth}\\
				 $ \ufrac{\Delta M}{\Delta t}=r_{M}(t)M-q_{P}(t)P,\;\;\;M\geq 0.08\;\mathrm{ha^{-1}}$ & \tagarray\label{Eq. Mammals dynamics} & Threatened species density at time $t$\\
				 $r_{M}(t)=-a_{M}+c_{M}\left[1-\exp\left\lbrace-d_MV\right\rbrace\right]$&  \tagarray\label{Eq. M growth} & Numerical response of  threatened species\\
				 $q_{P}(t)=h_P\left[1-\ufrac{g_{P}(t)}{k}\right]$ &  \tagarray\label{Eq. overall M death by P} & Fox predation rate on  threatened species\\
				 $h_P=\ufrac{\left[{k}/{w}\right]M}{M+D_{II}}$ &  \tagarray\label{Eq. M death by P} & Functional response of foxes to  threatened species\\
				\midrule
				\multicolumn{3}{l}{\textbf{Fox density ($\bm{P}$)}:}\hspace{0.015\textwidth}\\
				$\ufrac{\Delta P}{\Delta t}=r_{P}(t)P,\;\;\;P\geq 0.001\mathrm{ha^{-1}}$ & \tagarray\label{Eq. Fox dynamics} & Fox density at time $t$ \\
				$r_P(t)=-a_{P}+c_{P}\left[ 1-\exp\left\lbrace {-d_P(M+P)}\right\rbrace \right]$&  \tagarray\label{Eq. P growth} & Numerical response of foxes\\
				\bottomrule
			\end{tabular}\vspace{5pt}
			\caption{\textbf{Equations of the ecosystem network model.} Model developed by Pech~and~Hood~\CitePechA.}
			\label{tb:network-equations}
	\end{table}}

	{ \newcolumntype{C}{>{\centering\arraybackslash}p{0.075\textwidth}}
		\newcolumntype{Y}{>{\centering\arraybackslash}p{0.16\textwidth}}			
		\newcolumntype{Z}{>{\arraybackslash}p{0.59\textwidth}}
		\begin{table}
		\setlength\extrarowheight{1pt}
			\begin{threeparttable}[p]
				\begin{tabular}{CCYZ}
					\toprule
					\textbf{Symbol}  & $\bm{\theta_{R}}$  &  \textbf{Units} & \textbf{Description} \\
					\toprule
					$a_N$	&	$	4.60	$	&	$\mathrm{quarter^{-1}}$	&	Maximum rate of decrease of rabbit population in absence of food	\\
					$c_N$	&	$	5.50	$	&	$\mathrm{quarter^{-1}}$	&	Maximum  rate of increase of rabbit population when food is abundant	\\
					$d_N$	&	$	0.0045	$	&	$\mathrm{ha}$	&	Demographic efficiency of  rabbit population	\\
					$a_M$	&	$	4.60	$	&	$\mathrm{quarter^{-1}}$	&	Maximum rate of decrease of threatened species population in absence of food	\\
					$c_M$	&	$	5.50	$	&	$\mathrm{quarter^{-1}}$	&	Maximum  rate of increase of threatened species population when food is abundant	\\
					$d_M$	&	$	0.0045	$	&	$\mathrm{ha}$	&	Demographic efficiency of  threatened species population	\\
					$a_P$	&	$	0.56	$	&	$\mathrm{quarter^{-1}}$	&	Maximum rate of decrease of fox population in absence of food	\\
					$c_P$	&	$	0.77	$	&	$\mathrm{quarter^{-1}}$	&	Maximum  rate of increase of fox population when food is abundant	\\
					$d_P$	&	$	3.20	$	&	$\mathrm{ha}$	&	Demographic efficiency of  fox population	\\
					$k$	&	$	100.01	$	&	$\mathrm{{kg}/{quarter}}$	&	Maximum food consumption rate of fox population	\\
					$D_{II}$	&	$	0.99	$	&	$\mathrm{ha^{-1}}$	&	Limiting fox consumption rate when rabbit population is low 	\\
					$D_{III}$	&	$	1.32	$	&	$\mathrm{ha^{-1}}$	&	Density of rabbits at the inflection point	\\
					$\nu$	&	$	6.21	$	&	$\mathrm{{kg}/{(kg^{{3}/{4}}quarter)}}$   	&	Maximum (satiation) body-weight adjusted consumption rate per animal	\\
					$f$	&	$	138.00	$	&	$\mathrm{{kg}/{ha}}$	&	Biomass at which the herbivore's intake is depressed to $63\%$ of $\nu w^{{3}/{4}}$	\\
					$w$	&	$	0.78	$	&	$\mathrm{kg}$	&	Mean body weight of rabbits eaten by foxes	\\
					$R^\dagger$	&	$	74.48	$	&	$\mathrm{mm}$	&	Quarterly-averaged rainfall 	\\
					$V_0$	&	$	300.00	$	&	$\mathrm{{kg}/{ha}}$	&	Initial  pasture biomass	\\
					$N_0$	&	$	0.080	$	&	$\mathrm{ha^{-1}}$	&	Initial rabbit population	\\
					$M_0$	&	$	0.080	$	&	$\mathrm{ha^{-1}}$	&	Initial threatened species population	\\
					$P_0$	&	$	0.001	$	&	$\mathrm{ha^{-1}}$	&	Initial fox population	\\
					\bottomrule				
				\end{tabular}
			\end{threeparttable}\vspace{5pt}
			\caption{\textbf{Reference parameter values  $\bm{\theta_{R}}$ for the ecosystem network model.} Parameter  values reported by Pech~and~Hood~\CitePechA. $^\dagger$Parameter $R$ is calculated as the mean of the quarterly rainfall statistics for Lerida reported by  Pech~and~Hood~\CitePechA.}
			\label{tb:network-reference-parameters}
	\end{table}}
	
	\invisiblesubsection{Cardiac action potential (AP) model}

		{\newcolumntype{C}{>{\raggedright\arraybackslash}m{0.606\textwidth}}
		\newcolumntype{Y}{>{\centering\arraybackslash}m{0.09\textwidth}}
		\newcolumntype{Z}{>{\arraybackslash}m{0.233\textwidth}}
		\begin{table}[p]
			\begin{tabular}{CYZ}
				\toprule
				\textbf{Equation} &  & \textbf{Description}  \\ 
				\toprule
				\multicolumn{3}{l}{\textbf{Membrane potential ($\bm{V_m}$):}}\hspace{0.015\textwidth}\\
				$  \ufrac{\mathrm{d}V_{m}}{\mathrm{d}t}=-\ufrac{1}{C_m}\left[ i_{K_1}+i_{x_1}+i_{Na}+i_{s}-i_{stim} \right]$ & \tagarray\label{Eq. Action Potential} & Dynamics of the membrane potential  \\
				 $i_{K_1}=A_{K_1}\left[\ufrac{4\left(\exp\lbrace 0.04(V_{m}+85)\rbrace-1\right)}{\exp\lbrace 0.08(V_{m}+53)\rbrace+\exp\lbrace 0.04(V_{m}+53)\rbrace}+\ufrac{0.2\left(V_{m}+23\right)}{1-\exp\lbrace -0.04(V_{m}+23)\rbrace}\right]$ & \tagarray\label{Eq. Outward t-K} & Time-independent outward potassium  ion ($K^{+}$) current   \\
				 $i_{x_1}=A_{x_1}x_1\left[\ufrac{\exp\lbrace 0.04(V_{m}+77\rbrace-1}{\exp\lbrace 0.04(V_{m}+35)\rbrace}\right]$ & \tagarray\label{Eq. Outward v-K} &  Time- and voltage-dependent outward potassium ion ($K^{+}$) current   \\
				 $i_{Na}=\left(g_{Na}m^3hj+g_{NaC}\right)\left(V_{m}-E_{Na}\right)$ & \tagarray\label{Eq. Inward Na} & Two inward sodium ion ($Na^{+}$) current (fast and background)   \\
				 $i_{s}=g_{s}df\left(V_{m}-E_{s}\right)$ & \tagarray\label{Eq. Inward Ca} &  Slow inward calcium ion ($Ca^{2+}$) current   \\
				 $i_{stim}=\left\lbrace\begin{array}{ll}
	            A_{s}  &  t_{on} \leq t< t_{on}+t_{dur}\\
	            0 & \mbox{otherwise}
	            \end{array}\right.$ & \tagarray\label{Eq.Stim} &  Applied stimulus current   \\
				\midrule
				\multicolumn{3}{l}{\textbf{Intracellular calcium ion concentration  ($\bm{[Ca]_i}$):}}\hspace{0.015\textwidth}\\
				 $  \ufrac{\mathrm{d}[Ca]_i}{\mathrm{d}t}=-r_{Ca}i_{s}+k_{up}\left([Ca]_{SR-[Ca]_i}\right)$ & \tagarray\label{Eq. Free calcium} & Dynamics of intracellular calcium  ion ($Ca^{2+}$) concentration   \\
				 $E_{s}=-82.3-13.0287\ln{[Ca]_i}$ & \tagarray\label{Eq. V Ca} &  Equilibrium (Nernst) potential of calcium ion ($Ca^{2+}$)   \\
				\midrule
				\multicolumn{3}{l}{\textbf{Ion channel gating variables  ($\bm{x_1}$, $\bm{m}$, $\bm{h}$, $\bm{j}$, $\bm{d}$ and $\bm{f}$):}}\hspace{0.015\textwidth}\\[1ex]
                 $  \ufrac{\mathrm{d}x_1}{\mathrm{d}t}=\left(1-x_1\right)\left[\ufrac{0.0005\exp\lbrace 0.083(V_{m}+50)\rbrace}{\exp\lbrace 0.057(V_{m}+50)\rbrace+1}\right]-x_1\left[\ufrac{0.0013\exp\lbrace -0.06(V_{m}+20)\rbrace}{\exp\lbrace -0.04(V_{m}+20)\rbrace+1}\right]$ & \tagarray\label{Eq. x1} & Dynamics of potassium ion ($K^{+}$)  activation gate  \\
				 $ \ufrac{\mathrm{d}m}{\mathrm{d}t}=\left(1-m\right)\left[\ufrac{-\left(V_{m}+47\right)}{\exp\lbrace -0.1(V_{m}+47)\rbrace-1}\right]-m\left[40\exp\lbrace -0.056(V_{m}+72)\rbrace\right]$ & \tagarray\label{Eq. m} & Dynamics of  sodium ion ($Na^{+}$) activation gate   \\
				  $\ufrac{\mathrm{d}h}{\mathrm{d}t}=\left(1-h\right)\left[0.126\exp\lbrace -0.25(V_{m}+77)\rbrace\right]-h\left[\ufrac{1.7}{\exp\lbrace -0.082(V_{m}+22.5)\rbrace+1}\right]$ & \tagarray\label{Eq. h} &  Dynamics of  sodium ion ($Na^{+}$) fast inactivation gate  \\
				 $\ufrac{\mathrm{d}j}{\mathrm{d}t}=\left(1-j\right)\left[\ufrac{0.055\exp\lbrace -0.25(V_{m}+78)\rbrace}{\exp\lbrace -0.2(V_{m}+78)\rbrace+1}\right]-j\left[\ufrac{0.3}{\exp\lbrace -0.1(V_{m}+32)\rbrace+1}\right]$ & \tagarray\label{Eq. j} &  Dynamics of  sodium ion ($Na^{+}$) slow inactivation gate  \\
				 $\ufrac{\mathrm{d}d}{\mathrm{d}t}=\left(1-d\right)\left[\ufrac{0.095\exp\lbrace -0.01(V_{m}-5)\rbrace}{\exp\lbrace -0.072(V_{m}-5)\rbrace+1}\right]-d\left[\ufrac{0.07\exp\lbrace -0.017(V_{m}+44)\rbrace}{\exp\lbrace 0.05(V_{m}+44)\rbrace+1}\right]$ & \tagarray\label{Eq. d} &  Dynamics of  calcium ion ($Ca^{2+}$) activation gate \\
				 $\ufrac{\mathrm{d}f}{\mathrm{d}t}=\left(1-f\right)\left[\ufrac{0.012\exp\lbrace -0.008(V_{m}+28)\rbrace}{\exp\lbrace 0.15(V_{m}+28)\rbrace+1}\right]-f\left[\ufrac{0.065\exp\lbrace -0.02(V_{m}+30)\rbrace}{\exp\lbrace -0.2(V_{m}+30)\rbrace+1}\right]$ & \tagarray\label{Eq. f} &  Dynamics of  calcium ion ($Ca^{2+}$) inactivation gate \\
				\bottomrule
			\end{tabular}\vspace{5pt}
			\caption{\textbf{Equations of the cardiac action potential (AP) model~\CiteDokos.} Model developed by Beeler and Reuter~\CiteBeeler.}
			\label{tb:cardiac-equations}
	\end{table}}

		{\newcolumntype{B}{>{\centering\arraybackslash}p{0.02\textwidth}}
		\newcolumntype{C}{>{\centering\arraybackslash}p{0.07\textwidth}}
		\newcolumntype{D}{>{\centering\arraybackslash}p{0.10\textwidth}}
		\newcolumntype{Y}{>{\centering\arraybackslash}p{0.1\textwidth}}			
		\newcolumntype{Z}{>{\arraybackslash}p{0.59\textwidth}}
		\begin{table}
		 \aboverulesep=0ex
        \belowrulesep=0ex
        \setlength\extrarowheight{2pt}
				\begin{tabular}{B|CDYZ}
					\cmidrule[1pt]{1-5} 
					 \multicolumn{2}{r}{\textbf{Symbol}}  & $\bm{\theta_{R}}$  &  \textbf{Units} & \textbf{Description} \\
					\cmidrule[1pt]{1-5}
					\multirow{11}{\linewidth}{\rotatebox[origin=c]{90}{\textbf{Estimated model parameters}}} & $r_{Ca}$	&	$1\times 10^{-7}$	&	$\mathrm{M\;cm^2/nC}$	& Calcium ion  ($Ca^{2+}$) intracellular uptake rate		\\
					& $[Ca]_{SR}$	&	$1\times 10^{-7}$	&	$\mathrm{M}$	& Free calcium ion  ($Ca^{2+}$) concentration in the sarcoplasmic reticulum (SR)	\\
					& $k_{up} $	&	$0.07$	& 	$\mathrm{ms^{-1}}$	& Calcium ion  ($Ca^{2+}$) uptake rate  by the sarcoplasmic reticulum	\\
					& $A_{K_1}$	&	$0.35$	&	$\mathrm{\mu A/cm^2}$	& Maximal current density of time-independent outward potassium ion ($K^{+}$) current	\\
					& $A_{x_1}$	&	$0.8$	&	$\mathrm{\mu A/cm^2}$	& Maximal current density of time- and voltage-dependent outward potassium ion ($K^{+}$) current	\\
					& $E_{Na}$	&	$50$	&	$\mathrm{mV}$ & Equilibrium potential of sodium ion ($Na^{+}$) 		\\
					& $g_{Na}$	&	$4$	&	$\mathrm{mS/cm^2}$	& Conductance of background inward sodium ion ($Na^{+}$) current		\\
					& $g_{NaC}$	&	$0.003$	&	$\mathrm{mS/cm^2}$	&	Conductance of fast inward sodium ion ($Na^{+}$) current	\\
					& $g_s$	&	$0.09$	&	$\mathrm{mS/cm^2}$	& Conductance of slow inward calcium ion ($Ca^{2+}$) current \\
					\cmidrule{1-5} 
				\multirow{12}{\linewidth}{\rotatebox[origin=c]{90}{\textbf{Fixed model parameters}}}	& $A_s$	&	$40$	&	$\mathrm{\mu A/cm^2}$	&	Stimulus current	\\
					& $t_{on}$	&	$50$	&	$\mathrm{ms}$	& Stimulus start time	\\
					& $t_{dur}$	&	$1$	&	$\mathrm{ms}$	&	Stimulus duration \\
					& $C_m$	&	$1$	&	$\mathrm{\mu F/cm^2}$	&	Membrane capacitance	\\
					& $V_m(0)$	&	$-83.3$	&	$\mathrm{mV}$	& Initial membrane potential		\\
					& $[Ca]_i(0)$	&	$1.87\times10^{-7}$	&	$\mathrm{M}$	& Initial intracellular calcium ion ($Ca^{2+}$) concentration	\\
					& $x_1(0)$	&	$0.1644$	&	$\mathrm{Unitless}$	&	Initial value of the	potassium ion ($K^{+}$)  activation gate		\\
					& $m(0)$	&	$0.01$	&	$\mathrm{Unitless}$	&	Initial value of the sodium ion ($Na^{+}$) activation gate\\
					& $h(0)$	&	$0.9814$	&	$\mathrm{Unitless}$	& Initial value of the sodium ion ($Na^{+}$) fast inactivation gate		\\
					& $j(0)$	&	$0.9673$	&	$\mathrm{Unitless}$	&	Initial value of the sodium ion ($Na^{+}$) slow inactivation gate	\\
					& $d(0)$	&	$0.0033$	&	$\mathrm{Unitless}$	&	Initial value of the calcium ion ($Ca^{2+}$) activation gate 	\\
					& $f(0)$	&	$0.9884$	&	$\mathrm{Unitless}$	&	Initial value of the calcium ion ($Ca^{2+}$) inactivation gate	\\
					\cmidrule[1pt]{1-5} 				
				\end{tabular}
			\caption{\textbf{Reference (true) parameter values  $\bm{\theta_{R}}$ for the cardiac action potential (AP) model~\CiteDokos.} Parameter values originally reported by Beeler and Reuter~\CiteBeeler. Estimated model parameters correspond to those whose values are estimated in this work via Bayesian inference or  maximum likelihood estimation. Fixed model parameters are set to their reference value, and thus they are not estimated via our model-data fitting techniques.}
			\label{tb:cardiac-reference-parameters}
	\end{table}}